\documentclass[12pt]{article}

\usepackage{slashed,cite}
\usepackage{latexsym}
\usepackage{amsmath}
\usepackage{amssymb}
\usepackage{lscape}

\newcommand{\trace}{\mathop{\rm Tr}\nolimits}
\newcommand{\rank }{\mathop{\rm rank}\nolimits}

\def\nn{\nonumber}

\csname @addtoreset\endcsname{equation}{section}
\newcommand{\ft}[2]{{\textstyle\frac{#1}{#2}}}
\def\rmi{{\,\rm i\,}}
\newsavebox{\uuunit}
\sbox{\uuunit}
    {\setlength{\unitlength}{0.825em}
     \begin{picture}(0.6,0.7)
        \thinlines
        \put(0,0){\line(1,0){0.5}}
        \put(0.15,0){\line(0,1){0.7}}
        \put(0.35,0){\line(0,1){0.8}}
       \multiput(0.3,0.8)(-0.04,-0.02){12}{\rule{0.5pt}{0.5pt}}
     \end {picture}}
\newcommand {\unity}{\mathord{\!\usebox{\uuunit}}}
\newcommand{\J}[3]{J^{#1} {}_{#2} {}^{#3}}

\def\g{\gamma}

\def\d{\delta}
\def\D{\Delta}
\def\e{\epsilon}
\def\ve{\varepsilon}
\def\f{\phi}
\def\vf{\varphi}

\def\p{\psi}

\def\l{\lambda}
\def\L{\Lambda}
\def\m{\mu}
\def\n{\nu}
\def\r{\rho}
\def\s{\sigma}

\def\o{\omega}
\def\O{\Omega}

\newcommand{\SU}{\mathop{\rm SU}}
\newcommand{\SO}{\mathop{\rm SO}}
\newcommand{\U}{\mathop{\rm {}U}}
\newcommand{\USp}{\mathop{\rm {}USp}}

\newcommand{\Symp}{\mathop{\rm {}Sp}}
\newcommand{\Sl}{\mathop{\rm {}S}\ell }
\newcommand{\Gl}{\mathop{\rm {}G}\ell }

\newtheorem{lemma}{Lemma}

\newcommand{\QED}{{\hspace*{\fill}\rule{2mm}{2mm}}}
\def\Ric{R}
\newcommand{\covder}{\mathfrak{D}}
\begin{document}

\begin{titlepage}

\begin{flushright}
  UG-02-37 \\
  KUL-TF-02/04 \\
  SPIN-2002/14 \\
  ITP-UU-02/22 \\
  hep-th/0205230
\end{flushright}

\begin{center}
\vspace{.3cm} \baselineskip=16pt {\LARGE \bf Superconformal $N=2$, $D=5$
matter \\[3mm] with and without actions}
 \vskip 0.6 cm {\large
  Eric Bergshoeff$^1$, Sorin Cucu$^2$, Tim de Wit$^1$, \\[2mm]
Jos Gheerardyn$^2$, Rein Halbersma$^1$, Stefan Vandoren$^3$ \\[2mm]
and Antoine Van Proeyen$^2$
} \\
\vskip 6 mm {\small
  $^1$ Center for Theoretical Physics, University of Groningen,\\
       Nijenborgh 4, 9747 AG Groningen, The Netherlands. \\
       \{e.bergshoeff, t.c.de.wit, r.halbersma\}@phys.rug.nl \\[3mm]
  $^2$ Instituut voor Theoretische Fysica, Katholieke Universiteit Leuven,\\
       Celestijnenlaan 200D B-3001 Leuven, Belgium. \\
       \{sorin.cucu, jos.gheerardyn, antoine.vanproeyen\}@fys.kuleuven.ac.be \\[3mm]
  $^3$ Institute for Theoretical Physics, Utrecht University, \\
 Leuvenlaan 4, 3508 TA Utrecht, The Netherlands. \\
      s.vandoren@phys.uu.nl
}
\end{center}

\centerline{ABSTRACT}
\bigskip

We investigate $N=2$, $D=5$ supersymmetry and matter-coupled supergravity
theories in a superconformal context. In a first stage we do not require
the existence of a lagrangian. Under this assumption, we already find  at
the level of rigid supersymmetry, i.e.~\emph{before} coupling to
conformal supergravity, more general matter couplings than have been
considered in the literature. For instance, we construct new
vector-tensor multiplet couplings, theories with an \emph{odd} number of
tensor multiplets,  and hypermultiplets whose scalar manifold geometry is
\emph{not} hyperk{\"a}hler.

Next, we construct rigid superconformal lagrangians. This requires some
extra ingredients that are not available for all dynamical systems.
However, for the generalizations with tensor multiplets mentioned above,
we find corresponding new actions and scalar potentials. Finally, we
extend the supersymmetry to local superconformal symmetry, making use of
the Weyl multiplet. Throughout the paper, we will indicate the various
geometrical concepts that arise, and as an application we compute the
non-vanishing components of the Ricci tensor of hypercomplex group
manifolds. Our results can be used as a starting point to obtain more
general matter-couplings to Poincar{\'e} supergravity.

\end{titlepage}

\tableofcontents
\section{Introduction}

Recently, much attention has been given to $D=5$ matter-coupled
supergravity theories~\cite{Gunaydin:1999zx,Ceresole:2000jd}, thereby
generalizing the earlier results
of~\cite{Gunaydin:1984bi,Gunaydin:1985ak}. This is mainly due to the fact
that matter couplings in five dimensions play an important role in
theories with large extra
dimensions~\cite{Arkani-Hamed:1998rs,Antoniadis:1998ig,Randall:1999vf,Randall:1999ee}.
In particular, the properties of the scalar potential determine whether
or not a supersymmetric Randall-Sundrum (RS)
scenario~\cite{Randall:1999vf,Randall:1999ee} is possible. The
possibility of such a supersymmetric RS scenario relies on the existence
of a domain-wall solution containing a warp factor with the correct
asymptotic behaviour such that gravity is suppressed in the transverse
direction. It turns out that constructing such a domain-wall solution is
nontrivial.

With only vector multiplets and no singular source insertions, a no-go
theorem was established for smooth domain-wall
solutions~\cite{Kallosh:2000tj,Behrndt:2000tr}. It has been shown that
solutions acceptable for a supersymmetric RS scenario can be found
provided one allows for branes as singular
insertions~\cite{Bergshoeff:2000zn}. Another approach is to include
hypermultiplets~\cite{Behrndt:2000km,Behrndt:2000ph,Behrndt:2001qa}. The
general mixing of vector and hypermultiplets was considered
in~\cite{Ceresole:2001wi}, and its possibilities were further analysed
in~\cite{Alekseevsky:2001if}. It seems that with such general matter
couplings there is no \emph{a priori} obstruction for a supersymmetric RS
scenario, although an acceptable smooth solution has not yet been found.
Improvements in the last year involve curved
branes~\cite{LopesCardoso:2001rt,Chamseddine:2001ga,Chamseddine:2001hx,Cardoso:2002ec}
and the use of non-homogeneous quaternionic spaces~\cite{Behrndt:2001km}.

Matter-coupled $D=5$ supergravity theories also play an important role in
$AdS_6/CFT_5$ \cite{Nishimura:2000wj} and
$AdS_5/CFT_4$~\cite{D'Auria:2000ad} correspondences. In particular, the
$D=5$ domain-wall solutions describe the renormalization group flow of
the corresponding four-dimensional field theory. The geometrical warp
factor now plays the role of an energy scale. The structure of the domain
wall is determined by the properties of the scalar potential. Finally,
domain wall solutions have been applied to cosmology in the context of
e.g.~inflation~\cite{Kallosh:2001gr} and
quintessence~\cite{Townsend:2001ea}. In this context, it is important to
find out what the detailed properties of the scalar potential are, and
which kind of domain walls they give rise to.

The reasons given above motivated us to reconsider matter couplings in
five dimensions, to independently derive the most general $D=5$ matter
couplings of~\cite{Ceresole:2000jd} and, perhaps, to find more general
matter couplings. Our strategy was to use the so-called conformal
approach~\cite{Kaku:1977pa, Ferrara:1977ij,Kaku:1978ea,Kaku:1978nz}. An
advantage of the conformal construction is that, by past experience, it
leads to insights into the structure of the matter couplings. A recent
example is the insight in relations between hyperk{\"a}hler cones and
quaternionic manifolds, based on the study of superconformal matter
couplings with hypermultiplets~\cite{deWit:1999fp,deWit:2001dj}.

In~\cite{Bergshoeff:2001hc,Fujita:2001kv}, the first step in the
conformal programme has been performed by constructing the Weyl
multiplets of $N=2$ conformal supergravity in five dimensions. The
purpose of this paper is to take the next step in the conformal programme
and introduce the different $D=5$ matter multiplets with 8 conformal
supersymmetries together with the corresponding actions (when they
exist). Similar steps have been performed
in~\cite{Kugo:2000hn,Kugo:2000af,Fujita:2001kv}. These authors also
constructed off-shell superconformal multiplets. We will be able to
generalize their results by not restricting ourselves to off-shell
multiplets. Especially for the hypermultiplets this is important, as
general quaternionic manifolds are not obtained from an off-shell
calculus. In this context we should also mention earlier work on
(non-conformal) on-shell multiplets by
Zucker~\cite{Zucker:1999ej,Zucker:1999fn}. In a next paper, we will take
the last step in the conformal programme and impose different
gauge-fixings. This will give us the $D=5$ matter couplings we are aiming
at. It was recently~\cite{Kugo:2002js} shown how this method can be
applied in the context of the RS scenario, for coupling the $D=5$ bulk
supergravity to $D=4$ brane matter multiplets in a superconformal
invariant way. We hope that our more general results may also be helpful
in these investigations.

There is a rather different, more general, motivation of why the $D=5$
matter-coupled supergravities are interesting to study. The  reason is
that they belong to the class of theories with eight
supersymmetries~\cite{VanProeyen:2001wr}. Such theories are especially
interesting since the geometry, determined by the kinetic terms of the
scalars, contains undetermined functions. Theories with 32
supersymmetries have no matter multiplets while the geometry of those
with 16 supersymmetries is completely determined by the number of matter
multiplets. Of course, theories with 4 supersymmetries allow for more
general geometries. The restricted class of geometries, in the case of 8
supersymmetries, makes these theories especially interesting and
manageable. For instance, the work of Seiberg and
Witten~\cite{Seiberg:1994rs,Seiberg:1994aj} heavily relies on the
presence of 8 supersymmetries. Theories with 8 supersymmetries are thus
the maximally supersymmetric theories that, on the one hand, are not
completely determined by the number of matter multiplets in the model
and, on the other hand, allow arbitrary functions in their definition,
i.e.~continuous deformations of the metric of the manifolds.

The geometry related to supersymmetric theories with 8 real supercharges
is  called `special geometry'. Special geometry was first found
in~\cite{deWit:1984rz,deWit:1984pk} for local supersymmetry and
in~\cite{Sierra:1983cc,Gates:1984py} for rigid supersymmetry. It occurs
in Calabi-Yau compactifications of type II superstrings as the moduli
space of these
manifolds~\cite{Seiberg:1988pf,Cecotti:1989qn,Ferrara:1989vp,Candelas:1991pi,
Candelas:1991qd,Strominger:1990pd}. Special geometry was a very useful
tool in the investigation of supersymmetric black
holes~\cite{Kallosh:1992ii,Ferrara:1995ih}. The work of Seiberg and
Witten~\cite{Seiberg:1994rs,Seiberg:1994aj} was based on the use of
(rigid) special geometry. Later, the AdS/CFT
correspondence~\cite{Aharony:1999ti} gave new applications of special
geometry. So far, special geometry had been mainly investigated in the
context of four dimensions. In the context of M theory compactifications
on a Calabi-Yau~\cite{Cadavid:1995bk}, and with the advent of the
brane-world scenarios~\cite{Randall:1999vf,Randall:1999ee}, also the
$D=5$ variant of special geometry~\cite{Gunaydin:1984bi}, called `very
special geometry', received a lot of attention. The connection to special
geometry was made in~\cite{deWit:1992cr}. Last but not least,
mathematicians got interested in special geometry due to its relation
with quaternionic geometry~\cite{Cecotti:1989qn}, which lead to new
results on the classification of homogeneous quaternionic
spaces~\cite{deWit:1992nm,Cortes}.

We mentioned already that a conformal tensor calculus for $D=5$ matter
multiplets with 8 supersymmetries has already been introduced
in~\cite{Kugo:2000hn,Kugo:2000af,Fujita:2001kv}. However, there are still
some ingredients missing: in particular the geometrical features have not
been discussed at the most general level. In this paper, we use
superconformal methods to fill this gap. We start with listing the basic
superconformal matter multiplets: vector/tensor multiplets, linear
multiplets and hypermultiplets. Some of these multiplets are off-shell,
others imply equations of motion that define dynamical models. The
closure of the algebra leads to equations that determine the evolution of
the fields. In fact, by now we are used to handle theories without
starting from a bona-fide action. Indeed, this is the way in which we
often work with IIB supergravity, or theories with self-dual
antisymmetric tensor fields. Therefore, rather than starting to analyse
the most general matter couplings from looking for invariant actions, we
first can start the analysis of the multiplets, which in some cases
already gives dynamical systems. The latter allow more general matter
couplings than those constructed from a lagrangian.

In particular, we will not only introduce vector multiplets in the
adjoint representation but, more generally, so-called `vector-tensor'
multiplets in arbitrary representations. This includes couplings with an
\emph{odd} number of tensor multiplets. This may generalize the analysis
made e.g.\ recently in~\cite{Ellis:2001xd}. Furthermore, as far as the
hypermultiplets are concerned, we will introduce more general geometries
than hyperk{\"a}hler for rigid supersymmetry, or quaternionic-K{\"a}hler for
supergravity. We can find dynamical theories also without the need of an
action, i.e.\ in hypercomplex geometry, which is hyperk{\"a}hler geometry
where there may not be a metric. Also in $N=8$ theories in 5 dimensions,
more general possibilities were found in~\cite{Andrianopoli:2000fi} by
considering theories where the dynamical equations are considered without
the necessity of an action.

In a second step, we construct rigid superconformal lagrangians. This
will require an extra ingredient, namely the existence of a certain
covariant tensor, that is not available for all dynamical systems and
leads to a restriction on the possible geometries. In a last step, we
will extend the supersymmetry to a local conformal supersymmetry, making
use of the Weyl multiplet constructed
in~\cite{Bergshoeff:2001hc,Fujita:2001kv}.

The first two steps discussed above only deal with the case of rigid
conformal supersymmetry. This case is sufficient to explain most of the
subtleties concerning the possible geometrical structures. It is only at
the last step that we introduce the full complications of coupling the
matter multiplets to conformal supergravity.

The paper is organised as follows. First, in section~\ref{ss:multRigid},
we perform step one and list the basic superconformal matter multiplets.
We construct and discuss the possible matter couplings in the absence of
a lagrangian. Next, in section~\ref{ss:actions}, we perform the second
step and construct rigid superconformal lagrangians. We discuss the
restrictions on the possible geometries that follow from the requirement
of a lagrangian. Finally, in section~\ref{ss:localmult}, we perform the
last step and extend the supersymmetry to local superconformal symmetry,
making use of the Weyl multiplet constructed
in~\cite{Bergshoeff:2001hc,Fujita:2001kv}. Our aim is twofold: we want to
determine and deduce the various restrictions from supersymmetry, and we
want to determine the independent geometrical quantities that are needed
for constructing superconformal matter theories. Our results can be used
as a starting point to obtain more general matter couplings to Poincar{\'e}
supergravity.

In a first appendix, we mention the linear multiplet, which does not play
a big role in our paper. Appendix~\ref{hyperappendix} gives a summary of
the properties of hypercomplex manifolds and their place in the family of
quaternionic-like manifolds. Explicit examples of hypercomplex manifolds
that are not hyperk{\"a}hler are given in appendix~\ref{app:exHypercplxGr}.
In that last appendix we calculate explicitly the non-vanishing
antisymmetric Ricci tensor for these manifolds, which is also a new
result.

The conventions that we use are given
in~\cite[appendix~A]{Bergshoeff:2001hc}.

\section{Multiplets of rigid conformal supersymmetry}
\label{ss:multRigid}

In this section, we will introduce the basic superconformal matter
multiplets. We start with giving a short review of rigid conformal
supersymmetry in the first subsection. For a more extended discussion, see
e.g.~\cite{VanProeyen:1999ni}. In the remaining subsections, we will
discuss the various multiplets: the vector-tensor multiplet, the linear
multiplet and the hypermultiplet.

\subsection{Definition of rigid conformal (super-)symmetry}
\label{ss:defnRigidConf}

We first introduce conformal symmetry and in a second step extend this to
conformal supersymmetry. Given a spacetime with a metric tensor
$g_{\mu\nu}(x)$, the conformal transformations are defined as the general
coordinate transformations that leave ``angles'' invariant. The
parameters of these special coordinate transformations define a conformal
Killing vector $k^\mu(x)$. The defining equation for this conformal
Killing vector is given by
\begin{equation}
\d_{\rm g.c.t.} (k) g_{\mu\nu}(x) \equiv \nabla_\mu k_\nu(x) +
\nabla_\nu  k_\mu (x) = \omega(x) g_{\mu\nu}(x)\, , \label{cve}
\end{equation}
where $\omega(x)$ is an arbitrary function, $k_\mu = g_{\mu\nu}k^\nu$ and
the covariant derivative is given by $\nabla_\mu k_\nu = \partial_\mu
k_\nu - \Gamma_{\mu\nu}^\rho k_\rho$. In flat $D$-dimensional Minkowski
spacetime,~(\ref{cve}) implies
\begin{equation}
\partial_{(\mu}k_{\nu)}(x) - \frac 1D \eta_{\mu\nu}\partial_\rho k^\rho (x)
= 0\,. \label{eq:conf_killing_eqn}
\end{equation}
In dimensions $D>2$, the conformal algebra is finite-dimensional. The
solutions of~(\ref{eq:conf_killing_eqn}) are given by
\begin{equation}
k^\mu(x)=\xi^\mu +\lambda_M^{\mu\nu}x_\nu+\lambda_D x^\mu
+\left(x^2\Lambda_K^\mu-2x^\mu x\cdot \Lambda_K\right). \label{ximu}
\end{equation}
Corresponding to the parameters $\xi^\mu$ are the translations $P_\mu$,
the parameters $\lambda_M^{\mu\nu}$ correspond to Lorentz rotations
$M_{\mu\nu}$, to $\lambda_D$ are associated the dilatations $D$, and
$\Lambda_K^\mu$ are the parameters of `special conformal transformations'
$K_\mu$. Thus, the full set of conformal transformations $\delta_C$ can
be expressed as follows:
\begin{equation}
\delta_C= \xi^\mu  P_\mu + \lambda_M^{\mu\nu}M_{\mu\nu}+ \lambda_D D +
\Lambda_K^\mu K_\mu \,.
\end{equation}
The commutators between different generators define the conformal algebra
which is isomorphic to the algebra of $\SO(D,2)$.

We wish to consider representations of the conformal algebra on fields
$\phi^\alpha(x)$ where $\alpha$ stands for a collection of internal
indices referring to the stability subalgebra of $x^\mu=0$. From the
expression~(\ref{ximu}) for the conformal Killing vector, we deduce that
this algebra is isomorphic to the algebra generated by $M_{\mu\nu}, D$
and $K_\mu$. We denote the generators of this stability subalgebra by
$\Sigma_{\mu\nu}, \Delta$ and $\kappa_\mu$. Applying the theory of
induced representations, it follows that any representation $(\Sigma,
\Delta, \kappa)$ of the stability subalgebra induces a representation of
the full conformal algebra with the following transformation rules (we
suppress any internal indices):
\begin{eqnarray}
\d_P \phi (x)&=&\xi^\mu \partial_\mu \phi(x)\, ,\nonumber\\
 \d_M \phi
(x)&=&\frac 12 \lambda_M^{\mu\nu}(x_\nu\partial_\mu - x_\mu
\partial_\nu) \phi(x) + \d_\Sigma(\lambda_M)\phi(x)\, ,\nonumber\\
 \d_D
\phi(x)&=&\lambda_D x^\lambda \partial_\lambda \phi (x) + \d_\Delta
(\lambda_D)\phi(x)\, , \nonumber\\ \d_K
\phi(x)&=&\lambda_K^\mu(x^2\partial_\mu - 2 x_\mu
x^\lambda\partial_\lambda)\phi(x)\ + \nonumber\\ &&{}+\Bigl (\d_\D (-2
x\cdot \L_K) + \d_\Sigma (-4x_{[\mu}\lambda_{K\nu]}) + \d_\kappa
(\lambda_K)\Bigr ) \phi(x)\, . \label{xdepP}
\end{eqnarray}

We now look at the non-trivial representation $(\Sigma, \Delta, \kappa)$
that we use in this paper. First, concerning the Lorentz representations,
in this paper we will encounter anti-symmetric tensors $\phi_{a_1\cdots
a_n}(x)\ (n=0,1,2, \ldots)$ and spinors $\psi_\alpha(x)$:
\begin{eqnarray}
\d_\Sigma (\l_M) \phi_{a_1\cdots a_n}(x) &=& -n (\l_M)_{[a_1}{}^b
\phi_{|b|a_2 \cdots a_n]}(x) \,, \nonumber\\
\d_\Sigma (\l_M) \psi(x) &=& -\frac 14 \l^{ab}_M \g_{ab} \psi(x) \,.
\end{eqnarray}

Second, we consider the dilatations. For most fields, the $\Delta $
transformation is just determined by a number $w$, which is called the
Weyl weight of $\phi^\alpha$:
\begin{equation}
\d_\Delta(\l_D) \phi^\alpha(x) = w \l_D \phi^\alpha(x) \,.
\label{simplediltr}
\end{equation}
For scalar fields, it is often convenient to consider the set of fields
 $\phi^\alpha$ as the coordinates of a scalar manifold with
affine connection $\Gamma_{\alpha\beta}{}^\gamma$. With this
understanding, the transformation of $\phi^\alpha$ under dilatations can
be characterized by:
\begin{equation}
\delta_\Delta(\l_D) \phi^\alpha = \lambda_D k^\alpha (\phi) \,.
\label{diltr}
\end{equation}
Requiring dilatational invariance of kinetic terms determined by a metric
$g_{\alpha \beta }$, the vector $k^\alpha $ should be a homothetic
Killing vector, i.e.\ it should satisfy the conformal Killing
equation~(\ref{cve}) for \emph{constant}  $\omega(x)$:
\begin{equation}
\covder_\alpha k_\beta + \covder_\beta k_\alpha = (D-2) g_{\alpha
\beta}\, , \label{hkv}
\end{equation}
where $D$ denotes the spacetime dimension and $\covder_\alpha k_\beta =
\partial_\alpha k_\beta - \Gamma_{\alpha\beta}{}^\gamma k_\gamma$.
However,~(\ref{xdepP}) shows that the $\Delta $-transformation also
enters in the special conformal transformation. It turns out that
invariance of the kinetic terms under these special conformal
transformations restricts $k^\alpha(\phi)$ further to a so-called
\emph{exact} homothetic Killing vector, i.e.,
\begin{equation}
k_\alpha = \partial_\alpha \chi\, , \label{exact}
\end{equation}
for some function $\chi(\phi )$. One can show that the
restrictions~(\ref{hkv}) and~(\ref{exact}) are equivalent~to
\begin{equation}
\covder_\alpha k^\beta \equiv \partial_\alpha k^\beta +
\Gamma_{\alpha\gamma}{}^\beta  k^\gamma  = w \delta_\alpha{}^\beta\,.
\label{homothetic}
\end{equation}
The constant $w$ is identified with the Weyl weight of $\phi^\alpha$ and
is in general $w = (D-2)/2$, i.e.\ $3/2$ in our case. The proof of the
necessity of~(\ref{homothetic}) can be extracted
from~\cite{Sezgin:1995th}, see also~\cite{deWit:1998zg,Michelson:1999zf}.
In these papers the conditions for conformal invariance of a sigma model
with either gravity or supersymmetry are investigated. By restricting the
proof to rigid conformal symmetry (without supersymmetry) we find the
same conditions.

Note that the condition~(\ref{homothetic}) can be formulated
\emph{independent} of a metric. Only an affine connection is necessary.
Indeed, we will find the same condition from the closure of the
superconformal algebra before any metric and/or action has been
introduced. In four spacetime dimensions, this was done
in~\cite{deWit:1998zg}.

For the special case of a zero affine connection, the homothetic Killing
vector is given by $k^\alpha = w \phi^\alpha$ and the transformation
rule~(\ref{diltr}) reduces to $\delta_\Delta(\l_D) \phi^\alpha = w
\lambda_D \phi^\alpha$. Note that the homothetic Killing vector $k^\alpha
= w \phi^\alpha$ is indeed exact with $\chi$ given by
\begin{equation}
\chi = \frac1{(D-2)}k^\alpha g_{\alpha\beta} k^\beta\, . \label{valuechi}
\end{equation}

Finally, all fields that we will discuss in this paper are invariant
under the internal special conformal transformations, i.e.~$\delta_\kappa
\phi^\alpha = 0$.

We next consider the extension to conformal supersymmetry. The parameters
of these supersymmetries define a conformal Killing spinor
$\epsilon^i(x)$ whose defining equation is given~by
\begin{equation}
 \nabla_\mu \e^i(x) -\frac 1D
\gamma_\mu \gamma^\nu\nabla_\nu \e^i(x)=0 \, .
\end{equation}
In $D$-dimensional Minkowski spacetime this equation implies
\begin{equation}
\partial_\mu \epsilon^i(x) - \frac 1D \gamma_\mu \slashed{\partial}\epsilon^i(x) = 0\, .
\end{equation}
The solution to this equation is given by
\begin{equation}
\epsilon^i(x) = \epsilon^i + \rmi x^\mu \gamma_\mu \eta^i\, ,
\label{susyc}
\end{equation}
where the (constant) parameters $\epsilon^i$  correspond to ``ordinary''
supersymmetry transformations $Q_\alpha^i$ and the parameters $\eta^i$
define special conformal supersymmetries generated by $S_\alpha^i$. The
conformal transformation~(\ref{ximu}) and the
supersymmetries~(\ref{susyc}) do not form a closed algebra. To obtain
closure, one must introduce additional R-symmetry generators. In
particular, in the case of 8 supercharges $Q_\alpha^i$ in $D=5$, there is
an additional $\SU(2)$ R-symmetry with generators $U_{ij}= U_{ji}\
(i=1,2)$. Thus, the full set of superconformal transformations $\delta_C$
is given by:
\begin{equation}
\delta_C= \xi ^\mu  P_\mu + \lambda_M^{\mu\nu}M_{\mu\nu}+\lambda_D D +
\Lambda_K^\mu K_\mu + \Lambda^{ij} U_{ij} + \rmi \bar{\e} Q + \rmi
\bar{\eta} S \,.
\end{equation}
We refer to~\cite{Bergshoeff:2001hc} for the full superconformal algebra
$F^2(4)$ formed by (anti-)commutators between the (bosonic and fermionic)
generators.

To construct field representations of the superconformal algebra, one can
again apply the method of induced representations. In this case one must
use superfields $\Phi^a(x^\mu,\theta_\alpha^i)$, where $a$ stands for a
collection of internal indices referring to the stability subalgebra of
$x^\mu = \theta_\alpha^i = 0$. This algebra is isomorphic to the algebra
generated by $M_{\mu\nu}, D, K_\mu, U_{ij}$ and~$S_\alpha^i$.

An additional complication, not encountered in the bosonic case, is that
the representation one obtains is reducible. To obtain an irreducible
representation, one must impose constraints on the superfield. It is at
this point that the transformation rules become nonlinear in the fields.
In this paper, we will follow a different approach. Instead of working
with superfields we will work with the component ``ordinary'' fields. The
different nonlinear transformation rules are obtained by imposing the
superconformal algebra.

In the supersymmetric case, we must specify the $\SU(2)$-properties of the
different fields as well as the behaviour under $S$-supersymmetry.
Concerning the $\SU(2)$, we will only encounter scalars $\phi$, doublets
$\psi^i$ and triplets $\phi^{(ij)}$ whose transformations are given by
\begin{eqnarray}
\d_{\rm SU(2)} (\Lambda^{ij}) \phi &=& 0\, , \nonumber\\
\d_{\rm SU(2)} (\Lambda^{ij}) \psi^i(x) &=& -\Lambda^i{}_j \psi^j (x) \,,
\nonumber\\ \d_{\rm SU(2)} (\Lambda^{ij}) \phi^{ij} (x) &=& -2
\Lambda^{(i}{}_k\phi^{j)k}(x)\, .
\end{eqnarray}
The scalars of the hypermultiplet will also have an $\SU(2)$
transformation despite the absence of an $i$ index. We refer for that to
section~\ref{ss:schyper}.

This leaves us with specifying how a given field transforms under the
special supersymmetries generated by $S_\alpha^i$. In superfield language
the full $S$-transformation is given by a combination of an $x$-dependent
translation in superspace, with parameter $\epsilon^i(x) = \rmi x^\mu
\gamma_\mu \eta^i$, and an internal $S$-transformation. This is in
perfect analogy to the bosonic case. In terms of component fields, the
same is true. The $x$-dependent contribution is obtained by making the
substitution
\begin{equation}
  \epsilon^i \rightarrow \rmi\slashed{x}\eta^i
 \label{xdepQ}
\end{equation}
in the $Q$-supersymmetry rules. The internal $S$-transformations can be
deduced by imposing the superconformal algebra. In the next three
subsections, we will give the explicit form of these internal
$S$-transformations for different matter multiplets.

Finally, we give below some of the commutators of the (rigid)
superconformal algebra expressed in terms of commutators of variations of
the fields. These commutators are realized on \emph{all} matter
multiplets discussed in the next subsections. The commutators between
$Q$- and $S$-supersymmetry are given by
\begin{eqnarray}
[ \d_Q(\e_1) , \d_Q(\e_2)] &=& \d_{P} \left(\frac 12 \bar{\e}_2 \g_\m
\e_1\right) ,  \label{QQcomm}  \\
 \left[\d_S(\eta),\d_Q(\e)\right] &=& \d_D\left( \frac 12\rmi \bar\e\eta \right) + \d_M\left( \frac 12\rmi \bar\e\g^{ab} \eta\right) +
   \d_U\left(  -\frac 32\rmi \bar\e^{(i} \eta^{j)} \right), \label{SQcomm} \\
\left[ \d_S (\eta_1), \d_S(\eta_2) \right] &=& \d_K\left( \frac 12
\bar\eta_2 \g^a\eta_1 \right). \label{SScomm}
\end{eqnarray}
For later use we list a few more commutators:
\begin{eqnarray}
\left[\delta_D(\Lambda_D),\delta_Q(\epsilon^i)\right] &=&
\delta_Q\left(\frac 12\epsilon^i\Lambda_D \right), \label{DQcomm}\\
\left[\delta_{SU(2)}(\Lambda^{ij}),\delta_Q(\epsilon^k)\right] &=&
\delta_Q\left(\epsilon^j\Lambda_j{}^i\right), \label{SU2Qcomm}\\
\left[\delta_K(\Lambda_K), \delta_Q(\epsilon^i)\right]
&=&\delta_S\left(\rmi\slashed{\Lambda}_K \epsilon^i\right).
\label{KQcomm}
\end{eqnarray}
Note that to verify these commutators one should use not only the
internal but the {\sl full} superconformal transformation rules including
the $x$-dependent translations (see~(\ref{xdepP})) and
$Q$-supersymmetries (see~(\ref{xdepQ})).

Now it's clear how generic fields transform under the superconformal
group, we briefly give the field content and properties of the basic
superconformal multiplets in five dimensions. They will be used for
studying matter couplings in the remainder of this article. The linear
multiplet will only be used as the multiplet of the equations of motion
for the vector multiplet.
\begin{table}[htb]
\begin{minipage}{\linewidth}
\renewcommand{\thefootnote}{\thempfootnote}
\begin{center}
\begin{tabular}{||c|ccc||}
\hline \rule[-1mm]{0mm}{6mm}
Field       & $\SU(2)$ & $w$ & \# d.o.f.  \\
\hline
 & \multicolumn{3}{c||}{off-shell vector multiplet} \\
\rule[-1mm]{0mm}{6mm}
$A_\m^I$      & 1 & 0             & $4 n$ \\
\rule[-1mm]{0mm}{6mm}
$Y^{ij I}$    & 3 & 2             & $3 n$ \\
\rule[-1mm]{0mm}{6mm}
$\s^I$        & 1 & 1             & $1 n$ \\
\hline \rule[-1mm]{0mm}{6mm}
$\p^{i I}$      & 2 & $3/2$         & $8 n$ \\[1mm]
\hline \hline
 & \multicolumn{3}{c||}{on-shell tensor multiplet} \\
\rule[-1mm]{0mm}{6mm}
$B_{\m\n}^M$      & 1 & 0         & $3 m$ \\
\rule[-1mm]{0mm}{6mm}
$Y^{ij M}$    & 3 & 2             & $0$ \\
\rule[-1mm]{0mm}{6mm}
$\sigma ^M$        & 1 & 1             & $1 m$ \\
\hline \rule[-1mm]{0mm}{6mm}
$\psi ^{i M}$      & 2 & $3/2$         & $4 m$ \\[1mm]
\hline \hline
 & \multicolumn{3}{c||}{on-shell hypermultiplet
 }   \\
\rule[-1mm]{0mm}{6mm}
$q^X$      & 2 & $3/2$        & $4 r$ \\
\hline \rule[-1mm]{0mm}{6mm}
$\zeta^A$      & 1 & 2       & $4 r$ \\[1mm]
\hline \hline
 & \multicolumn{3}{c||}{off-shell linear multiplet} \\
\rule[-1mm]{0mm}{6mm}
$L^{ij}$   & 3 & 3             & 3 \\
\rule[-1mm]{0mm}{6mm}
$E_a$      & 1 & 4             & 4 \\
\rule[-1mm]{0mm}{6mm}
$N$        & 1 & 4             & 1 \\
\hline \rule[-1mm]{0mm}{6mm}
$\vf^i$    & 2 & $7/2$ & 8 \\[1mm]
\hline
\end{tabular}
\caption{\it The  $D=5$ matter multiplets. We introduce $n$ vector
multiplets, $m$ tensor multiplets and $r$ hypermultiplets. Indicated are
their degrees of freedom, the Weyl weights and the $\SU(2)$
representations, including the linear multiplet for completeness.
  }
\label{tbl:multiplets}
\end{center}
\end{minipage}
\end{table}

\subsection{The vector-tensor multiplet}

In this section, we will discuss superconformal vector multiplets that
transform in arbitrary representations of the gauge group. From work on
$N=2$, $D=5$ Poincar{\'e} matter couplings~\cite{Gunaydin:1999zx} it is known
that vector multiplets transforming in representations other than the
adjoint have to be dualized to tensor fields. We define a vector-tensor
multiplet to be a vector multiplet transforming in a reducible
representation that contains the adjoint representation as well as
another, arbitrary representation.

We will show that the analysis of~\cite{Gunaydin:1999zx} can be extended
to superconformal vector multiplets. In doing this we will generalize the
gauge transformations for the tensor fields~\cite{Gunaydin:1999zx} by
allowing them to transform into the field-strengths for the adjoint gauge
fields. These more general gauge transformations are consistent with
supersymmetry, even after breaking the conformal symmetry.

The vector-tensor multiplet contains \emph{a priori} an arbitrary number
of tensor fields. The restriction to an even number of tensor fields is
not imposed by the closure of the algebra. If one demands that the field
equations do not contain tachyonic modes, an even number is
required~\cite{Townsend:1984xs}. Closely related to this is the fact that
one can only construct an action for an even number of tensor multiplets.
But supersymmetry without an action allows the more general possibility.
Note that these main results are independent of the use of superconformal
or super-Poincar{\'e} algebras.

To make contact with other results in the literature we will break the
rigid conformal symmetry by using a vector multiplet as a compensating
multiplet for the superconformal symmetry. The adjoint fields of the
vector-tensor multiplet are given constant expectation values, and the
scalar expectation values will play the role of a mass parameter. This
will reduce the superconformal vector-tensor multiplet, for the case of
two tensor multiplets, to the massive self-dual complex tensor multiplet
of~\cite{Townsend:1984xs}.

\subsubsection{Adjoint representation} \label{ss:adjoint}

We will start with giving the transformation rules for a vector multiplet
in the adjoint representation~\cite{Fujita:2001kv}. An off-shell vector
multiplet has $8+8$ real degrees of freedom whose $\SU(2)$ labels and
Weyl weights we have indicated in table~\ref{tbl:multiplets}.

The gauge transformations that we consider satisfy the commutation
relations ($I=1, \ldots ,n$)
\begin{equation}
\left[\d_G(\L_1^I), \d_G(\L_2^J)\right] = \d_G(\L_3^K ) \,, \qquad \L_3^K
= g \L_1^I \L_2^J f_{IJ}{}^K \,. \label{Lambda3}
\end{equation}
The gauge fields $A_\mu^I$\ ($\mu = 0,1, \ldots ,4$) and general matter
fields of the vector multiplet as e.g.\ $X^I$ transform under gauge
transformations with parameters $\Lambda^I$ according to
\begin{equation}
\d_G(\L^J) A_\mu^I = \partial_\mu \L^I + g A_\mu^J f_{JK}{}^I \L^K \,,
\qquad \d_G(\L^J) X^I = -g \L^J f_{JK}{}^I X^K \,, \label{eq:gaugecov}
\end{equation}
where $g$ is the coupling constant of the group G. The expression for the
gauge-covariant derivative of $X^I$  and the field-strengths are given by
\begin{equation}
{\cal D}_\m X^I = \partial_\m X^I + g A_\m^J f_{JK}{}^I X^K \,, \qquad
F_{\mu\nu}^I = 2 \partial_{[\mu} A_{\nu]}^I + g f_{JK}{}^I A_\mu^J
A_\nu^K\,. \label{eq:F}
\end{equation}
The field-strength satisfies the Bianchi identity
\begin{equation}
{\cal D}_{[\m} F_{\n\l]}^I = 0 \,. \label{eq:BIvec}
\end{equation}

The rigid $Q$- and $S$-supersymmetry transformation rules for the
off-shell Yang-Mills multiplet are given by~\cite{Fujita:2001kv}
\begin{eqnarray}
\d A_\m^I
&=& \frac 12 \bar{\e} \g_\m \p^I \,, \nonumber\\
\d Y^{ij I}
&=& -\frac 12 \bar{\e}^{(i} \slashed{\cal D} \p^{j) I} -  \frac {1}{2} \rmi g \bar \e^{(i} f_{JK}{}^I \sigma^J \psi^{j) K} + \frac 12 \rmi \bar{\eta}^{(i} \p^{j)I} \,, \nonumber \\
\d \p^{i I}
&=& - \frac 14 \g \cdot F^I \e^i -\frac 12\rmi \slashed{\cal D} \s^I \e^i - Y^{ij I} \e_j + \s^I \eta^i \,, \nonumber\\
\d \s^I &=& \frac 12 \rmi \bar{\e} \p^I \,. \label{ymflat}
\end{eqnarray}
The commutator of two $Q$-supersymmetry transformations yields a
translation with an extra $G$-transformation
\begin{equation}
[ \d(\e_1) , \d(\e_2)] = \d_{P} \left(\frac 12 \bar{\e}_2 \g_\m
\e_1\right) + \d_{G} \left(-\frac 12\rmi \s \bar\e_2 \e_1 \right).
\label{softrigid}
\end{equation}
Note that even though we are considering rigid superconformal symmetry,
the al\-ge\-bra~(\ref{softrigid}) contains a field-dependent term on the
righthand side. Such soft terms are commonplace in local superconformal
symmetry but here they already appear at the rigid level. In hamiltonian
language, it means that the algebra is satisfied modulo constraints.

\subsubsection{Reducible representation} \label{ss:reducVT}

Starting from $n$ vector multiplets we  now wish to consider a more
general set of fields ${\cal H}^{\widetilde I}_{\mu\nu}\ ({\widetilde I}
= 1, \ldots , n+m)$. We write ${\cal H}^{\widetilde I}_{\mu\nu} =
\{F_{\m\n}^I , B_{\m\n}^M\}$ with $\widetilde{I} = (I,M)\ (I=1,\ldots ,n;
M = n+1, \ldots n+m)$. The first part of these fields corresponds to the
generators in the adjoint representation. These are the fields that we
used in subsection~\ref{ss:adjoint}. The other fields may belong to an
arbitrary, possibly reducible, representation:
\begin{equation}
(t_I)_{\widetilde J}{}^{\widetilde K} =
\begin{pmatrix}
(t_I)_J{}^K & (t_I)_J{}^N \\
(t_I)_M{}^K & (t_I)_M{}^N
\end{pmatrix}, \qquad
\left\{
\begin{array}{ccl}
I,J,K &=& 1, \ldots ,n \\
M,N &=& n+1, \ldots ,n+m\,.
\end{array}
\right.  \label{reducible}
\end{equation}
It is understood that the $(t_I)_J{}^K $ are in the adjoint
representation, i.e.\
\begin{equation}
(t_I)_J{}^K  = f_{IJ}{}^K  \,.
 \label{adjointt}
\end{equation}
If $m \neq 0$, then the representation $(t_I)_{\widetilde
J}{}^{\widetilde K}$ is reducible. We will see that this representation
can be more general than assumed so far in treatments of vector-tensor
multiplet couplings. The requirement that $m$ is even will only appear
when we demand the existence of an action in section~\ref{ss:actvt}, or
if we require absence of tachyonic modes. The matrices $t_I$ satisfy
commutation relations
\begin{equation}
[t_I, t_J] = -f_{IJ}{}^{K} t_K \,, \qquad \mbox{or}\qquad t_{I \widetilde
N}{}^{\widetilde M} t_{J \widetilde M}{}^{\widetilde L} - t_{J \widetilde
N}{}^{\widetilde M} t_{I \widetilde M}{}^{\widetilde L} = -f_{I J}{}^K
t_{K \widetilde N}{}^{\widetilde L}\,. \label{commutt}
\end{equation}
If the index $\tilde L$ is a vector index, then this relation is
satisfied using the matrices as in~(\ref{adjointt}).

Requiring the closure of the superconformal algebra, we find $Q$- and
$S$-supersymmetry transformation rules for the vector-tensor multiplet
and a set of constraints. The transformations are
\begin{eqnarray}
\d {\cal H}_{\m\n}^{\widetilde I} &=& - \bar \e \g_{[\m} {\cal D}_{\n]}
\p^{\widetilde I} + \rmi g \bar{\e} \g_{\m\n} t_{({\widetilde
J}{\widetilde K)}}{}^{\widetilde I}
\s^{\widetilde J} \p^{\widetilde K} + \rmi \bar{\eta}\g_{\m\n}\p^{\widetilde I} \,, \nonumber \\
\d Y^{ij {\widetilde I}} &=& -\frac 12 \bar{\e}^{(i} \slashed{\cal D}
\p^{j) {\widetilde I}} -\frac {1}{2} \rmi g \bar \e^{(i}
\left(t_{[{\widetilde J}{\widetilde K]}}{}^{\widetilde I} - 3
t_{({\widetilde J}{\widetilde K)}}{}^{\widetilde I} \right)
\sigma^{\widetilde J} \psi^{j) {\widetilde K}} + \frac 12 \rmi \bar{\eta}^{(i} \p^{j)\widetilde I} \,, \nonumber \\
\d \p^{i {\widetilde I}} &=& - \frac 14 \g \cdot {\cal H}^{\widetilde I}
\e^i -\frac 12\rmi \slashed{\cal D} \s^{\widetilde I} \e^i - Y^{ij
{\widetilde I}} \e_j +\frac 12 g t_{({\widetilde J}{\widetilde
K)}}{}^{\widetilde I} \s^{\widetilde J}
\s^{\widetilde K} \e^i + \s^{\widetilde I} \eta^i \, , \nonumber\\
\d \s^{\widetilde I} &=& \frac 12 \rmi \bar{\e} \p^{\widetilde I} \,.
\label{tensflat}
\end{eqnarray}
The curly derivatives denote gauge-covariant derivatives as
in~(\ref{eq:F}) with the replacement of structure constants by general
matrices $t_I$ according to~(\ref{adjointt}). We have extended the range
of the generators from $I$ to $\widetilde I$ in order to simplify the
transformation rules with the understanding that
\begin{equation}
(t_M)_{\widetilde J}{}^{\widetilde K} = 0\, .\label{tM0}
\end{equation}
We use a convention where (anti)symmetrizations are done with total
weight~1. We find that the supersymmetry algebra~(\ref{softrigid}) is
satisfied provided the representation matrices are restricted to
\begin{equation}
t_{(\widetilde{J}\widetilde{K})}{}^I = 0\,, \label{constrT}
\end{equation}
and provided the following two constraints on the fields are imposed:
\begin{eqnarray}
L^{ij \widetilde I} &\equiv& t_{({\widetilde J}{\widetilde
K)}}{}^{\widetilde I} \left(2 \s^{\widetilde J} Y^{ij {\widetilde K}} -
\frac 12 \rmi \bar{\p}^{i {\widetilde J}} \p^{j {\widetilde K}}\right)=0
\,,
\label{tensEOM} \\
 E_{\mu\nu\lambda}^{\widetilde I} &\equiv& \frac{3}{g} {\cal D}_{[\m}
{\cal H}_{\n\l]}{}^{\widetilde I} -  \ve_{\m\n\l\r\s} t_{({\widetilde
J}{\widetilde K)}}{}^{\widetilde I} \left(\s^{\widetilde J} {\cal
H}^{\r\s {\widetilde K}} + \frac 14 \rmi \bar{\p}^{\widetilde J}
\g^{\r\s} \p^{\widetilde K} \right)=0  \,. \label{tensBI}
\end{eqnarray}
For ${\widetilde I}=I$, the constraint~(\ref{tensBI}) reduces to the
Bianchi identity~(\ref{eq:BIvec}). The tensor $F_{\m\n}^{ I}$ can
therefore be seen as the curl of a gauge vector $A_\mu^I$. Moreover, the
constraint~(\ref{tensEOM}) is trivially satisfied for $\widetilde{I}=I$.
We conclude that the fields with indices $\widetilde{I}=I$ form an
off-shell vector multiplet in the adjoint representation of the gauge
group.

On the other hand, when  ${\widetilde I}=M$, the
constraint~(\ref{tensBI}) does not permit the fields $B_{\m\n}^{ M}$ to
be written as the curl of a gauge field and they should be seen as
independent tensor fields. Instead, the constraint~(\ref{tensBI}) is a
massive self-duality condition that puts the tensors $B_{\m\n}^{ M}$
\emph{on-shell}. The constraint~(\ref{tensEOM}) implicitly allows us to
eliminate the fields  $Y^{ij M}$ altogether. The general vector-tensor
multiplet can then be interpreted as a set of $m$ on-shell tensor
multiplets in the background of $n$ off-shell vector multiplets.

Using~(\ref{constrT}) we have reduced the representation matrices $t_I$
to the following block-upper-triangular form:
\begin{equation}
(t_I)_{\widetilde J}{}^{\widetilde K} =
\begin{pmatrix}
f_{IJ}{}^K & (t_I)_{J}{}^N \\
0 & (t_I)_M{}^N
\end{pmatrix}.
\label{uppertriang}
\end{equation}
In~\cite{Gunaydin:1999zx} it is mentioned that, ``since terms of the form
$B^M \wedge F^I \wedge A^J$ appear to be impossible to supersymmetrize in
a gauge invariant way (except possibly in very special cases) we shall
also assume that $C_{MIJ}=0$''. This corresponds, as we will see below,
to the assumption that the representation is completely reducible,
i.e.~$t_{IJ}{}^N=0$, meaning that gauge transformations do not mix the
pure Yang-Mills field-strengths and the tensor fields. However, we  find
that off-diagonal generators {\sl are} allowed, both when requiring
closure of the superconformal algebra and when writing down an action. We
thus allow reducible, but not necessarily completely reducible
representations.

Recall that every \emph{unitary} reducible representation of a Lie group
is also completely reducible, and that every representation of a
\emph{compact} Lie group is equivalent to a unitary representation.
Hence, every reducible representation of a compact Lie group is also
completely reducible. Non-compact Lie groups, on the other hand, have no
non-trivial and finite-dimensional unitary representations. However,
every reducible representation of a \emph{connected}, \emph{semi-simple},
non-compact Lie group or a semi-simple, non-compact Lie \emph{algebra} is
also completely reducible. See~\cite{Cornwell:1997ke} for an exposition
of these theorems.

This leaves us with the class of non-compact Lie algebras that contain an
abelian invariant subalgebra. 
Examples of non-diagonal terms can thus be given for $t_I$ of the form
\begin{equation}
(t_I)_{\widetilde J}{}^{\widetilde K} =
\begin{pmatrix}
0 & (t_I)_{J}{}^M \\
0 & 0
\end{pmatrix}.
\label{eq:reducible}
\end{equation}
The simplest one is thus with one gauge multiplet and a number of tensor
multiplets, with only the off-diagonal parts $t_{11}{}^M$ non-vanishing.
But more examples are possible, e.g.\ the lower right corner does not
have to be zero.

The constraints~(\ref{tensEOM}) and~(\ref{tensBI}), with $\widetilde I =
M$, do not form a supersymmetric set: they are invariant under
$S$-supersymmetry but under $Q$-super\-sym\-me\-try they lead to a
constraint on the spinors $\psi^{iM}$ which we will call $\varphi^{iM}$:
\begin{equation}
  \delta L^{ijM} = \rmi \bar \epsilon ^{(i}\varphi^{j)M}\,, \qquad \delta E_{\mu\nu\rho}^M =  \bar\epsilon \gamma_{\mu\nu\rho}\varphi^M \,.
\end{equation}
The expression for this constraint is given by
\begin{eqnarray}
\varphi^{iM}&\equiv& t_{\left(\widetilde J\widetilde K\right)}{}^M\left[
  \rmi \sigma ^{\widetilde J}\slashed{\cal D}\psi ^{i\widetilde K}
  +\frac 12\rmi \left(\slashed{\cal D}\sigma ^{\widetilde J}\right)
  \psi ^{i\widetilde K}+Y^{ik\widetilde J}\psi _k^{\widetilde K}
  -\frac 14\gamma \cdot {\cal H}^{\widetilde J}\psi ^{i\widetilde K}
  \right]- \nonumber\\
  &&-\, g \left(  \left[ t_{\left[\widetilde J\widetilde K\right]}{}^{\widetilde L}
                  -3t_{\left(\widetilde J\widetilde K\right)}{}^{\widetilde L}
            \right] t_{\left(\widetilde I \widetilde L\right)}{}^M
           +\frac 12t_{\widetilde I\widetilde J}{}^{\widetilde L}t_{\left(\widetilde K \widetilde L\right)}{}^M
           \right)\sigma^{\widetilde I}\sigma ^{\widetilde J}\psi ^{i\widetilde
  K}  \nonumber \\
&=& 0 \,. \label{deltaSij}
\end{eqnarray}
The second line can be rewritten, by splitting the indices in tensor
versus vector parts, as
\begin{equation}
  +\frac 12 g\sigma ^I\sigma ^J\psi ^{\tilde K} (t_It_J)_{\tilde K}{}^M+\frac 14 g\sigma ^I\sigma
  ^{\tilde K}\psi ^J\left(t_It_J+2t_Jt_I\right)_{\tilde K}{}^M\,.
 \label{rewritettGamma}
\end{equation}

Varying the new constraint $\varphi^{i\, M}$ under $Q$-and
$S$-supersymmetry, one finds at first sight two more constraints, $E_a^M$
and $N^M$, of which the first one turns out to be dependent (see below):
\begin{eqnarray}
\d \varphi^{i\, M} &=& - \frac 12 \rmi \slashed {\cal D} L^{ij \, M}
\epsilon_j - \frac 12 \rmi \g^a E_a^M \epsilon^i + \frac 12 N^M
\epsilon^i
+ \frac 12 g t_{\widetilde J\widetilde K}{}^M \s^{\widetilde J} L^{ij \widetilde K} \epsilon_j - \nn \\
&&{} - \frac {1}{12} \rmi g t_{\left(\widetilde J\widetilde K\right)}{}^M
\g^{abc} \s^{\widetilde J} E_{abc}^{\widetilde K} \epsilon^i + 3 L^{ijM}
\eta_j\,.
\end{eqnarray}
The constraint $N^M$ is given by
\begin{eqnarray}
N^M
&\equiv& t_{(\widetilde J\widetilde K)}{}^M \left(\s^{\widetilde J} \Box \s^{\widetilde K} + \frac 12 {\cal D}^a \s^{\widetilde J} {\cal D}_a \s^{\widetilde K} -\frac 14 {\cal H}_{ab}^{\widetilde J} {\cal H}^{ab \widetilde K} - \frac 12 \bar{\p}^{\widetilde J} \slashed{\cal D} \p^{\widetilde K} + Y^{ij \widetilde J} Y_{ij}{}^{\widetilde K} \right) - \nonumber \\
&&{}-\rmi g \left[ -\frac 12\, t_{\left[\widetilde J \widetilde K\right]}{}^{\widetilde L}
t_{\left(\widetilde I \widetilde L\right)} {}^M
+ 2\, t_{\left(\widetilde I \widetilde J\right)} {}^{\widetilde L}
t_{\left(\widetilde K \widetilde L\right)}{}^M  \right] \s^{\widetilde I} \bar{\p}^{\widetilde J} \p^{\widetilde K} + \nonumber \\
&&{}+ \frac 12 g^2 \left( t_I t_J t_K \right)_{\widetilde L} {}^M\s^I
\s^J\s^K
\s^{\widetilde L} \nonumber \\
&=& 0 \,, \label{scalareom}
\end{eqnarray}
and for $E_a^M$ we find
\begin{equation}
E_a^M \equiv  t_{(\widetilde J\widetilde K)}{}^M \left(  {\cal D}^b
\left(\s^{\widetilde J} {\cal H}_{ba}{}^{\widetilde K} + \frac 14 \rmi
\bar{\p}^{\widetilde J} \g_{ba} \p^K \right) - \frac 18 \ve_{abcde} {\cal
H}^{bc {\widetilde J}} {\cal H}^{de {\widetilde K}} \right) =0\,.
\label{vectoreom}
\end{equation}
We made use of identities as
\begin{equation}
  t_{K\widetilde I}{}^{\widetilde L}t_{\left(\widetilde J\widetilde L\right)}{}^M
  +t_{K\widetilde J}{}^{\widetilde L}t_{\left(\widetilde I\widetilde L\right)}{}^M
  -t_{\left(\widetilde I\widetilde J\right)}{}^{\widetilde L}t_{K\widetilde L}{}^M=0\,,
 \label{ttid1}
\end{equation}
which follow from the commutator relation~(\ref{commutt}), and the
restrictions~(\ref{tM0}) and~(\ref{constrT}).

We find that the expression for $E_a$ is related to the one corresponding
to $E_{abc}^M$ as follows:
\begin{equation}
E_a^M = -\frac 1{12} \ve_{abcde} {\cal D}^b E^{cde M}\,.
\end{equation}

By now we have found a set of constraints that under $Q$- and
$S$-supersymmetry transform to each other. These constraints do not seem
to form a multiplet by themselves.

\subsubsection{The massive self-dual tensor multiplet}

To obtain the massive self-dual tensor multiplet
of~\cite{Townsend:1984xs}, we consider a vector-tensor multiplet for
general $n$ and $m$. Our purpose is to use the  vector multiplet as a
compensating multiplet for the superconformal symmetry. Thus, we impose
conditions on the fields that break the conformal symmetry, and preserve
$Q$-supersymmetry. We  give the fields of the vector multiplets the
following vacuum expectation values
\begin{equation}
F_{\mu\nu}^I = Y^{ij I} = \p^{i I} = 0 \,, \qquad \s^I = \frac {2 m^I}
{g} \,, \label{vvacuum}
\end{equation}
where $m^I$ are constants. Note that these conditions break the conformal
group to the Poincar{\'e} group, and break $S$-supersymmetry ($\eta =0$).
This is an example of a compensating multiplet in rigid supersymmetry.
The breaking of conformal symmetry is characterized by the mass
parameters $m^I$ in~(\ref{vvacuum}). If we substitute~(\ref{vvacuum})
into the expression~(\ref{tensEOM}) for $L^{ij M}$, then we find that we
can eliminate the field $Y^{ij M}$
\begin{equation}
Y^{ij M} = 0 \,.
\end{equation}
Moreover, we can also substitute~(\ref{vvacuum}) into the constraints
$E_{\mu\nu\lambda}^M$, $\varphi^{i M}$ and $N^M$ obtaining
\begin{eqnarray}
3 \partial_{[\mu} B_{\nu\lambda]}^M &=&  \frac 12\varepsilon_{\mu\nu\lambda\rho\sigma}{\cal M}_N{}^M B^{\rho\sigma \, N}\,,  \nonumber\\
\slashed{\partial} \psi^{i M} &=& \rmi {\cal M}_N{}^M \psi^{i N} \,, \nonumber \\
\Box \s^M &=& - \left({\cal M}^2\right){}_N{}^M \s^N - \frac 4g
t_{IJ}{}^N m^I m^J {\cal M}_N{}^M\,. \label{gfixeom}
\end{eqnarray}
The mass-matrix ${\cal M}_N{}^M$ is defined as
\begin{equation}
{\cal M}_N{}^M \equiv g \s^I (t_I)_N{}^M = 2 m^I  (t_I)_{N}{}^M \,,
\end{equation}
and has been assumed to be invertible. The last term of~(\ref{gfixeom})
can be eliminated by redefining $\sigma ^M$ with a constant shift. In
order for the tensor fields to have no tachyonic modes, the mass-matrix
needs to satisfy a symplectic condition which can only be satisfied if
the number of tensor fields is even~\cite{Townsend:1984xs}. We denote the
number of tensor multiplets by~$m=2k$.

In the particular gauge~(\ref{vvacuum}) and
representation~(\ref{eq:reducible}) the mass matrix ${\cal M}$ is zero.
The last two equations in~(\ref{gfixeom}) are not present and the first
one becomes the usual Bianchi identity  for a set of $m$ abelian vectors.
Thus, we are dealing with $n+m$ off-shell gauge vectors.

To obtain the massive self-dual tensor multiplet
of~\cite{Townsend:1984xs} we consider the case $n=1$, $m=2$, i.e.\ two
(real) tensor multiplets
 $\{B_{\mu\nu}^M, \l^{i M}, \f^M \}$
($M,N = 2,3$) in the background of one vector multiplet $\{F_{\mu\nu} \,
\p^i, \s \}$, which has been given the vacuum expectation
value~(\ref{vvacuum}). In what follows we will use a complex notation:
\begin{equation}
B_{\mu\nu} = B_{\mu\nu}^2 + \rmi B_{\mu\nu}^3 \,, \qquad
\overline{B}_{\mu\nu} = B_{\mu\nu}^2 - \rmi B_{\mu\nu}^3\, .
\end{equation}
The generators $(t_1)_{\widetilde I}{}^{\widetilde J}$ must form a
representation of U$(1) \simeq \SO(2)$. Under a U(1) transformation the
field-strength $F_{\mu\nu}$ is invariant and the tensor field gets a phase
\begin{equation}
B_{\mu\nu}^\prime = e^{\rmi \theta} B_{\mu\nu} \rightarrow
\begin{pmatrix}
B_{\mu\nu}^2 \cr B_{\mu\nu}^3
\end{pmatrix}^\prime
= \begin{pmatrix} \cos \theta & -\sin \theta \cr \sin \theta & \cos
\theta
\end{pmatrix}
\cdot
\begin{pmatrix}
B_{\mu\nu}^2 \cr B_{\mu\nu}^3
\end{pmatrix}.
\end{equation}
{}From this we obtain the generator
\begin{equation}
(t_1)_{\widetilde I}{}^{\widetilde J} =
\begin{pmatrix}
0 & 0  & 0 \cr 0 & 0  & -1 \cr 0 & 1  & 0
\end{pmatrix}.
\label{eq:O2rep}
\end{equation}

After substituting the conditions~(\ref{vvacuum}) into the transformation
rules we obtain
\begin{eqnarray}
\d {B}_{\m\n}{}
&=& -\bar \e \g_{[\m} \partial_{\n]} \l - m \bar{\e} \g_{\m\n} \l \,, \nonumber \\
\d \l^i
&=& - \frac 14 \g \cdot B \e^i - \frac 12\rmi \slashed{\partial} \f \e^i -\rmi m \f \e^i \,, \nonumber\\
\d \f &=& \frac 12 \rmi \bar{\e} \l \,,
\end{eqnarray}
and
\begin{equation}
3 \partial_{[\m} B_{\n\l]} - \rmi m \ve_{\m\n\l\r\s} B^{\r\s} = 0 \,.
\label{massBI}
\end{equation}
This reproduces the massive self-dual tensor multiplet
of~\cite{Townsend:1984xs}. Note that the commutator of two
$Q$-supersymmetries yields a translation plus a (rigid)
U(1)-transformation whose parameter can be obtained from the general
$G$-transformation in the superconformal algebra, see~(\ref{softrigid}),
by making the substitution~(\ref{vvacuum}).

{}From a six-dimensional point of view the interpretation of the mass
parameter $m$ is that it is the label of the $m$-th Kaluza-Klein mode in
the reduction of the $D=6$ self-dual tensor multiplet. The zero-mode of
the reduced tensor multiplet corresponds to a vector multiplet as can be
seen from~(\ref{massBI}) which becomes a Bianchi identity for a
field-strength when $m=0$.

\subsection{The hypermultiplet} \label{ss:hypermultiplet}

In this subsection, we discuss hypermultiplets in five dimensions. As for
the tensor multiplets, there is in general no known off-shell formulation
with a finite number of auxiliary fields. Therefore, the supersymmetry
algebra already leads to the equations of motion.

A single hypermultiplet contains four real scalars and two spinors subject
to the symplectic Majorana reality condition. For $r$ hypermultiplets, we
introduce real scalars $q^X(x)$, with $X=1,\dots ,4r$, and spinors
$\zeta^A(x)$ with $A=1,\dots ,2r$. To formulate the symplectic Majorana
condition, we introduce two matrices $\rho_A {}^B$ and $E_i{}^j$, with
\begin{equation}
 \rho \rho^{\ast} = - \unity_{2r}\,,\qquad E E^{\ast} = -\unity_2\,. \label{defrho}
\end{equation}
This defines symplectic Majorana conditions for the fermions and
supersymmetry transformation parameters~\cite{VanProeyen:2001ng}:
\begin{equation}
  \alpha {\cal C}\gamma _0\zeta ^B \rho_B{}^A=\left(\zeta ^A\right)^*, \qquad
\alpha {\cal C}\gamma _0\epsilon ^j E_j{}^i=\left(\epsilon ^i\right)^*,
 \label{symplMaj}
\end{equation}
where ${\cal C}$ is the charge conjugation matrix, and $\alpha $ is an
irrelevant number of modulus 1. We can always adopt the basis where
$E_i{}^j=\varepsilon_{ij}$, and will further restrict to that.

The scalar fields are interpreted as coordinates of some target space,
and requiring the on-shell closure of the superconformal algebra imposes
certain conditions on the target space, which we derive below.
Superconformal hypermultiplets in four spacetime dimensions were
discussed in~\cite{deWit:1999fp}; our discussion is somehow similar, but
we extend it to the case where an action is not needed, in the spirit
explained in~\cite{VanProeyen:2001wr}.

\subsubsection{Rigid supersymmetry}

We will show how the closure of the supersymmetry transformation laws
leads to a `hypercomplex manifold'. The closure of the algebra on the
bosons leads to the defining equations for this geometry, whereas the
closure of the algebra on the fermions and its further consistency leads
to equations of motion in this geometry, independent of an action.

The supersymmetry transformations (with $\epsilon^i$ constant parameters)
of the bosons $q^X\!(x)$, are parametrized by arbitrary functions
$f^X_{iA}(q)$. Also for the transformation rules of the fermions we write
the general form compatible with the supersymmetry algebra. This
introduces other general functions $f_X^{iA}(q)$ and
$\omega_{XB}{}^A(q)$:\footnote{In fact, one can write down a more general
supersymmetry transformation rule for the fermions than
in~(\ref{SUzeta}), but using Fierz relations and simple considerations
about the supersymmetry algebra, one can bring its form into the one
written above.}
\begin{eqnarray}
\delta (\epsilon) q^X
&=& - \rmi \bar\epsilon^i \zeta^A f_{iA}^X \,,\nonumber\\
\delta (\epsilon) \zeta^A &=& \frac 12 \rmi \slashed{\partial} q^X
f_X^{iA} \epsilon_i -\zeta^B \omega_{XB}{}^A \big( \delta (\epsilon) q^X
\big)\,. \label{SUzeta}
\end{eqnarray}
The functions satisfy reality properties consistent with reality of $q^X$
and the symplectic Majorana conditions, e.g.:
\begin{equation}
\left(f_X^{iA}\right)^{\ast} = f_X^{jB} E_j {}^i \rho_B {}^A\,, \qquad
 \left( \omega_{XA}{}^B\right)^*= \left(\rho ^{-1}\omega _X\rho
 \right)_A{}^B\,.
 \label{realfunctions}
\end{equation}
A priori the functions $f_{iA}^X$ and $f^{iA}_X$ are independent, but the
commutator of two supersymmetries on the scalars only gives a translation
if one imposes
\begin{eqnarray}\label{cov_const}
f^{iA}_Y f^X_{iA} &=& \delta_Y^X \,,\qquad f^{iA}_Xf^X_{jB}=\delta^i_j
\delta^A_B\,,\nonumber\\
\covder_Y f_{iB}^X &\equiv&
\partial_Y f_{iB}^X - \omega_{YB}^{\quad A} f_{iA}^X + \Gamma _{ZY}^{\quad
X} f_{iB}^Z = 0\,,
\end{eqnarray}
where $\Gamma _{XY} {}^Z$ is some object, symmetric in the lower indices.
This means that $f_{iA}^X$ and $f^{iA}_X$ are each others inverse and are
covariantly constant with connections $\Gamma$ and $\omega$. It also
implies that $\rho $ is covariantly constant. The
conditions~(\ref{cov_const}) encode all the constraints on the target
space that follow from imposing the supersymmetry algebra. Below, we show
that there are no further geometrical constraints coming from the fermion
commutator; instead this commutator defines the equations of motion for
the on-shell hypermultiplet. \bigskip

The supersymmetry transformation rules are covariant with respect to two
kinds of reparametrizations. The first ones are the target space
diffeomorphisms, $q^X\rightarrow {\widetilde q}^X(q)$, under which
$f^X_{iA}$ transforms as a vector, $\omega_{XA}{}^B$ as a one-form, and
$\Gamma_{XY}{}^Z$ as a connection. The second set are the
reparametrizations which act on the tangent space indices $A,B,\ldots$ On
the fermions, they act as
\begin{equation}
\zeta^A \rightarrow {\widetilde \zeta}^A(q)=\zeta^B
U_B{}^A(q)\,,\label{ferm-equiv}
\end{equation}
where $U_A{}^B(q)$ is an invertible matrix, and the reality conditions
impose $U^*=\rho ^{-1}U\rho $, defining $\Gl(r,\mathbb H)$. In general,
such a transformation brings us into a basis where the fermions depend on
the scalars $q^X$. In this sense, the hypermultiplet is written in a
special basis where $q^X$ and $\zeta^A$ are independent fields. The
supersymmetry transformation rules~(\ref{SUzeta}) are covariant
under~(\ref{ferm-equiv}) if we transform $f^{iA}_X(q)$ as a vector and
$\omega_{XA}{}^B$ as a connection,
\begin{equation}
\omega_{XA}{}^B\rightarrow {\widetilde \omega}_{XA}{}^B=[(\partial_X
U^{-1})U+ U^{-1}\omega_X U]_A{}^B\,.
\end{equation}
These considerations lead us to define the covariant variation of the
fermions:
\begin{equation}
{\widehat \delta} \zeta^A\equiv \delta \zeta^A+\zeta^B\omega_{XB}{}^A
\delta q^X\,, \label{cov-var}
\end{equation}
for any transformation $\delta$ (supersymmetry, conformal
transformations,\ldots). Two models related by either target space
diffeomorphisms or fermion reparametrizations of the
form~(\ref{ferm-equiv}) are equivalent; they are different coordinate
descriptions of the same system. Thus, in  a covariant formalism, the
fermions can be functions of the scalars. However, the expression
$\partial _X\zeta ^A$ makes only sense if one compares different bases.
But in the same way also the expression $\zeta^B \omega_{XB}{}^A$ makes
only sense if one compares different bases, as the connection has no
absolute value. The only covariant object is the covariant derivative
\begin{equation}
  \covder_X \zeta ^A\equiv \partial_X \zeta ^A+\zeta ^B\omega
  _{XB}{}^A\,.
 \label{calDXzeta}
\end{equation}
The covariant transformations are also a useful tool to calculate any
transformation on e.g.\ a quantity $W_A(q)\zeta ^A$:
\begin{eqnarray}
  \delta \left(W_A(q)\zeta ^A\right)&=& \partial _X \left(W_A\zeta ^A\right) \delta q^X +
   W_A\left.\delta \zeta ^A\right|_{q} \nonumber\\
   &=& \covder_X \left(W_A\zeta ^A\right)\delta q^X+ W_A\left(\widehat \delta \zeta
   ^A - \covder_X\zeta ^A\delta q^X\right) \nonumber\\
   &=& \left( \covder_XW_A\right)\delta q^X \zeta ^A + W_A\,\widehat \delta \zeta
   ^A \,.
 \label{simpledelWzeta}
\end{eqnarray}
We will frequently use the covariant transformations~(\ref{cov-var}). It
can similarly be used on target-space vectors or tensors. E.g.\ for a
quantity $\Delta^X$:
\begin{equation}
  \widehat \delta \Delta ^X= \delta \Delta ^X + \Delta ^Y \Gamma _{ZY}{}^X\,\delta
  q^Z\,.
 \label{tildelDel}
\end{equation}
\bigskip

The geometry of the target space is that of a \emph{hypercomplex}
manifold. It is a weakened version of hyperk{\"a}hler geometry where no
hermitian covariantly constant metric is defined. We refer the reader to
appendix~\ref{hyperappendix} for an introduction to these manifolds,
references and the mathematical context in which they can be situated.

The crucial ingredient is a triplet of complex structures, the
hypercomplex structure, defined as
\begin{equation}
\J\alpha XY \equiv -\rmi f_X^{iA}(\sigma ^\alpha )_i{}^j f_{jA}^Y\,.
\label{defJf}
\end{equation}
Using~(\ref{cov_const}), they are covariantly constant and satisfy the
quaternion algebra
\begin{equation}
J^\alpha J^\beta =- \unity _{4r} \delta^{\alpha \beta} +
\varepsilon^{\alpha \beta \gamma} J^\gamma \,. \label{defJ}
\end{equation}
At some places we also use a doublet notation, for which
\begin{equation}
J_X{}^Y{}_i{}^j \equiv \rmi
 \J\alpha XY \left(\sigma ^\alpha \right)_i{}^j=2f^{jA}_Xf^Y_{iA}-
\delta_i^j\delta_X^Y\,.\label{doublet-J}
\end{equation}
The same transition between doublet and triplet notation is used also for
other $\SU(2)$-valued quantities.

The holonomy group of such a space is contained in
$\Gl(r,\mathbb{H})=\SU^*(2r)\times $U$(1)$, the group of transformations
acting on the $A,B$-indices. This follows from the integrability
conditions on the covariantly constant vielbeins $f^{iA}_X$, which
relates the curvatures of the $\omega _{XA}{}^B$ and $\Gamma _{XY}{}^Z$
connections (see appendix~\ref{app:convCurv} for conventions on the
curvatures),
\begin{equation}
R_{XYZ}{}^W=f_{iA}^Wf^{iB}_Z{\cal R}_{XYB} {}^A \,,\qquad
\delta_j^i\,{\cal R}_{XYB}{}^A = f^{iA}_W f^Z_{jB}\,R_{XYZ}{}^W\,,
\label{curv-rel}
\end{equation}
such that the Riemann curvature only lies in $\Gl(r,\mathbb{H})$.
Moreover, from the cyclicity properties of the Riemann tensor, it follows
that
\begin{eqnarray}
&&  f_{Ci}^X f_{jD}^Y{\cal R}_{XYB} {}^A= - \frac 12 \varepsilon_{ij}
W_{CDB}{}^A\,,\nonumber\\
&& W_{CDB}{}^A \equiv  f^{iX}_C f^Y_{iD} {\cal R}_{XY}{}_B {}^A = \frac
12f^{iX}_C f^Y_{iD} f_{jB}^Z f_W^{Aj} R_{XYZ}{}^W\,, \label{def-W}
\end{eqnarray}
where $W$ is symmetric in all its three lower indices. For a more detailed
discussion on hypercomplex manifolds and their curvature relations, we
refer to appendix~\ref{hyperappendix}. We show there that, in contrast
with hyperk{\"a}hler manifolds, hypercomplex manifolds are not necessarily
Ricci flat; instead, the Ricci tensor is antisymmetric and defines a
closed two-form. \bigskip

We have so far only used the commutator of supersymmetry on the
hyperscalars, and this lead us to the geometry of hypercomplex manifolds.
Before continuing, we want to see what are the independent objects that
determine the theory, and what are the independent constraints. We start
in the supersymmetric theory from the vielbeins $f_X^{iA}$. They have to
be real in the sense of~(\ref{realfunctions}) and invertible. With these
vielbeins, we can construct the complex structures as in~(\ref{defJf}).
In the developments above, the only remaining independent equation is the
covariant constancy of the vielbein in~(\ref{cov_const}). This equation
contains the affine connection $\Gamma _{XY}{}^Z$ and the
$\Gl(r,\mathbb{H})$-connection $\omega _{XA}{}^B$. These two objects can
be determined from the vielbeins if and only if the (`diagonal')
Nijenhuis tensor~(\ref{Nijenhuisdiag}) vanishes. Indeed, for vanishing
Nijenhuis tensor, the `Obata'-connection~\cite{Obata}
\begin{equation}
\Gamma_{XY}{}^Z=-\frac 16\left( 2 \partial_{(X}\J\alpha {Y)}W+\varepsilon
^{\alpha \beta \gamma }\J{\beta  }{(X}U
\partial_{|U|} \J{\gamma }{Y)}W\right) \J\alpha WZ \,, \label{Obata}
\end{equation}
leads to covariantly constant complex structures. Moreover, one can show
that any torsionless connection that leaves the complex structures
invariant is equal to this Obata connection (similar to the fact that a
connection that leaves a metric invariant is the Levi-Civita connection).
With this connection one can then construct the
$\Gl(r,\mathbb{H})$-connection
\begin{equation}
  \omega _{XA}{}^B=\frac 12 f_Y^{iB}\left( \partial _X f_{iA}^Y+\Gamma
  _{XZ}^Yf_{iA}^Z\right) \,,
 \label{determineOm}
\end{equation}
such that the vielbeins are covariantly constant.

\paragraph{Dynamics.}
Now we consider the commutator of supersymmetry on the fermions, which
will determine the equations of motion for the hypermultiplets.

Using~(\ref{cov_const}),~(\ref{curv-rel}) and~(\ref{def-W}), we compute
this commutator on the fermions, and~find\footnote{To obtain this result,
we use Fierz identities expressing that only the cubic fermion
combinations of~\cite[(A.11)]{Bergshoeff:2001hc} are independent:
\[ \zeta^{(B}\bar{\zeta}^C \gamma_a \zeta^{D)} =- \gamma_a\zeta^{(B} \bar{\zeta}^C \zeta^{D)}\,.
\]}
\begin{equation}
[\delta ( \epsilon_1),\delta ( \epsilon_2)] \zeta^A =\frac 12\partial_a
\zeta^A \bar{\epsilon}_2 \gamma^a \epsilon_1 + \frac 14 \Gamma ^A
\bar{\epsilon}_2 \epsilon_1 -\frac 14 \gamma_a \Gamma ^A \bar{\epsilon}_2
\gamma^a \epsilon_1 \,. \label{FinalCommzeta}
\end{equation}

The $\Gamma ^A$ are the non-closure functions, and define the equations
of motion for the fermions,
\begin{equation}
\Gamma^A = \slashed{\covder} \zeta^A + \frac 12 W_{CDB} {}^A
 \zeta^B\bar{\zeta}^D \zeta^C\,, \label{eqmozeta}
\end{equation}
where we have introduced the covariant derivative with respect to the
transformations~(\ref{cov-var})
\begin{equation}
  \covder_\mu  \zeta^A \equiv \partial_\mu  \zeta^A + (\partial_\mu  q^X)
\zeta^B\omega_{XB} {}^A\,.
 \label{defDzeta}
\end{equation}

By varying the equations of motion under supersymmetry, we derive the
corresponding equations of motion for the scalar fields:
\begin{equation}
\widehat \delta(\epsilon ) \Gamma^A =  \frac 12 \rmi f_{X}^{iA}
\epsilon_i \Delta ^X \,,\label{delQGamma3}
\end{equation}
where
\begin{equation}
\label{covscal} \Delta^X=\Box q^X -\frac 12 \bar{\zeta}^B \gamma_a
\zeta^D
\partial^a q^Y f_Y^{iC}f_{iA}^X W_{BCD} {}^A-\frac{1}{4}\covder_Y W_{BCD}{}^A \bar{\zeta}^E \zeta^D
\bar{\zeta}^C \zeta^B f_E^{iY}f_{iA}^X\,,
\end{equation}
and the covariant laplacian is given by
\begin{equation}
\Box q^X = \partial_a \partial^a q^X+ \left( \partial_a q^Y\right)
\left(\partial^a q^Z \right) \Gamma_{YZ} {}^X\,.
\end{equation}

In conclusion, the supersymmetry algebra imposes the hypercomplex
constraints~(\ref{cov_const}) and the equations of
motion~(\ref{eqmozeta}) and~(\ref{covscal}). These form a multiplet,
as~(\ref{delQGamma3}) has the counterpart
\begin{equation}
  \widehat \delta (\epsilon ) \Delta^X= -\rmi\bar{\epsilon}^i \slashed\covder \Gamma^A f_{iA}^X+2\rmi\bar{\epsilon}^i \Gamma^B \bar{\zeta}^C \zeta^D f_{Bi}^Y {\cal R}^X
{}_{YCD}\,, \label{susydelta}
\end{equation}
where the covariant derivative of $\Gamma^A$ is defined similar
to~(\ref{defDzeta}). In the following, we will derive further constraints
on the target space geometry from requiring the presence of conformal
symmetry.

\subsubsection{Superconformal symmetry}\label{ss:schyper}

Now we define transformation rules for the hypermultiplet under the full
(rigid) superconformal group. The scalars do not transform under special
conformal transformations and special supersymmetry, but under
dilatations and $\SU(2)$ transformations, we parametrize
\begin{eqnarray}
\delta_D(\Lambda_D) q^X&=& \Lambda_D k^X(q) \,, \nonumber\\
\delta_{\SU(2)}(\Lambda^{ij}) q^X&=& \Lambda^{ij} k_{ij}^X(q) \,,
\label{SUzetacov}
\end{eqnarray}
for some unknown functions $k^X(q)$ and $k_{ij}^X(q)$.

To derive the appropriate transformation rules for the fermions, we first
note that the hyperinos should be invariant under special conformal
symmetry. This is due to the fact that this symmetry changes the Weyl
weight with one. If we realize the commutator~(\ref{KQcomm}) on the
fermions $\zeta^A$, we read off the special supersymmetry transformation
\begin{equation}
\delta_S(\eta^i)\zeta^A=-k^X f_{X}^{iA} \eta_i\,.
\end{equation}
To proceed, we consider the commutator of regular and special
supersymmetry~(\ref{SQcomm}). Realizing this on the scalars, we determine
the expression for the generator of $\SU(2)$ transformations in terms of
the dilatations and complex structures,
\begin{equation}\label{dil-su2}
k_{ij}^X= \frac 13 k^Y J_Y {}^X {}_{ij} \qquad \mbox{or}\qquad k^{\alpha
X}=\frac 13 k^Y\J\alpha YX\,.
\end{equation}
Realizing~(\ref{SQcomm}) on the hyperinos, we determine the covariant
variations
\begin{equation}
\widehat \delta_D \zeta^A = 2 \Lambda_D \zeta^A \,, \qquad \widehat
\delta_{\SU(2)} \zeta^A =0\,,
\end{equation}
and furthermore the commutator~(\ref{SQcomm}) only closes if we impose
\begin{equation}\label{conf-constr}
\covder_Y k^X 
= \frac 32 \delta_Y{}^X\,,
\end{equation}
which also implies
\begin{equation}
 \covder_Y k^{\alpha X}=\frac 12\J\alpha Y X \,.
 \label{DkSU2}
\end{equation}
Note that~(\ref{conf-constr}) is imposed by supersymmetry. In a more
usual derivation, where one considers symmetries of the lagrangian, we
would find this constraint by imposing dilatation invariance of the
action, see~(\ref{homothetic}). Our result, though, doesn't require the
existence of an action. The relations~(\ref{conf-constr})
and~(\ref{dil-su2}) further restrict the geometry of the target space,
and it is easy to derive that the Riemann tensor has four zero
eigenvectors,
\begin{equation}
k^XR_{YZX}{}^W=0\,,\qquad  k^{\alpha X}\,R_{YZX}{}^W=0\,. \label{konRis0}
\end{equation}
Also, under dilatations and $\SU(2)$ transformations, the hypercomplex
structure is scale invariant and rotated into itself,
\begin{eqnarray}
\Lambda_D \left( k^Z \partial_Z \J\alpha XY -\partial_Z k^Y \J\alpha
XZ+\partial_X k^Z \J\alpha ZY\right) &=&0\,, \nonumber\\
\Lambda^\beta  \left( k^{\beta Z}  \partial_Z \J\alpha XY -\partial_Z
k^{\beta Y} \J\alpha XZ+\partial_X k^{\beta Z} \J\alpha ZY\right)
&=&-\epsilon^{\alpha\beta\gamma}\Lambda^\beta \J\gamma XY\,.
\end{eqnarray}

All these properties are similar to those derived from superconformal
hypermultiplets in four spacetime
dimensions~\cite{deWit:1980gt,deWit:1999fp}. There, the $\Symp(1) \times
\Gl(r,\mathbb{H})$ sections, or simply, hypercomplex sections, were
introduced
\begin{equation}
A^{iB}(q)\equiv k^Xf_X^{iB}\,, \qquad
(A^{iB})^*=A^{jC}E_j{}^i\rho_C{}^B\,,
\end{equation}
which allow for a coordinate independent description of the target space.
This means that all equations and transformation rules for the sections
can be written without the occurrence of the $q^X$ fields. For example,
the hypercomplex sections are zero eigenvectors of the
$\Gl(r,\mathbb{H})$ curvature,
\begin{equation}
A^{iB}W_{BCD}{}^E=0\,,
\end{equation}
and have supersymmetry, dilatation and $\SU(2)$ transformation laws.
\begin{equation}
\widehat \delta A^{iB} = \frac 32f^{iB}_X\delta q^X = -\frac 32 \rmi{\bar
\epsilon}^i\zeta^B + \frac 32 \Lambda_D A^{iB} - \Lambda^i{}_jA^{jB}\,,
\label{deltaA}
\end{equation}
where $\widehat \delta$ is understood as a covariant variation, in the
sense of~(\ref{cov-var}).

\subsubsection{Symmetries}\label{ss:symmhyper}

We now assume the action of a symmetry group on the hypermultiplet. We
have no action, but the `symmetry' operation should leave invariant the
set of equations of motion. The symmetry algebra must commute with the
supersymmetry algebra (and later with the full superconformal algebra).
This leads to hypermultiplet couplings to a non-abelian gauge group $G$.
The symmetries are parametrized by
\begin{eqnarray}
\delta_G q^X&=& -g \Lambda_G^I k_I^X(q)\,, \nonumber\\
\widehat \delta_G \zeta^A &=& -g \Lambda^I_G t_{IB}{}^A(q) \zeta^B\,.
\label{gauge-tr}
\end{eqnarray}
The vectors $k_I^X$ depend on the scalars and generate the algebra of $G$
with structure constants $f_{IJ}{}^K$,
\begin{equation} \label{G-alg}
k_{[I|}^Y \partial_Y k^X_{|J]}=-\frac 12 f_{IJ}{}^K k_K^X \,.
\end{equation}
The commutator of two gauge transformations~(\ref{Lambda3}) on the
fermions requires the following constraint on the field-dependent matrices
$t_I(q)$,
\begin{equation} \label{tt-comm}
[t_I,t_J]_B{}^A=-f_{IJ}{}^Kt_{KB}{}^A -2k^X_{[I|}\covder_Xt_{|J]B}{}^A
+k^X_Ik^Y_J{\cal R}_{XYB}{}^A\,.
\end{equation}

Requiring the gauge transformations to commute with supersymmetry leads
to further relations between the quantities $k_I^X$ and $t_{IB}{}^A$.
Vanishing of the commutator on the scalars yields
\begin{equation} \label{GG-q}
t_{IB}{}^Af^X_{iA}=\covder_Yk^X_If^{Y}_{iB}\,.
\end{equation}
These constraints determine $t_I(q)$ in terms of the vielbeins $f^{iA}_X$
and the vectors $k^X_I$,
\begin{equation} \label{def-t}
t_{IA}{}^B=\frac 12f^Y_{iA}\covder_Yk^X_If^{iB}_X\,,
\end{equation}
and furthermore
\begin{equation} \label{symm-ff}
f^{Y(i}_Af^{j)B}_X \covder_Y k^X_I=0\,.
\end{equation}
The relations~(\ref{symm-ff}) and~(\ref{def-t}) are equivalent
to~(\ref{GG-q}). We interpret~(\ref{def-t}) as the definition for
$t_{IA}{}^B$. The vanishing of an $(ij)$-symmetric part in an equation
as~(\ref{symm-ff}) can be expressed as the vanishing of the commutator of
$\covder_Y k^X_I$ with the complex structures:\footnote{This can be seen
directly from lemma~\ref{lem:McommJ} in appendix~\ref{hyperappendix}.}
\begin{equation}
  \left(\covder_X k^Y_I\right)J^\alpha {}_Y{}^Z= J^\alpha {}_X{}^Y\left(\covder_Y
  k^Z_I\right)\,.
 \label{commDkJ}
\end{equation}
Extracting affine connections from this equation, it can be written as
\begin{equation}
\label{intJ} \left( {\cal L}_{k_I} J^\alpha\right) _X{}^Y\equiv  k^Z_I
\partial_Z \J\alpha XY -\partial_Z k^Y_I \J\alpha XZ+\partial_X
k^Z_I \J\alpha ZY=0\,.
\end{equation}
The left-hand side is the Lie derivative of the complex structure in the
direction of the vector $k_I$. In part~\ref{app:symmetries} of the
appendix, it is mentioned that~(\ref{intJ}) is a special case of the
statement that the vector $k_I$ normalizes the hypercomplex structures.
The latter would allow that this Lie derivative is proportional to a
complex structure. Killing vectors which normalize the hypercomplex
structure can be decomposed in an $\SU(2)$ part and a $\Gl(r,\mathbb{H})$
part. The vanishing of this Lie derivative, or~(\ref{symm-ff}), is
expressed by saying that the gauge transformations act
\emph{triholomorphic}. Thus, it says that all the symmetries are embedded
in $\Gl(r,\mathbb{H})$.
\bigskip

Vanishing of the gauge-supersymmetry commutator on the fermions requires
\begin{equation}
\covder_Y t_{IA}{}^B= k^X_I {\cal R}_{YXA} {}^B\,. \label{t-id}
\end{equation}
Using~(\ref{GG-q}) this implies a new constraint,
\begin{equation}
\label{DDk} \covder_X\covder_Y k^Z_I=R_{XWY}{}^Zk_I^W\,.
\end{equation}
Note that this equation is in general true for any Killing vector of a
metric. As we have no metric here, we could not rely on this fact, but
here the algebra imposes this equation. It turns out that~(\ref{symm-ff})
and~(\ref{DDk}) are sufficient for the full commutator algebra to hold.
In particular,~(\ref{t-id}) follows from~(\ref{DDk}), using the
definition of $t$ as in~(\ref{def-t}), and~(\ref{curv-rel}).

A further identity can be derived: substituting~(\ref{t-id})
into~(\ref{tt-comm}) one gets
\begin{equation}
[t_I,t_J]_B{}^A=-f_{IJ}{}^Kt_{KB}{}^A-k^X_Ik^Y_J{\cal R}_{XYB}{}^A\,.
\end{equation}
This identity can also be obtained from substituting~(\ref{def-t}) in the
commutator on the left hand side, and then
using~(\ref{G-alg}),~(\ref{symm-ff}),~(\ref{DDk}) and~(\ref{curv-rel}).
\bigskip

The group of gauge symmetries should also commute with the superconformal
algebra, in particular with dilatations and $\SU(2)$ transformations.
This leads to
\begin{equation}
\label{comDG} k^Y \covder_Y k_I^X=\frac {3}{2} k_I^X\,, \qquad k^{\alpha
Y}\covder_Yk_I^X=\frac 12 k_I^Y\J\alpha YX\,,
\end{equation}
coming from the scalars, and there are no new constraints from the
fermions or from other commutators. Since $\covder_Yk_I^X$ commutes with
$\J\alpha YX$, the second equation in~(\ref{comDG}) is a consequence of
the first one.
\bigskip

In the above analysis, we have taken the parameters $\Lambda^I$ to be
constants. In the following, we also allow for local gauge
transformations. The gauge coupling is done by introducing vector
multiplets and defining the covariant derivatives
\begin{eqnarray}
\covder_{\mu} q^X&\equiv& \partial_{\mu} q^X +g A_{\mu}^I k_I^X \,, \nonumber\\
\covder_{\mu} \zeta^A &\equiv& \partial_\mu \zeta^A + \partial_\mu q^X
\omega_{XB}{}^A \zeta^B + g A_{\mu}^I t_{IB} {}^A \zeta^B\,.
\label{covderg}
\end{eqnarray}
The commutator of two supersymmetries should now also contain a local
gauge transformation, in the same way as for the multiplets of the
previous sections, see~(\ref{softrigid}). This requires an extra term in
the supersymmetry transformation law of the fermion,
\begin{equation}
\widehat \delta (\epsilon) \zeta^A = \frac 12 \rmi \slashed{\covder} q^X
f_X^{iA} \epsilon_i +\frac {1}{2}g \sigma^I k_I^X f_{iX}^A \epsilon^i\,.
\end{equation}
With this additional term, the commutator on the scalars closes, whereas
on the fermions, it determines the equations of motion
\begin{equation}
\Gamma^A\equiv \slashed{\covder} \zeta^A +\frac 12 W_{BCD} {}^A
\bar{\zeta}^C \zeta^D \zeta^B-g(\rmi k_I^X f_{iX}^A \psi^{iI}+ \rmi
\zeta^B\sigma^I t_{IB} {}^A )=0\,,
\end{equation}
with the same conventions as in~(\ref{FinalCommzeta}).

Acting on $\Gamma^A$ with supersymmetry determines the equation of motion
for the scalars
\begin{eqnarray}
\Delta^X&=&\Box q^X -\frac 12 \bar{\zeta}^B \gamma_a \zeta^D \covder^a
q^Y f_Y^{iC}f_{iA}^X W_{BCD} {}^A-\frac {1}{4} \covder_Y W_{BCD}{}^A
\bar{\zeta}^E \zeta^D \bar{\zeta}^C \zeta^B f_E^{iY}f_{iA}^X- \\
&&-\,g \left(  2\rmi \bar{\psi}^{iI} \zeta^B t_{IB}{}^A f_{iA}^X- k_I^Y
J_Y{}^X {}_{ij} Y^{ijI} -\rmi
f^X_{iA}f^{Yi}_{B}\bar{\zeta}^B\zeta^C\sigma ^I \covder_Y
t_{IC}{}^A\right) +g^2 \sigma^I \sigma^J \covder_Y k_I^X
k_J^Y\,.\nonumber
\end{eqnarray}
The first line is the same as in~(\ref{covscal}), the second line
contains the corrections due to the gauging. The gauge-covariant
laplacian is here given by
\begin{equation}
\Box q^X=\partial_a \covder^a q^X +g \covder_a q^Y \partial_Y k_I^X
A^{aI}+\covder_a q^Y \covder^a q^Z \Gamma_{YZ}^X\,.
\end{equation}
The equations of motions $\Gamma^A $ and $\Delta ^X$ still satisfy the
same algebra with~(\ref{delQGamma3}) and~(\ref{susydelta}).

\section{Rigid superconformal actions} \label{ss:actions}

In this section, we will present rigid superconformal actions for the
multiplets discussed in the previous section. We will see that demanding
the existence of an action is more restrictive than only considering
equations of motion. For the different multiplets, we find that new
geometric objects have to be introduced.

\subsection{Vector multiplet action}

The coupling of Poincar{\'e}-supergravity to $n$ vector multiplets (having
$n$ scalars $\varphi^x$) is completely determined by an
$(n+1)$-dimensional constant symmetric tensor
$C_{IJK}$~\cite{Gunaydin:1984bi}. The reason for the difference in the
number of scalars and the rank of $C_{IJK}$ is that the graviton
multiplet also contains a vector field called the graviphoton.

The tensor $C_{IJK}$ appears directly in the $A \wedge F\wedge F$
Chern--Simons couplings, and indirectly in all other terms of the action.

In particular, the manifold parametrized by the scalars $\varphi^x$ of
the vector multiplets can be viewed as an $n$-dimensional hypersurface in
an $(n+1)$-dimensional space parametrized by $n+1$ coordinates
$\s^I(\varphi^x)$:
\begin{equation}
  C_{IJK} \s^I \s^J \s^K = 1 \,.
 \label{cisone}
\end{equation}
 The resulting geometry goes under the name of
``very special geometry''. For every value of $n$ there are many
different ``very special real'' manifolds: a classification of such
spaces that are homogeneous was given in~\cite{deWit:1992cr}. This
classification includes the previously found symmetric
spaces~\cite{Gunaydin:1984bi, Gunaydin:1986fg}.

{}From the viewpoint of superconformal symmetry, the
equation~(\ref{cisone}) looks like a gauge-fixing condition for
dilatation invariance. Indeed, it turns out that the coupling of $n$
vector multiplets (with $n$ scalars $\s^I$) in rigid supersymmetry (or in
conformal supergravity as we will give the generalization in
section~\ref{ss:localmult}) is also completely determined by the tensor
$C_{IJK}$, but in contrast to the case of Poincar{\'e} supergravity, this
tensor will multiply the complete action, not just the Chern--Simons
term.

The rigidly superconformal invariant action describing $n$ vector
multiplets was obtained from tensor calculus using an intermediate linear
multiplet in~\cite{Kugo:2000af}. The abelian part can be obtained by just
taking the (cubic) action of one vector multiplet as given
in~\cite{Bergshoeff:2001hc}, adding indices $I,J,K$ on the fields and
multiplying with the symmetric tensor $C_{IJK}$. For the non-abelian
case, we need conditions expressing the gauge invariance of this tensor:
\begin{equation}
  f_{I(J}{}^HC_{KL)H}=0\,.
 \label{cyclic}
\end{equation}
Moreover one has to add a few more terms, e.g.\ to complete the
Chern--Simons term to its non-abelian form. This leads to the action
\begin{eqnarray}
{\cal L}_{\rm vector} &=&
 \Bigl[ \left(- \frac 14 F_{\m\n}^{ I} F^{\m\n {
J}} - \frac 12  \bar{\p}^{ I} \slashed{\cal D} \p^{ J} - \frac 12 {\cal
D}_a \s^{ I} {\cal D}^a \s^{ J}
+  Y_{ij}^{ I} Y^{ij { J}}  \right) \s^{ K}-  \nn \\
&&
       \hphantom{\Bigl[} -\frac 1{24} \ve^{\m\n\l\r\s} A_\m^I \left( F_{\n\l}^J F_{\r\s}^K + \frac 12 g [A_\n, A_\l]^J F_{\r\s}^K + \frac 1{10} g^2 [A_\n, A_\l]^J [A_\r, A_\s]^K \right)- \nn \\
&&
       \hphantom{\Bigl[} -\frac 18 \rmi \bar{\p}^{ I} \g \cdot F^{ J} \p^{ K} -\frac 12
\rmi \bar{\p}^{i { I}} \p^{j { J}} Y_{ij}^{ K}  + \frac 14 \rmi g
\bar{\p}^{ L}\psi ^H\sigma ^{{I}}\sigma ^{{J}} f_{LH}{}^{K} \Bigr]
C_{IJK}  \,. \label{eq:Vectoraction}
\end{eqnarray}
The equations of motion for the fields of the vector multiplet following
from the action~(\ref{eq:Vectoraction}) are
\begin{equation}
0 = L^{ij}_I = \varphi^i_I = E^a_I = N_I \,,
\end{equation}
where we have defined
\begin{eqnarray}
L^{ij}_I &\equiv& C_{IJK} \left(2 \s^{ J} Y^{ij { K}} - \frac 12 \rmi
\bar{\p}^{i { J}} \p^{j { K}}\right),
\nonumber\\
\varphi^{i}_I &\equiv& C_{IJK} \left(
  \rmi \sigma ^{ J}\slashed{\cal D}\psi ^{i K}
  +\frac 12\rmi(\slashed{\cal D}\sigma ^{ J})
  \psi ^{i K}+Y^{ik J}\psi _k^{ K}
  -\frac 14\gamma \cdot F^{ J}\psi ^{i K}\right)- \nonumber \\
  && -\,g C_{IJK} f_{LH}{}^{K} \sigma^{ J} \sigma ^{ L}\psi ^{iH} \,, \nonumber\\
E_{a I} &\equiv&  C_{IJK} \left[{\cal D}^b \left(\s^{ J} F_{ba}{}^{ K} +
\frac 14 \rmi \bar{\p}^{ J} \g_{ba} \p^K \right) - \frac 18 \ve_{abcde}
F^{bc {J}}
F^{de { K}} \right]- \nonumber\\
&& -\,\frac 12 g C_{JKL} f_{IH}{}^J \s^K \bar\p^L \g_a \p^H - g C_{JKH}
f_{IL}{}^J \s^K \s^L {\cal D}_a \s^H
\,,\nonumber\\
 N_I
&\equiv& C_{IJK} \left(\s^{ J} \Box \s^{ K} + \frac 12 {\cal D}^a \s^{ J} {\cal D}_a \s^{ K}
-\frac14 F_{ab}^{ J} F^{ab  K} - \frac 12 \bar{\p}^{ J} \slashed{\cal D} \p^{ K} + Y^{ij  J} Y_{ij}{}^{ K} \right) + \nonumber \\
&& +\, \frac 12 \rmi g C_{IJK} f_{LH}{}^K \s^{J} \bar{\p}^{ L} \p^H \,.
 \label{eq:vectoreom}
\end{eqnarray}
We have given these equations of motion the names $L^{ij}_I, \phi^i_I,
E_{a I}, N_I$ since they form a linear multiplet in the adjoint
representation of the gauge group for which the transformation rules have
been given in~(\ref{eq:linmultiplet}).

\subsection{The vector-tensor multiplet action} \label{ss:actvt}

We will now generalize the vector action~(\ref{eq:Vectoraction}) to an
action for the vector-tensor multiplets (with $n$ vector multiplets and
$m$ tensor multiplets) discussed in section~\ref{ss:reducVT}.

The supersymmetry transformation rules for the vector-tensor
multiplet~(\ref{tensflat}) were obtained from those for the vector
multiplet~(\ref{ymflat}) by replacing all contracted indices by the
extended range of tilde indices. In addition, extra terms of ${\cal
O}(g)$ had to be added to the transformation rules. Similar
considerations apply to the generalization of the action, as we will see
below.

To obtain the generalization of the Chern-Simons (CS) term, it is
convenient to rewrite this CS-term as an integral in six dimensions which
has a boundary given by the five-dimensional Minkowski spacetime. The
six-form appearing in the integral is given by
\begin{equation}
I_{\rm vector} = C_{IJK} F^I F^J F^K \,, \label{YM6}
\end{equation}
where we have used form notation. This six-form is both gauge-invariant
and closed, by virtue of~(\ref{cyclic}) and the Bianchi
identities~(\ref{eq:BIvec}). It can therefore be written as the exterior
derivative of a five-form which is gauge-invariant up to a total
derivative. The spacetime integral over this five-form is the CS-term
given in the second line of~(\ref{eq:Vectoraction}).

We now wish to generalize~(\ref{YM6}) to the case of vector-tensor
multiplets. It turns out that the generalization of~(\ref{YM6}) is
somewhat surprising. We find
\begin{equation}
I_{\rm vec-tensor} = C_{\widetilde I\widetilde J\widetilde K} {\cal
H}^{\widetilde I}{\cal H}^{\widetilde J}{\cal H}^{\widetilde K} - \frac3g
\Omega_{MN} {\cal D}B^M {\cal D}B^N \,. \label{eq:WZ}
\end{equation}
The tensor $\Omega_{MN}$ is antisymmetric and invertible, and it
restricts the number of tensor multiplets to be \emph{even}
\begin{equation}
\O_{MN} = -\O_{NM} \,, \qquad \O_{MP} \O^{MR} = \d_P{}^R\,.
\end{equation}
The covariant derivative of the tensor field is
\begin{eqnarray}
  {\cal D}_\lambda B_{\rho\sigma}^N&=& \partial _\lambda B_{\rho\sigma}^N +
g\,  A_\lambda ^I t_{I\widetilde J}{}^N{\cal H}_{\rho\sigma}^{\widetilde
J}\nonumber\\ &=&
\partial _\lambda B_{\rho\sigma}^N +g\,
  A_\lambda ^I t_{IJ}{}^NF_{\rho\sigma}^J
   + g\, A_\lambda ^I t_{IP}{}^NB_{\rho\sigma}^P\,.
 \label{covDB}
\end{eqnarray}
When this is reduced to 5 dimensions, one of the ${\cal H}$ factors of
the first term of~(\ref{eq:WZ}) should correspond to a vector field
strength $F^I$ in order that it can be written as a 5-form $A^I{\cal
H}^{\widetilde J}{\cal H}^{\widetilde K}$. Thus, the components of $C$
can have only three different forms, namely $C_{IJK}$, $C_{IJM}$ and
$C_{IMN}$ (and permutations).

To see why~(\ref{eq:WZ}) is a closed six-form, we write out the first
term of~(\ref{eq:WZ})
\begin{equation}
C_{\widetilde I\widetilde J\widetilde K} {\cal H}^{\widetilde I}{\cal
H}^{\widetilde J}{\cal H}^{\widetilde K} = C_{IJK} F^I F^J F^K + 3
C_{IJM} F^I F^J B^M + 3 C_{IMN} F^I B^M B^N \,. \label{3terms}
\end{equation}
Since the $B^M$ tensors in~(\ref{3terms}) do not satisfy a Bianchi
identity, we also need the second term in~(\ref{eq:WZ}) to render it a
closed six-form. This requirement of closure leads to the following
relations between the $C$ and $\Omega$ tensors:
\begin{equation}
 C_{IJM}= t_{(IJ)}{}^N \Omega_{NM}\,, \qquad C_{IMN}=\frac 12 t_{IM}{}^P\Omega _{PN}\,.
 \label{valuesCM}
\end{equation}

We stress that the tensor $C_{\widetilde I\widetilde J\widetilde K}$ is
not a fundamental object: the essential data for the vector-tensor
multiplet are the representation matrices $t_{I \widetilde
J}{}^{\widetilde K}$, the Yang-Mills components $C_{IJK}$, and the
symplectic matrix $\Omega_{MN}$. The tensor components of the $C$ tensor
are derived quantities, and we can summarize~(\ref{valuesCM}) as
\begin{equation}
C_{M\widetilde J \widetilde K} = t_{(\widetilde{J} \widetilde{K})}{}^P
\Omega_{ PM}\,. \label{cot}
\end{equation}

{}From~(\ref{valuesCM}), we deduce that the second term of~(\ref{3terms})
only depends on the off-diagonal (between vector and tensor multiplets)
transformations. The first term of~(\ref{3terms}) will induce the usual
five-dimensional CS-term. The generalized CS-term induced by the third
term of~(\ref{3terms}) was given in~\cite{Gunaydin:1999zx}. With our
extension to also allow for the off-diagonal term in~(\ref{uppertriang}),
we also get CS-terms induced by the $C_{IJM}$ components, which were not
present in~\cite{Gunaydin:1999zx}.

Gauge invariance of the first term of~(\ref{eq:WZ}) requires that the
tensor $C$ satisfies a modified version of~(\ref{cyclic})
\begin{equation}
f_{I(J}{}^HC_{KL)H} = t_{I(J}{}^Mt_{KL)}{}^N\Omega _{MN}\,.
\label{fullinvC}
\end{equation}
In addition to this, the second term of~(\ref{eq:WZ}) is only gauge
invariant if the tensor $\Omega$ satisfies
\begin{equation}
  t_{I[M}{}^P\Omega _{N]P}=0\,,
 \label{giOmega}
\end{equation}
such that the last one of~(\ref{valuesCM}) is consistent with the
symmetry $(MN)$. The two conditions~(\ref{fullinvC}) and~(\ref{giOmega})
combined with the definition~(\ref{cot}) imply the following
generalization of~(\ref{cyclic})
\begin{equation}
t_{I(\widetilde J}{}^{\widetilde M} C_{\widetilde K\widetilde
L)\widetilde M} = 0 \,. \label{gaugeinvC}
\end{equation}

The superconformal action for the combined system of $m=2k$ tensor
multiplets and $n$ vector multiplets contains the CS-term induced
by~(\ref{eq:WZ}) and the generalization of the vector
action~(\ref{eq:Vectoraction}) to the extended range of indices. Some
extra terms are necessary to complete it to an invariant action: we need
mass terms and/or Yukawa coupling for the fermions at ${\cal O}(g)$, and
a scalar potential at ${\cal O}(g^2)$. We thus find the following action:
\begin{eqnarray}
{\cal L}_{\rm vec-tensor}\!\! &=& \left(- \frac 14 {\cal
H}_{\m\n}^{\widetilde I} {\cal H}^{\m\n {\widetilde J}} - \frac 12
\bar{\p}^{\widetilde I} \slashed{\cal D} \p^{\widetilde J} - \frac 12
{\cal D}_a \s^{\widetilde I} {\cal D}^a \s^{\widetilde J} +
Y_{ij}^{\widetilde I} Y^{ij {\widetilde J}}  \right) \s^{\widetilde K}
C_{{\widetilde I}{\widetilde J}{\widetilde K}}+
\nn \\
&&{}+ \frac {1}{16 g}  \ve^{\m\n\l\r\s} \Omega_{MN} B_{\m\n}^M \left(
\partial _\lambda B_{\rho\sigma}^N +2g
  t_{IJ}{}^NA_\lambda ^I F_{\rho\sigma}^J
   + g\, t_{IP}{}^N A_\lambda ^I B_{\rho\sigma}^P\right)- \nonumber\\
 & &  -\frac1{24} \ve^{\m\n\l\r\s} C_{IJK} A_\m^I\!\left(\! F_{\n\l}^J F_{\r\s}^K + f_{FG}{}^J A_\n^F A_\l^G \!\left(\!- \frac 12 g \, F_{\r\s}^K + \frac1{10} g^2 f_{HL}{}^K A_\r^H A_\s^L \!\right)\!\right)\!\!-
\nn \\&& {}- \frac{1}{8} \varepsilon^{\mu\nu\lambda\rho\sigma} \Omega_{MN} t_{IK}{}^M t_{FG}{}^N A_{\mu}^I A_\nu^F A_\lambda^G
\left(-\frac 12 g \, F_{\rho\sigma}^K  + \frac{1}{10} g^2 f_{HL}{}^K A_\rho^H A_\sigma^L \right)+ \nn \\
& & + \left(-\frac18 \rmi \bar{\p}^{\widetilde I} \g \cdot {\cal
H}^{\widetilde J} \p^{\widetilde K} -\frac 12 \rmi \bar{\p}^{i
{\widetilde I}} \p^{j {\widetilde J}} Y_{ij}^{\widetilde K}  \right)
C_{{\widetilde I}{\widetilde J}{\widetilde K}}+
\nn \\
& & + \frac14 \rmi g \bar{\p}^{\widetilde I}\psi ^{\widetilde{J}}\sigma
^{\widetilde{K}}\sigma ^{\widetilde{L}}\left(
t_{[\widetilde{I}\widetilde{J}]}{}^{\widetilde{M}}C_{\widetilde{M}\widetilde{K}\widetilde{L}}
-4
t_{(\widetilde{I}\widetilde{K})}{}^{\widetilde{M}}C_{\widetilde{M}\widetilde{J}\widetilde{L}}
\right)- \nonumber\\
&&- \frac 12 g^2 \s^K C_{KMN} t_{I{\widetilde L}}{}^M \s^I \s^{\widetilde
L} t_{J{\widetilde P}}{}^N \s^J \s^{\widetilde P} \,. \label{LtensAnsatz}
\end{eqnarray}
To check the supersymmetry of this action, one needs all the relations
between the various tensors given above. Another useful identity implied
by the previous definitions is
\begin{equation}
t_{(\widetilde I \widetilde J)}{}^M C_{\widetilde K \widetilde L M} = -
t_{(\widetilde K \widetilde L)}{}^M C_{\widetilde I \widetilde J M} \,.
\end{equation}

The terms in the action containing the fields of the tensor multiplets
can also be obtained from the field equations~(\ref{deltaSij}). They are
now related to the action as
\begin{equation}
  \frac{\delta  S_{\rm vec-tensor}}{\delta  \bar \psi ^{iM}}=\rmi
  \varphi^N_i \Omega _{NM}\,,
 \label{normStens}
\end{equation}
and the remaining bosonic terms can be obtained from comparing with $N^M$
in~(\ref{scalareom}). One may then further check that also the field
equations~(\ref{tensEOM}) and~(\ref{tensBI}) follow from this action.

Note however that the equations of motion for the vector multiplet
fields, obtained from this action, are similar to those given
in~(\ref{eq:vectoreom}), but with the contracted indices running over the
extended range of vector and tensor components. Furthermore, the
$A_\mu^I$ equation of motion gets corrected by a term proportional to the
self-duality equation for $B_{\mu\nu}^M$:
\begin{equation}
\frac{\delta  S_{\rm vec-tensor}}{\delta A_a^I} = E^a_I + \frac {1}{12} g
\varepsilon^{abcde} A_b^J E_{cde}{}^M t_{JI}{}^N \Omega_{MN}\,.
\label{eq:vectoreom_vt}
\end{equation}

Finally, we remark that the action~(\ref{LtensAnsatz}) is invariant under
supersymmetry for the completely general form~(\ref{uppertriang}) of the
representation matrices $(t_I)_{\widetilde J}{}^{\widetilde K}$.

We thus conclude that in order to write down a superconformal action for
the vector-tensor multiplet, we need to introduce another geometrical
object, namely a gauge-invariant anti-symmetric invertible tensor
$\Omega_{MN}$. This symplectic matrix will restrict the number of tensor
multiplets to be even. We can still allow the transformations to have
off-diagonal terms between vector and tensor multiplets, if we
adapt~(\ref{cyclic}) to~(\ref{gaugeinvC}). In this way, we have
constructed more general matter couplings than were known so far. Terms
of the form $A \wedge F \wedge B$ did not appear in previous papers. We
see that such terms appear generically in our lagrangian by allowing for
these off-diagonal gauge transformations for the tensor fields. In some
cases these may disappear after field redefinitions.

\subsection{The hypermultiplet}
Until this point, the equations of motion we derived, found their origin
in the fact that we had an open superconformal algebra. The non-closure
functions $\Gamma^A$, together with their supersymmetric partners
$\Delta^X$ yielded these equations of motion. We discovered a
hypercomplex scalar manifold $\mathcal{M}$, where $\Gamma_{XY}{}^Z$ was
interpreted as an affine connection. We also needed a $\Gl
(r,\mathbb{H})$-connection $\omega_{XA}{}^B$ on a vector bundle and
discovered that the manifold also admitted a trivial $\SU(2)$-vector
bundle.

Now, we will introduce an action to derive the field equations of the
hypermultiplet. An important point to note is that the necessary data for
the scalar manifold we had in the previous section, are not sufficient
any more. This is not specific to our setting, but is a general property
of non-linear sigma models.

In such models, the kinetic term for the scalars is multiplied by a
scalar-dependent symmetric tensor $g_{\alpha\beta}(\phi)$,
\begin{equation}
S=-\frac 12 \int d^Dx \; g_{\alpha\beta}(\phi) \partial_\mu  \phi^\alpha
\partial^\mu  \phi^\beta\,,
\end{equation}
in which $\alpha$ and $\beta$ run over the dimensions of the scalar
manifold. The tensor $g$ is interpreted as the metric on the target space
$\mathcal{M}$. As the field equations for the scalars should now be also
covariant with respect to coordinate transformations on the target
manifold, the connection on the tangent bundle $T \mathcal{M}$ should be
the Levi-Civita connection. Only in that particular case, the field
equations for the scalars will be covariant. In other words, in $\Box
\phi^\alpha+\dots =0$ the Levi-Civita connection on $T \mathcal{M}$ will
be used in the covariant box.

To conclude, we will need to introduce a metric on the scalar manifold,
in order to be able to write down an action. This metric will also
restrict the possible target spaces for the theory.

Observe that most steps in this section do not depend on the use of
superconformal symmetry.\footnote{Of course, the form of the field
equations does reflect the superconformal symmetry.} Only at the end of
section~\ref{isoaction}, we make explicitly use of this symmetry.

\subsubsection{Without gauged isometries}

To start with, we take the non-closure functions $\Gamma^A$ to be
proportional to the field equations for the fermions $\zeta^A$. In other
words, we ask
\begin{equation}
\frac{\delta S_{\rm hyper}}{\delta \bar{\zeta}^A}=2C_{AB}\Gamma ^B\,.
\end{equation}

In general, the tensor $C_{AB}$ could be a function of the scalars and
bilinears of the fermions. If we try to construct an action with the
above Ansatz, it turns out that the tensor  has to be anti-symmetric in
$AB$ and
\begin{eqnarray}
\frac{\delta C_{AB}}{\delta \zeta^C}&=&0\,, \label{Czeta}\\
\covder_X C_{AB}&=&0\,. \label{Cq}
\end{eqnarray}
This means that the tensor does not depend on the fermions and is
covariantly constant.\footnote{This can also easily be seen by using the
Batalin-Vilkovisky formalism.}

This tensor $C_{AB}$ will be used to raise and lower indices according to
the NW--SE convention similar to $\varepsilon_{ij}$:
\begin{equation}
A_A = A^BC_{BA} \,,\qquad A^A = C^{AB} A_B\,,
 \label{NWSE}
\end{equation}
where $\varepsilon^{ij}$ and $C^{AB}$ for consistency should satisfy
\begin{equation}
  \varepsilon_{ik}\varepsilon^{jk} = \delta _i{}^j\,,\qquad
 C_{AC} C^{BC} =   \delta_A{}^B \,.
 \label{epsCup}
\end{equation}

We may choose $C_{AB}$ to be constant. To prove this, we look at the
integrability condition for~(\ref{Cq})
\begin{equation}
\left[\covder_X,\covder_Y\right]C_{AB}=0=-2{\cal R}_{XY[A}{}^CC_{B]C}\,.
\end{equation}
This implies that the anti-symmetric part of the connection
$\omega_{XAB}\equiv \omega _{XA}{}^CC_{CB}$ is pure gauge, and can be
chosen to be zero. If we do so, the covariant constancy condition for
$C_{AB}$ reduces to the equation that $C_{AB}$ is just constant. For this
choice, the connection $\omega_{XAB}$ is symmetric, so the structure
group $\Gl (r,\mathbb{H})$ breaks to $\USp(2r-2p,2p)$. The signature is
the signature of $d_{CB}$, which is defined as $C_{AB}=\rho_A {}^C
d_{CB}$ where $\rho_A {}^C$ was given in~(\ref{defrho}). However, we will
allow $C_{AB}$ also to be non-constant, but covariantly constant.

We now construct the metric on the scalar manifold as
\begin{equation}
  g_{XY} = f_X^{iA} C_{AB} \varepsilon_{ij}f_Y^{jB}\,.
 \label{defg}
\end{equation}
The above-mentioned requirement that the Levi-Civita connection should be
used (as $\Gamma_{XY}{}^Z$) is satisfied due to~(\ref{Cq}). Indeed, this
guarantees that the metric is covariantly constant, such that the affine
connection is the Levi-Civita one. On the other hand we have seen already
that for covariantly constant complex structures we have to use the Obata
connection. Hence, the Levi-Civita and Obata connection should coincide,
and this is obtained from demanding~(\ref{Cq}) using the Obata
connection. This makes us conclude that we can only write down an action
for a hyperk{\"a}hler scalar manifold.

We can now write down the action for the rigid hypermultiplets. It has
the following form:
\begin{equation} \label{abaction}
S_{\rm hyper}= \int d^5 x  \left(-\frac 12 g_{XY} \partial_a q^X
\partial^a q^Y + \bar{\zeta}_A \slashed{\covder} \zeta^A-\frac 14 W_{ABCD}
\bar{\zeta}^A \zeta^B \bar{\zeta}^C \zeta^D\right),
\end{equation}
where the tensor $W_{ABCD}$ can be proven to be completely symmetric in
all of its indices (see appendix~\ref{hyperappendix}). The field
equations derived from this action are
\begin{eqnarray}\label{feh}
\frac{\delta S_{\rm hyper}}{\delta \bar{\zeta}^A} &=& 2C_{AB}\Gamma^B \,, \nonumber\\
\frac{\delta S_{\rm hyper}}{\delta q^X}&=&g_{XY} \Delta^Y -
2\bar{\zeta}_A \Gamma^B \omega_{XB} {}^A\,.
\end{eqnarray}
Also remark that due to the introduction of the metric, the expression of
$\Delta^X$ simplifies to
\begin{equation}
\Delta^X=\Box q^X - \bar{\zeta}^A \slashed{\partial} q^Y \zeta^B  {\cal
R}^X {}_{YAB}-\frac {1}{4} \covder^X W_{ABCD}\bar{\zeta}^A \zeta^B
\zeta^C \zeta^D\,.
\end{equation}

Let us mention that we could also have followed a slightly different
route. We could have introduced the metric $g_{XY}$ first, and shown that
the connection $\Gamma_{XY}{}^Z$ is the Levi-Civita connection with
respect to this metric, as pointed out in the introduction of this
section. Then, the identification of the vielbeins $f_{iA}^X$ of the
tangent bundle $T\mathcal{M}$ with the $\Gl(r,\mathbb{H}) \otimes \SU(2)$
vector bundle would enable us to find a standard antisymmetric tensor $C
\otimes \varepsilon$ on the latter bundle. As the metric is covariantly
constant, this should be inherited by $C \otimes \varepsilon$, reflecting
the possibility to choose it to be constant. The result of the
introduction of a metric is that the scalar manifold should be
hyperk{\"a}hler.

\paragraph{Conformal invariance.} Due to the presence of the metric, the
condition for the homothetic Killing vector~(\ref{conf-constr}) implies
that $k_X$ is the derivative of a scalar function as in~(\ref{exact}).
This scalar function $\chi (q)$ is called the hyperk{\"a}hler
potential~\cite{Swann,deWit:1998zg,deWit:1999fp}. It determines the
conformal structure, but should be restricted to
\begin{equation}
  \covder_X\covder_Y\,\chi =\frac 32 g_{XY}\,.
 \label{DDchiisg}
\end{equation}
The relation with the homothetic Killing vector is
\begin{equation}
  k_X=\partial _X \chi \,,\qquad \chi =\frac 13 k_Xk^X\,.
 \label{kchi}
\end{equation}
Note that this implies that, when $\chi $ and the complex structures are
known, one can compute the metric with~(\ref{DDchiisg}), using the
formula for the Obata connection~(\ref{Obata}).

\subsubsection{With gauged isometries}\label{isoaction}

With a metric, the symmetries of section~\ref{ss:symmhyper} should be
isometries, i.e.\
\begin{equation}
  \covder_X k_{YI} + \covder_Y k_{XI}=0\,.
 \label{Killingeq}
\end{equation}
This makes the requirement~(\ref{DDk}) superfluous, but we still have to
impose the tri\-holo\-morphicity expressed by either~(\ref{symm-ff})
or~(\ref{commDkJ}) or~(\ref{intJ}).

In order to integrate the equations of motion to an action we have to
define (locally) triples of `moment maps', according to
\begin{equation}
\partial_X P_I^\alpha =-\frac 12J^\alpha{}_{XY} k_I^Y\,. \label{momentmap}
\end{equation}
The integrability condition that makes this possible is the
triholomorphic condition.

In the kinetic terms of the action, the derivatives should now be
covariantized with respect to the new transformations. We are also forced
to include some new terms proportional to $g$ and $g^2$
\begin{eqnarray}
S_{\rm hyper}^g&=& \int d^5x \biggl(\!-\frac 12 g_{XY} \covder_a q^X
\covder^a q^Y+ \bar{\zeta}_A \slashed{\covder} \zeta^A-\frac 14 W_{ABCD}
\bar{\zeta}^A \zeta^B \bar{\zeta}^C \zeta^D-
\label{Sna}\\
&&\hphantom{\int d^5x \biggl(} -g\left( 2 \rmi k_I^X f_{iX}^A
\bar{\zeta}_A \psi^{iI}+\rmi \sigma^I t_{IB} {}^A \bar{\zeta}_A \zeta^B-2
P_{Iij} Y^{Iij}\right)-g^2\frac 12 \sigma^I \sigma^J k_I^X k_{JX}\!
\biggr) \,,\quad \nonumber
\end{eqnarray}
[where the covariant derivatives $\covder$ now also include
gauge-covariantization proportional to $g$ as in~(\ref{covderg})], while
the field equations have the same form as in~(\ref{feh}). Supersymmetry
of the action imposes
\begin{equation}
  k_I^X J^\alpha{}_{XY} k_J^Y= 2f_{IJ}{}^K P_K^\alpha \,.
 \label{constraintfP}
\end{equation}
As only the derivative of $P$ appears in the defining
equation~(\ref{momentmap}), one may add an arbitrary constant to $P$. But
that changes the right-hand side of~(\ref{constraintfP}). One should then
consider whether there is a choice of these coefficients such
that~(\ref{constraintfP}) is satisfied. This is the question about the
center of the algebra, which is discussed
in~\cite{D'Auria:1991fj,Andrianopoli:1997cm}. For simple groups there is
always a solution.\footnote{We thank Gary Gibbons for a discussion on
this subject.} For abelian theories the constant remains undetermined.
This free constant is the so-called Fayet--Iliopoulos term.

In a conformal invariant theory, the Fayet--Iliopoulos term is not
possible. Indeed, dilatation invariance of the action needs
\begin{equation}
3P^\alpha _I=k^X \partial_XP^\alpha _I\,.
\end{equation}
Thus, $P_I^\alpha $ is completely determined [using~(\ref{momentmap})
or~(\ref{comDG})] as (see also~\cite{Wit:2001bk})
\begin{equation}
  -6P^\alpha _I=k^XJ^\alpha{}_{XY}  k_I^Y= -\frac 23k^Xk^Z\J\alpha ZY \covder_Y k_{IX}\,.
\end{equation}
The proof of the invariance of the action under the complete
superconformal group, uses the equation obtained from~(\ref{comDG})
and~(\ref{momentmap}):
\begin{equation}
k^{X\alpha } \covder_X k_I^Y= \partial^Y P^\alpha _I\,.
\end{equation}
If the moment map $P^\alpha _I$ has the value that it takes in the
conformal theory, then~(\ref{constraintfP}) is satisfied due
to~(\ref{G-alg}). Indeed, one can multiply that equation with
$k_Xk^ZJ^\alpha {}_Z{}^W\covder_W$ and use~(\ref{commDkJ}),~(\ref{DDk})
and~(\ref{konRis0}). Thus, in the superconformal theory, the moment maps
are determined and there is no further relation to be obeyed, i.e.\ the
Fayet--Iliopoulos terms of the rigid theories are absent in this case.
\bigskip

To conclude, isometries of the scalar manifold that commute with
dilatations, see~(\ref{comDG}), can be gauged. The resulting theory has an
extra symmetry group $G$, its algebra is generated by the corresponding
Killing vectors.

\subsection{Potential} \label{ss:potential}

We complete this section with a discussion of the scalar potential for
the general matter-coupled (rigid) superconformal theory. Gathering
together our results~(\ref{LtensAnsatz}) and~(\ref{Sna}) the total
lagrangian describing the most general couplings of vector/tensor
multiplets to hypermultiplets with rigid superconformal symmetry is
\begin{equation}
{\cal L}_{\rm total} = {\cal L}_{\rm vec-tensor} + {\cal L}^g_{\rm
hyper}\,. \label{Ltotal}
\end{equation}

{}From this expression the explicit form of the total scalar potential can
be read off as
\begin{equation}
V(\s^{\widetilde I}, q^X) = \s^{\widetilde K} C_{{\widetilde
I}{\widetilde J}{\widetilde K}} Y_{ij}^{\widetilde I} Y^{ij\, {\widetilde
J}} +\frac 12 g^2 \s^K C_{KMN} t_{I{\widetilde L}}{}^M \s^I
\s^{\widetilde L} t_{J{\widetilde P}}{}^N \s^J \s^{\widetilde P} + \frac
12 g^2 \s^I \s^J
 k^X_I k_{JX} \,, \label{potential}
\end{equation}
where
\begin{equation}
Y^{ij\widetilde{J}}C_{I\widetilde{J}\widetilde K}\sigma ^{\widetilde
K}=-gP_I^{ij}\,,\qquad Y^{ij\widetilde{J}}C_{M\widetilde{J}\widetilde
K}\sigma ^{\widetilde K}=0\,.
 \label{valueY}
\end{equation}
Note that the auxiliary field $Y$ has been eliminated here. Secondly,
written as in~(\ref{potential}), the potential does not contain the
auxiliary field $Y^{\widetilde I}$ any more, but rather its solution of
the field equations. This explains the apparent wrong sign in the
$Y_{ij}^I Y^{ij \, J}$ term, and the field equation made use of the term
$2gP_{Iij} Y^{Iij}$ in~(\ref{Sna}). In fact, the first term
of~(\ref{potential}) is equal to $-gP_{Iij} Y^{Iij}$.

This potential reflects the general form in supersymmetry that it is the
square of the transformations of the fermions, where the definition of
`square' uses the fermion kinetic terms. The first term is the square of
the transformations of the gauginos, the second term depends on the
transformations of the fermionic partners of the antisymmetric tensors,
and the last one is the square of the transformation law of the
hyperinos. Note that off-diagonal terms between the contributions of
$Y^{ij}$ and the $t_{I{\widetilde L}}{}^M \s^I \s^{\widetilde L}$ terms
do not survive as these would be proportional to $\varepsilon
^{ij}Y_{ij}=0$.

The difference between our potential~(\ref{potential}) and the one in a
rigid limit of~\cite{Ceresole:2000jd}, is the generalization to
off-diagonal couplings of vectors and tensor multiplets in the first two
terms.

Summarizing, in this section the actions of rigid superconformal
vector/tensor-hy\-per\-mul\-ti\-plet couplings have been constructed. The
full answer is~(\ref{Ltotal}). We found that the existence of an action
requires the presence of additional tensorial objects.
Table~\ref{tbl:action} gives an overview of what are the independent
objects to know, either to determine the transformation laws, or to
determine the action.

In the next section we generalize our results to the local case.

\landscape
\begin{table}[htb]
\begin{center}
\begin{tabular}{||c||c|c||c|c|}
\hline &
 \multicolumn{2}{c||}{ALGEBRA (no action)}&
 \multicolumn{2}{c|}{ACTION}\\ \hline
 multiplets & objects & Def/restriction & objects & Def/restriction\\
\hline\hline\rule[-1mm]{0mm}{6mm}
 Vect.    & $f_{[IJ]}{}^K$ & Jacobi identities
            & $C_{(IJK)}$ & $f_{I(J}{}^H C_{KL)H} = 0\ \blacktriangle$\\[1mm]
\hline\rule[-1mm]{0mm}{11mm}
 Vect./Tensor &
\begin{tabular}{l}       $(t_I)_{{\widetilde J}}{}^{\widetilde K}$ \\[2mm]
{\scriptsize $\widetilde{I}=(I,M)$} \end{tabular}&
\begin{tabular}{l}
  $[t_I, t_J] = -f_{IJ}{}^{K} t_K$\\[2mm]
  $t_{IJ}{}^K=f_{IJ}{}^K \,,\qquad t_{IM}{}^J=0$\\
\end{tabular} &
$\Omega_{[MN]}$  &
\begin{tabular}{l}
invertible \\[2mm]
   $ f_{I(J}{}^HC_{KL)H}=t_{I(J}{}^Mt_{KL)}{}^N\Omega _{MN}$ \\[2mm]
$t_{I[M}{}^P \Omega_{N]P} = 0 $    \\
\end{tabular}
     \\[1mm]
\hline \rule[-1mm]{0mm}{12mm}
 Hyper     &  $f_X{}^{iA}$ &
\begin{tabular}{l}
  invertible and real using $\rho $  \\[2mm]
   Nijenhuis condition: $N_{XY}{}^Z=0$
\end{tabular}
 & $C_{[AB]}$ & $\covder_XC_{AB}=0$\\[1mm]
\hline \rule[-1mm]{0mm}{12mm}
 \begin{tabular}{l}Hyper  +\\ gauging \end{tabular}  & $k_I^X$ &
 \begin{tabular}{l} $\covder_X\covder_Y k^Z_I=R_{XWY}{}^Zk_I^W\ \blacktriangleleft$\\[2mm]
 $k_{[I|}^Y \partial_Y k^X_{|J]}=-\frac 12 f_{IJ}{}^K k_K^X $\\[2mm]
 ${\cal L}_{k_I} J^\alpha=0\ \blacktriangleleft $
  \end{tabular}
 & $P_I^\alpha\ \blacktriangle$ &   \begin{tabular}{l}
 $\covder_X k_{YI} + \covder_Y k_{XI}=0$\\[2mm]
 $\partial_X P_I^\alpha =-\frac 12 J^\alpha{}_{XY} k^Y_I\ \blacktriangle$\\[2mm]
 $ k_I^X J_{XY}^\alpha k_J^Y=2f_{IJ}{}^K P_K^\alpha\ \blacktriangle$
 \end{tabular}
 \\[1mm]
\hline
 \begin{tabular}{c}Hyper + \\ conformal \end{tabular}&$k^X\ \blacktriangleleft$ &
  $\covder_Y k^X =\frac32 \delta_Y{}^X\ \blacktriangleleft$&$\chi $ &
  $ \covder_X\covder_Y\, \chi =\frac32 g_{XY}$ \\ \hline
 \begin{tabular}{c}Hyper + \\ conformal +
 gauged \end{tabular} & &
 $k^Y \covder_Y k_I^X=\frac{3}{2} k_I^X$ & & \\ \hline
\end{tabular}
\caption{\it Various matter couplings with or without action. We indicate
which are the geometrical objects that determine the theory and what are
the independent constraints. The symmetries of the objects are already
indicated when they appear first. In general, the equations should also
be valid for the theories in the rows below (apart from the fact that
`hyper+gauging' and `hyper+conformal' are independent, but both are used
in the lowest row). However, the symbol $\blacktriangle$ indicates that
these equations are not to be taken over below. E.g.\ the moment map
$P_I^\alpha $ itself is completely determined in the conformal theory,
and it should thus not any more be given as an independent quantity. For
the rigid theory without conformal invariance, only constant pieces can
be undetermined by the given equations, and are the Fayet--Iliopoulos
terms. On the other hand, the equations indicated by $\blacktriangleleft$
have not to be taken over for the theories with an action, as they are
then satisfied due to the Killing equation or are defined by $\chi$.
\label{tbl:action} }
\end{center}
\end{table}
\endlandscape
\section{Local superconformal multiplets}\label{ss:localmult}

We are now ready to perform the last step in our programme, i.e.~extend
the supersymmetry to a local conformal supersymmetry. We will make use
here of the off-shell $32+32$ Weyl multiplet constructed
in~\cite{Bergshoeff:2001hc,Fujita:2001kv}, and in particular of the
`standard' Weyl multiplet. In fact, there exist two Weyl multiplets: the
`dilaton' Weyl multiplet and the `standard' Weyl multiplet. They contain
the same gauge fields but differ in their matter fields. We restrict
ourselves here to the standard Weyl multiplet, due to two considerations.
First, it turns out that with the standard Weyl multiplet we already find
a local generalization for any rigid theory. Second, the experience in
other similar situations has shown that two different sets of auxiliary
fields for theories with the same rigid limit do not lead to physically
different results. This has e.g.\ been investigated in full detail for
the old minimal, new minimal and non-minimal set of auxiliary fields for
$N=1$, $D=4$ in~\cite{Ferrara:1983dh}. We therefore expect that the
couplings to the dilaton Weyl multiplet are only those obtained from the
replacement of the fields of the standard Weyl multiplet by their
functions in terms of the dilaton Weyl multiplet given
in~\cite[eq.~(3.14)]{Bergshoeff:2001hc}. Whether the conformal
gauge-fixing program will also be insensitive to the choice of Weyl
multiplet, remains to be seen. For instance in~\cite{Fujita:2001kv}, the
connection between the dilaton Weyl multiplet and an inequivalent set of
auxiliary fields for Poincar{\'e} supergravity~\cite{Nishino:2000cz} was
discussed.

We have listed all the gauge fields and matter fields of the standard Weyl
multiplet in table~\ref{tbl:fieldsWeyls}. For the full details of the
standard Weyl multiplet, we refer to~\cite{Bergshoeff:2001hc}.

\bigskip

\begin{table}[ht]
\begin{center}
$  \begin{array}{||c|cccc||} \hline\hline
\mbox{Field} & \#  & \mbox{Gauge}&\SU(2)& w \\
\hline \rule[-1mm]{0mm}{6mm}& \multicolumn{4}{c||}{\mbox{Elementary gauge
fields}} \\
\rule[-1mm]{0mm}{6mm} e_\mu{}^a & 9 & P^a &1&-1\\
\rule[-1mm]{0mm}{6mm}
b_\mu   & 0 & D &1&0\\
\rule[-1mm]{0mm}{6mm}
V_\mu^{(ij)} &12  & \SU(2)\,&3&0\\
\hline \rule[-1mm]{0mm}{6mm} \psi _\mu ^i  & 24 & Q^i_\alpha &2&-\frac12
\\ [1mm] \hline\hline \rule[-1mm]{0mm}{6mm}&
\multicolumn{4}{c||}{\mbox{Dependent gauge
fields}} \\
\rule[-1mm]{0mm}{6mm} \omega_\mu{}^{ab} & - & M^{[ab]} &1&0\\
\rule[-1mm]{0mm}{6mm}
f_\mu{}^a   & - & K^a &1&1\\
\hline \rule[-1mm]{0mm}{6mm} \phi _\mu ^i  & - & S^i_\alpha &2&\frac12 \\
[1mm] \hline\hline \rule[-1mm]{0mm}{6mm} &
\multicolumn{4}{c||}{\mbox{Matter fields
}}\\
\rule[-1mm]{0mm}{6mm} T_{[ab]}  & 10 &  &1&1\\
\rule[-1mm]{0mm}{6mm} D & 1 & &1&2\\ \hline \rule[-1mm]{0mm}{6mm} \chi ^i
& 8 & &2&\frac32\\ [1mm] \hline\hline
\end{array}$
  \caption{\it Fields of the standard Weyl multiplet.
The symbol $\#$ indicates the off-shell degrees of freedom. The first
block contains the (bosonic and fermionic) gauge fields of the
superconformal algebra. The fields in the middle block are dependent
gauge fields. In the lower block are the extra matter fields that appear
in the standard Weyl multiplet.
 }\label{tbl:fieldsWeyls}
\end{center}
\end{table}

The procedure for extending the rigid superconformal transformation rules
for the various matter multiplets is to introduce covariant derivatives
with respect to the superconformal symmetries. These derivatives contain
the superconformal gauge fields which, in turn, will also transform to
additional matter fields (this is explained in detail
in~\cite{Bergshoeff:2001hc}).

Since in the previous sections we have explained most of the subtleties
concerning the possible geometrical structures, we can be brief here. We
will obtain our results in two steps. First, we require that the local
superconformal commutator algebra, as it is realized on the standard Weyl
multiplet (see~\cite[eqs.~(4.3)--(4.6)]{Bergshoeff:2001hc}) is also
realized on the matter multiplets (with possible additional
transformations under which the fields of the standard Weyl multiplet do
not transform, and possibly field equations if the matter multiplet is
on-shell). Note that this local superconformal algebra is a modification
of the rigid superconformal algebra~(\ref{SScomm}),~(\ref{QQcomm}) where
all modifications involve the fields of the standard Weyl multiplet.

Now we will apply a standard Noether procedure to extend the rigid
supersymmetric actions to a locally supersymmetric one. This will
introduce the full complications of coupling the matter multiplets to
conformal supergravity.

\subsection{Vector-tensor multiplet}

The local supersymmetry rules are given by
\begin{eqnarray}
\d A_\m^I
&=& \frac 12 \bar{\e} \g_\m \p^I -\frac 12 \rmi \s^I \bar\e \p_\mu \,, \nonumber\\
\d B_{ab}^M &=& - \bar \e \g_{[a} D_{b]} \p^M- \ft 12 \rmi \sigma^M \bar
\epsilon  \widehat{R}_{ab}(Q) + \rmi \bar \epsilon  \gamma _{[a} \gamma
\cdot T \gamma _{b]} \psi ^M+\nonumber\\
&& + \rmi g \bar{\e} \g_{ab} t_{({\widetilde J}{\widetilde K)}}{}^M
\s^{\widetilde J} \p^{\widetilde K} + \rmi \bar{\eta}\g_{ab}\p^M\,, \nonumber\\
\d Y^{ij \widetilde I}
&=& -\frac 12 \bar{\e}^{(i} \slashed{D} \p^{j) \widetilde I}
+\frac 12 \rmi \bar\e^{(i} \gamma \cdot T \p^{j) \widetilde I}
- 4 \rmi \s^{\widetilde I} \bar\e^{(i} \chi^{j)}- \nonumber \\
& & -\frac 12 \rmi g \bar \e^{(i} \left(t_{[{\widetilde J}{\widetilde
K]}}{}^{\widetilde I} - 3 t_{({\widetilde J}{\widetilde
K)}}{}^{\widetilde I} \right)
\sigma^{\widetilde J} \psi^{j) {\widetilde K}} + \frac 12 \rmi \bar{\eta}^{(i} \p^{j) \widetilde I} \,, \nonumber \\
\d \p^{i \widetilde I} &=& - \frac 14 \g \cdot \widehat{{\cal
H}}{}^{\widetilde I} \e^i -\frac 12\rmi \slashed{D} \s^{\widetilde I}
\e^i - Y^{ij \widetilde I} \e_j + \s^{\widetilde I} \gamma \cdot T \e^i
+\frac 12 g t_{({\widetilde J}{\widetilde K)}}{}^{\widetilde I}
\s^{\widetilde J}
\s^{\widetilde K} \e^i + \s^{\widetilde I} \eta^i \,, \nonumber\\
\d \s^{\widetilde I} &=& \frac 12 \rmi \bar{\e} \p^{\widetilde I} \,.
\label{tensorlocal}
\end{eqnarray}
The covariant derivatives are given by
\begin{eqnarray} \label{localderiv-tensor}
D_\mu\, \sigma^{\widetilde I} &=& {\cal D}_\m \s^{\widetilde I} - \frac
12\, \rmi \bar{\psi}_\mu \psi^{\widetilde I} \,, \nn \\ [2pt]
 {\cal D}_\m \s^{\widetilde I} &=& (\partial_\mu - b_\mu)
\sigma^{\widetilde I} + g t_{J\widetilde K}{}^{\widetilde I} A_\mu^J
\sigma^{\widetilde K} \,, \nn \\ [2pt]
 D_\mu \psi^{i \widetilde I} &=&
{\cal D}_\m \psi^{i \widetilde I} + \frac 14 \g \cdot \widehat{{\cal
H}}^{\widetilde I} \p_\mu^i + \frac 12\rmi \slashed{D} \s^{\widetilde I}
\p_\mu^i + Y^{ij \widetilde I} \p_{\mu\, j}
 - \s^{\widetilde I} \g \cdot T \p_\mu^i - \nonumber\\
 && -\frac 12 g
t_{({\widetilde J}{\widetilde K)}}{}^{\widetilde I} \s^{\widetilde J}
\s^{\widetilde K} \psi _\mu ^i - \s^{\widetilde I} \phi_\mu^i  \,, \nn \\
{\cal D}_\m \psi^{i \widetilde I} &=& (\partial_\mu - \frac 32 \, b_\mu +
\frac 14\, \g_{ab}\, \o_\mu {}^{ab}) \p^{i \widetilde I} - V_\mu^{ij}
\p_j^{\widetilde I} + g t_{J\widetilde K}{}^{\widetilde I} A_\mu^J
\psi^{i \widetilde K} \,.
 \end{eqnarray}
The covariant curvature $\widehat{{\cal H}}^{\widetilde I}_{\mu \nu }$
should be understood as having components $(\widehat F^I_{\mu \nu
},B^M_{\mu \nu })$ and
\begin{equation}
\widehat{F}_{\m\n}^I = 2 \partial_{[\mu } A_{\nu ]}^I + g f_{JK}{}^I
A_\mu^J A_\nu^K - \bar{\p}_{[\m} \g_{\n]} \p^I + \frac 12 \rmi \s^I
\bar{\p}_{[\m} \p_{\n]}
 \,.
\end{equation}

The locally superconformal constraints needed to close the algebra are
given by the following extensions of~(\ref{tensEOM}) and~(\ref{tensBI})
(which are non-zero only for $\widetilde I$ in the tensor multiplet
range)
\begin{eqnarray}
L^{ij M} &\equiv& t_{({\widetilde J}{\widetilde K)}}{}^M \left(2
\s^{\widetilde J} Y^{ij {\widetilde K}} - \frac 12 \rmi \bar{\p}^{i
{\widetilde J}} \p^{j {\widetilde K}}\right)=0 \,,\nonumber\\
 E_{abc}^M &\equiv& \frac3g D_{[a} B_{bc]}{}^M - \varepsilon _{abcde}
t_{({\widetilde J}{\widetilde K)}}{}^M \left(\sigma ^{\widetilde J}
\widehat{{\cal H}}^{de {\widetilde K}} - 8 \sigma ^{\widetilde J} \sigma
^{\widetilde K} T^{de} + \frac14 \rmi \bar{\psi }^{\widetilde J}
\gamma ^{de} \psi ^{\widetilde K} \right)- \nonumber\\
&&-\, \frac 3{2g} \bar{\psi }^M \gamma _{[a} \widehat{R}_{bc]}(Q) \nn\\
&=& 0 \,. \label{eomB}
\end{eqnarray}
Here, the supercovariant derivative on the tensor is defined as
\begin{eqnarray*}
D_{[a}B_{bc]}^M&=&\partial_{[a}B_{bc]}^M-2b_{[a}B^M_{bc]}-2\omega_{[ab}{}^dB_{c]d}^M+\bar{\psi}_{[a}\gamma_b
D_{c]}\psi^M+\ft12\rmi
\sigma^M\bar{\psi}_{[a}\widehat{R}_{bc]}(Q)-\nonumber\\&& -\rmi
\bar{\psi}_{[a}\gamma_b \gamma \cdot T \gamma_{c]}\psi^M-\rmi
\bar{\phi}_{[a}\gamma_{b c]}\psi^M-\nonumber\\&& -\rmi g
\bar{\psi}_{[a}\gamma_{bc]}\psi^{\widetilde{K}}\sigma^{\widetilde{J}}t_{(\widetilde{J}\widetilde{K})}{}^M+gt_{J\widetilde{K}}{}^MA_{[a}^J
\widehat{\mathcal{H}}_{bc]}^{\widetilde{K}}\,.
\end{eqnarray*}
Analogously to subsection~\ref{ss:reducVT}, the full set of constraints
could be obtained by varying these constraints under supersymmetry.

The action, invariant under local superconformal symmetry, can be
obtained by replacing the rigid covariant derivatives
in~(\ref{LtensAnsatz}) by the local covariant
derivatives~(\ref{localderiv-tensor}) and adding extra terms proportional
to gravitinos or matter fields of the Weyl multiplet, determined by
supersymmetry. It is more convenient to use a new tensor field
$\widetilde{B}_{\mu \nu}^M$ defined as
\begin{displaymath}
B_{\mu \nu}^M=\widetilde{B}_{\mu
\nu}^M-\bar{\psi}_{[\mu}\gamma_{\nu]}\psi^M+\ft12 \rmi \bar{\psi}_\mu
\psi_\nu \sigma^M\,,
\end{displaymath}
such that the symbol $\widehat{\mathcal{H}}^{\widetilde{I}}_{\mu \nu}$
now means
\begin{displaymath}
\widehat{\mathcal{H}}^{\widetilde{I}}_{\mu \nu}=(F_{\mu
\nu}^I-\bar{\psi}_{[\mu}\gamma_{\nu]}\psi^I+\ft12 \rmi \bar{\psi}_\mu
\psi_\nu \sigma^I,\widetilde{B}_{\mu
\nu}^M-\bar{\psi}_{[\mu}\gamma_{\nu]}\psi^M+\ft12 \rmi \bar{\psi}_\mu
\psi_\nu \sigma^M)\,.
\end{displaymath}
The action reads
\begin{eqnarray}\label{conf-VTaction}
e^{-1}\! {\cal L}_{\rm vec-ten}^{\rm conf}\!\! &=\!&\! \biggl[ \left(-
\frac14 {\cal\widehat H}_{\m\n}^{ {\widetilde I}} {\cal\widehat H}^{\m\n
{ {\widetilde J}}} - \frac 12 \bar{\p}^{ {\widetilde I}} \slashed{D} \p^{
{\widetilde J}} + \frac 13 \s^{\widetilde I} \Box^c \s^{\widetilde J} +
\frac 16 D_a \s^{ {\widetilde I}} D^a \s^{ {\widetilde J}}
+  Y_{ij}^{ {\widetilde I}} Y^{ij { {\widetilde J}}}  \right) \s^{ {\widetilde K}}- \nn \\
& &\hphantom{\! \biggl[} {-}\frac 43 \s^{\widetilde I} \s^{\widetilde J}
\s^{\widetilde K} \left(D + \frac{26}{3} T_{ab} T^{ab} \right)
+ 4 \s^{\widetilde I} \s^{\widetilde J} {\cal\widehat  H}_{ab}^{\widetilde K} T^{ab} - \nn \\
& &\hphantom{\! \biggl[} -\frac 18 \rmi \bar{\p}^{ {\widetilde I}} \g
\cdot {\cal\widehat H}^{ {\widetilde J}} \p^{ {\widetilde K}} -\frac 12
\rmi \bar{\p}^{i { {\widetilde I}}} \p^{j { {\widetilde J}}} Y_{ij}^{
{\widetilde K}}  + \rmi \s^{\widetilde I} \bar\p^{\widetilde J}
\g \cdot T \p^{\widetilde K} - 8 \rmi \s^{\widetilde I} \s^{\widetilde J} \bar\p^{\widetilde K} \chi+\nn \\
& &\hphantom{\! \biggl[}
{+} \frac 16 \s^{\widetilde I} \bar \p_\m \g^\m \!\left(\!\rmi \s^{\widetilde J} \slashed{D} \p^{\widetilde K} + \frac 12 \rmi \slashed{D} \s^{\widetilde J} \p^{\widetilde K} - \frac 14 \g
{\cdot} {\cal\widehat  H}^{\widetilde J} \p^{\widetilde K}
+ 2 \s^{\widetilde J} \g {\cdot} T \p^{\widetilde K} - 8 \s^{\widetilde J} \s^{\widetilde K} \chi \!\right)\! {-}\nn\\
&&\hphantom{\! \biggl[}
{-} \frac 16 \bar \p_a \g_b \p^{\widetilde I} \left(\s^{\widetilde J} {\cal\widehat  H}^{ab {\widetilde K}} -8 \s^{\widetilde J} \s^{\widetilde K} T^{ab} \right) -\frac 1{12} \s^{\widetilde I} \bar \p_\l \g^{\m\n\l} \p^{\widetilde J} {\cal\widehat  H}_{\m\n}^{\widetilde K} {+} \nn\\
&&\hphantom{\! \biggl[} {+}\frac 1{12} \rmi \s^{\widetilde I} \bar \p_a
\p_b  \left(\s^{\widetilde J} {\cal\widehat  H}^{ab {\widetilde K}} -8
\s^{\widetilde J} \s^{\widetilde K} T^{ab} \right) +\frac 1{48} \rmi
\s^{\widetilde I} \s^{\widetilde J} \bar \p_\l \g^{\m\n\l\r} \p_\r
{\cal\widehat  H}_{\m\n}^{\widetilde K} {-} \nn \\&& \hphantom{\!
\biggl[} {-} \frac 12 \s^{\widetilde I} \bar \p_\m^i \g^\m \p^{j
\widetilde J} Y_{ij}^{\widetilde K} +\frac 1{6} \rmi \s^{\widetilde I}
\s^{\widetilde J} \bar\p_\m^i \g^{\m\n} \p_\n^j Y_{ij}^{\widetilde K}
-\frac{1}{24} \rmi \bar \p_\m \g_\n \p^{\widetilde I} \bar \p^{\widetilde
J} \g^{\m\n} \p^{\widetilde K}{+} \nn \\&& \hphantom{\! \biggl[}
{+}\frac{1}{12} \rmi \bar \p_\mu^i \g^\mu \p^{j {\widetilde I}} \bar
\p_i^{\widetilde J} \p_j^{\widetilde K} -\frac{1}{48} \s^{\widetilde I}
\bar \p_\m \p_\n \bar \p^{\widetilde J} \g^{\m\n} \p^{\widetilde K}
+\frac {1}{24} \s^{\widetilde I} \bar\p_\m^i \g^{\m\n} \p_\n^j
\bar\p_i^{\widetilde J} \p_j^{\widetilde K} {-} \nn\\&& \hphantom{\!
\biggl[}
{-}\frac 1{12} \s^{\widetilde I} \bar \p_\l \g^{\m\n\l} \p^{\widetilde J} \bar{\p}_{\m} \g_{\n} \p^{\widetilde K} + \frac 1{24} \rmi
\s^{\widetilde I} \s^{\widetilde J} \bar \p_\l \g^{\m\n\l} \p^{\widetilde K} \bar{\p}_{\m} \p_{\n}+ \nn \\
& &\hphantom{\! \biggl[}
 {+} \frac 1{48} \rmi \s^{\widetilde I} \s^{\widetilde J} \bar \p_\l \g^{\m\n\l\r} \p_\r \bar{\p}_{\m} \g_{\n} \p^{\widetilde K} + \frac1{96}\s^{\widetilde I} \s^{\widetilde J} \s^{\widetilde K} \bar \p_\l \g^{\m\n\l\r} \p_\r \bar{\p}_{\m} \p_{\n}\biggr]
  C_{{\widetilde I}{\widetilde J}{\widetilde K}} + \nn\\
  &&+ \frac {1}{16 g} e^{-1} \varepsilon ^{\mu \nu \rho \sigma \tau} \Omega_{MN}
\widetilde{B}_{\mu \nu}^M \Big(
\partial _\rho \widetilde{B}_{\sigma \tau}^N +2g\,  t_{IJ}{}^NA_\rho^I F_{\sigma \tau}^J
   + g\, t_{IP}{}^N A_\rho^I \widetilde{B}_{\sigma \tau}^P\Big)-\nonumber\\
& &  -\frac 1{24} e^{-1} \ve^{\m\n\l\r\s} C_{IJK} A_\m^I\!
 \left(\!F_{\n\l}^J F_{\r\s}^K + f_{FG}{}^J \!A_\n^F A_\l^G \!  \left(\!{-} \frac 12 g \, F_{\r\s}^K + \frac1{10} g^2 f_{HL}{}^K A_\r^H A_\s^L  \!\right)\!\right)\!\! {-}
\nn \\
&& - \frac{1}{8} e^{-1} \varepsilon^{\mu\nu\lambda\rho\sigma} \Omega_{MN} t_{IK}{}^M t_{FG}{}^N A_{\mu}^I A_\nu^F A_\lambda^G \left(-\frac 12 g \, F_{\rho\sigma}^K
+ \frac{1}{10} g^2 f_{HL}{}^K A_\rho^H A_\sigma^L \right)+ \nn \\
 &&+ \frac14 \rmi g \bar{\p}^{\widetilde I}\psi
^{\widetilde{J}}\sigma ^{\widetilde{K}}\sigma ^{\widetilde{L}}\left(
t_{[\widetilde{I}\widetilde{J}]}{}^{\widetilde{M}}C_{\widetilde{M}\widetilde{K}\widetilde{L}}
-4t_{(\widetilde{I}\widetilde{K})}{}^{\widetilde{M}}C_{\widetilde{M}\widetilde{J}\widetilde{L}}
\right)- \nonumber\\
&& - \frac14 g \bar\psi _\mu  \gamma ^\mu \psi ^{\widetilde I} \sigma
^{\widetilde J} \sigma^{\widetilde K} \sigma^{\widetilde L}
t_{(\widetilde J \widetilde K)}{}^{\widetilde M} C_{\widetilde M
\widetilde I \widetilde L}- \nonumber\\
&&  - \frac 12 g^2 \s^I \s^J \s^K \s^{\widetilde M} \s^{\widetilde N}
t_{J{\widetilde M}}{}^P t_{K{\widetilde N}}{}^Q C_{IPQ} \,,
\end{eqnarray}
where the superconformal d'alembertian is defined as
\begin{eqnarray}
 \Box^c
\s^{\widetilde I}
&=& D^a D_a \s^{\widetilde I} \nn \\[2pt]
&=& \left( \partial^a  - 2 b^a + \o_b^{~ba} \right) D_a \s^{\widetilde I}
+ g t_{J \widetilde K}{}^{\widetilde I} A_a^J D^a \sigma^{\widetilde K} -
\frac{1}2\rmi \bar{\p}_\m D^\m \p^{\widetilde I} - 2 \s^{\widetilde I}
\bar{\p}_\m \g^\m \chi +
\nn \\[2pt]
& & + \frac12 \bar{\p}_\m \g^\m \g \cdot T \p^{\widetilde I} + \frac12
\bar{\f}_\m \g^\m \p^{\widetilde I} + 2 f_\m{}^\m \s^{\widetilde I}
-\frac 12 g \bar\p_\m \g^\m t_{ J \widetilde K}{}^{\widetilde I}\p^J
\s^{\widetilde K}\,.
\end{eqnarray}

\subsection{Hypermultiplet}

Imposing the local superconformal algebra we find the following
supersymmetry rules:
\begin{eqnarray}
\delta q^X
&=& - \rmi \bar\epsilon^i \zeta^A f_{iA}^X \,,\nonumber\\
\widehat \delta \zeta^A &=& \frac 12 \rmi \slashed{D} q^X f_X^{iA}
\epsilon_i
  - \frac13 \gamma \cdot T k^X f^A_{iX} \epsilon^i + \frac 12 g \sigma^I k_I^X f_{iX}^A \epsilon^i + k^X f^A_{iX} \eta^i\, .
\end{eqnarray}
The covariant derivatives are given by
\begin{eqnarray}
D_\mu q^X
&=& {\mathcal D}_\mu q^X + \rmi \bar{\psi}_\mu^i \zeta^A f_{iA}^X\, , \nonumber\\
{\mathcal D}_{\mu} q^X &=& \partial_\mu  q^X - b_\mu  k^X - V_\mu^{jk}
k_{jk}^X +
 g A_{\mu}^I k_I^X \,, \nn\\
D_\mu \zeta^A &=& {\mathcal D}_\mu \zeta^A - k^X f_{iX}^A \phi_\mu^i +
\frac 12 \rmi \slashed{D} q^X f_{iX}^A \psi_\mu^i + \frac13 \gamma \cdot
T k^X f_{iX}^A \psi_\mu^i -  \frac 12 g\sigma^I k_I^X f_{iX}^A \psi^i_{\mu} \nonumber\\
{\mathcal D}_\mu \zeta^A &=& \partial_\mu  \zeta^A + \partial_\mu  q^X
\omega_{XB} {}^A \zeta^B + \frac14 \omega_\mu {}^{bc} \gamma_{bc} \zeta^A
- 2 b_\mu  \zeta^A + g A_\mu^I t_{IB}{}^A \zeta^B\, .
\end{eqnarray}

Similar to section~\ref{ss:hypermultiplet}, requiring closure of the
commutator algebra on these transformation rules yields the equation of
motion for the fermions
\begin{eqnarray}
\Gamma_{\rm conf}^A &=& \slashed{D} \zeta^A + \frac 12 W_{CDB} {}^A
\zeta^B \bar{\zeta}^D \zeta^C - \frac 83 \rmi k^X f_{iX}^A \chi^i + 2
\rmi \gamma \cdot T \zeta^A -
\nn\\
&& -g\left(\rmi k_I^X f_{iX}^A \psi^{iI}+ \rmi \sigma^I t_{IB} {}^A
\zeta^B\right). \label{fermion_eom_local}
\end{eqnarray}
The scalar equation of motion can be obtained from
varying~(\ref{fermion_eom_local}):
\begin{equation}
\widehat \delta_Q \Gamma^A = \frac 12 \rmi f_X^{iA} \Delta^X \epsilon_i +
\frac14 \gamma^\mu \Gamma^A \bar{\epsilon} \psi_\mu - \frac14 \gamma^\mu
\gamma^\nu \Gamma^A \bar{\epsilon} \gamma_\nu \psi_\mu \,,
\end{equation}
where
\begin{eqnarray}
\Delta^X_{\rm conf} &=& \Box^c q^X -\frac 12 \bar{\zeta}^B \gamma_a
\zeta^D D^a q^Y f_Y^{iC}f_{iA}^X W_{BCD} {}^A-\frac {1}{4} \covder_Y
W_{BCD}{}^A \bar{\zeta}^E \zeta^D \bar{\zeta}^C \zeta^B f_E^{iY}f_{iA}^X+ \nonumber\\
&& + \frac89 T^2 k^X + \frac43 D k^X + 8 \rmi \bar{\chi}^i \zeta^A
f_{iA}^X -\\
&& -g \big( 2\rmi \bar{\psi}^{iI} \zeta^B t_{IB}{}^A f_{iA}^X - k_I^Y
J_Y{}^X {}_{ij} Y^{Iij}  -\rmi \bar{\zeta}^A\zeta^B \sigma ^I{\cal
D}^Xt_{IAB} \big)+ g^2 \sigma^I \sigma^J \mathfrak{D}_Y k_I^X
k_J^Y\,,\nonumber
\end{eqnarray}
and the superconformal d'alembertian is given by
\begin{eqnarray}
\Box^c q^X
&\equiv& D_a D^a q^X \nn \\
&=& \partial_a D^a q^X - \frac{5}{2} b_a \left(\partial
_Yk^X+\delta_Y{}^X\right)  D^a q^Y - V_a^{ij}(\partial_Yk^X_{ij}) D^a q^Y
+ \rmi \bar{\psi}_a^i
D^a \zeta^A f_{iA}^X + \nn\\
&& + 2 f_a{}^a k^X - 2\bar{\psi}_a \gamma^a \chi k^X + 4
\bar{\psi}_a^{(j} \gamma^a \chi^{k)} k_{jk}^X - \bar{\psi}_a^i \gamma^a
\gamma \cdot T \zeta^A f_{iA}^X- \nn \\
&&- \bar{\phi}_a^i \gamma^a \zeta^A f_{iA}^X + \omega_a {}^{ab} D_b q^X
-\frac 12 g \bar{\psi}^a \gamma_a \psi^I k_I^X- g D_a q^Y \partial_Y
k_I^X A^{aI}+ \nonumber\\&&+D_a q^Y D^a q^Z \Gamma_{YZ}^X\,.
\end{eqnarray}

Note that so far we didn't require the presence of an action. Introducing
a metric, the locally conformal supersymmetric action is given by
\begin{eqnarray}\label{conf-hyperaction}
e^{-1} \mathcal{L}_{\rm hyper}^{\rm conf}
&=& - \frac 12 g_{XY} \mathcal{D}_a q^X {\mathcal{D}^a} q^Y +\bar{\zeta}_A \slashed{D} \zeta^A + \frac 49 Dk^2 + \frac {8}{27} T^2 k^2
- \nn\\
&& - \frac{16}{3} \rmi \bar{\zeta}_A \chi^i k^X f_{iX}^A
+ 2 \rmi \bar{\zeta}_A \gamma \cdot T \zeta^A
-\frac14 W_{ABCD} \bar{\zeta}^A \zeta^B \bar{\zeta}^C \zeta^D - \nn\\
&& - \frac29 \bar{\psi}_a \gamma^a \chi k^2 + \frac 13 \bar{\zeta}_A \gamma^a \gamma \cdot T \psi_a^i k^X f_{iX}^A
+ \frac 12 \rmi \bar \zeta_A \gamma^a \gamma^b \psi_a^i \mathcal{D}_b q^X f_{iX}^A + \nn\\
&& + \frac23 f_a {}^a k^2 - \frac16 \rmi \bar{\psi}_a \gamma^{ab} \phi_b k^2
 - \bar{\zeta}_A \gamma^a \phi_a^i k^X f_{iX}^A +\nn\\
&& + \frac{1}{12} \bar{\psi}_a^i \gamma^{abc} \psi_b^j \mathcal{D}_c q^Y J_Y {}^X {}_{ij} k_X - \frac 19 \rmi \bar{\psi}^a \psi^b T_{ab} k^2 + \frac {1}{18} \rmi \bar{\psi}_a \gamma^{abcd} \psi_b T_{cd} k^2
-\nn\\
&& - g\biggl( \rmi \sigma^I t_{IB} {}^A  \bar{\zeta}_A \zeta^B + 2 \rmi k_I^X f_{iX}^A \bar{\zeta}_A \psi^{iI} +\frac 12 \sigma^I k_I^X f_{iX}^A \bar{\zeta}_A \gamma^a \psi_a^i
+ \nn\\
&&\hphantom{- g\biggl(} + \bar{\psi}_a^i \gamma^a \psi^{jI}
P_{Iij}-\frac{1}{2} \rmi \bar{\psi}_a^i \gamma^{ab} \psi^j_b \sigma^I
P_{Iij}\biggr)+
\nn\\
&&{}+2g Y^{ij}_I P_{ij}^I  -\frac 12 g^2 \sigma^I \sigma^J k_I^X
k_{JX}\,.
\end{eqnarray}
No further constraints, other than those given in
section~\ref{ss:hypermultiplet} were necessary in this local case. In
particular, the target space is still hypercomplex or, when an action
exists, hyperk{\"a}hler. This action leads to the following dynamical
equations
\begin{eqnarray}
\frac{\delta \mathcal{S}_{\rm hyper}^{\rm conf}}{\delta \bar{\zeta}^A}
&=& 2\, C_{AB} \Gamma^B_{\rm conf}\, ,\nonumber\\
\frac{\delta \mathcal{S}_{\rm hyper}^{\rm conf}}{\delta q^X} &=& g_{XY}
\left(\Delta^Y_{\rm conf} - 2 \bar{\zeta}_A \Gamma^B_{\rm conf} \omega^Y
{}_B {}^A - \rmi \bar{\psi}_a^i \gamma^a \Gamma^A_{\rm conf}
f_{iA}^Y\right).
\end{eqnarray}

The lagrangians~(\ref{conf-VTaction}) and~(\ref{conf-hyperaction}) are
the starting point for obtaining matter couplings to Poincar{\'e}
supergravity. This involves a gauge fixing of the local scale and
$\SU(2)$ symmetries, which will be studied in a forthcoming paper.

\section{Conclusions and discussion}

In this paper, we have analysed various multiplets in five spacetime
dimensions with $N=2$ supersymmetry in a superconformal context. Although
we have so far only considered rigid supersymmetry and superconformal
(both rigid and local) supersymmetry, we have found new couplings. The
main emphasis was on the vector-tensor multiplet and on the
hypermultiplet. Both these multiplets are on-shell and from the closure
of the supersymmetry algebra, one can read off the equations of motion
that determine the dynamics of the system. These equations of motion do
not necessarily follow from an action. The existence of an action
requires extra tensors which are needed to integrate the equations of
motion into an action. In this way we have generalized the work
of~\cite{Fujita:2001kv} where off-shell hypermultiplets were considered,
leading e.g.\ to a restricted class of quaternionic-K{\"a}hler manifolds.

For vector-tensor multiplets, we have written down equations of motion
with an \emph{odd} number of tensor multiplets in the background of an
arbitrary number of vector multiplets. This is in contrast with
formulations based on an action, where an even number of tensor
multiplets is always needed. Even in the case when an action exists, we
have found new couplings where vectors and tensors mix non-trivially due
to the off-diagonal structure of the representation matrices for the
gauge group. This introduces new terms in the scalar potential, such that
we have a broader class of models than in the existing literature so far.
We hope that these new potentials lead to interesting new physical
applications.

For hypermultiplets, it has been known that the geometry of the scalars
is hyperk{\"a}hler for rigid supersymmetry~\cite{Alvarez-Gaume:1981hm} or
quaternionic-K{\"a}hler for supergravity~\cite{Bagger:1983tt}. This was based
on an analysis of the requirements imposed by the existence of an
invariant action, and has been fully proved in~\cite{DeJaegher:1998ka}.
We have written down equations of motion without the need of a target
space metric (and thus a supersymmetric action), but which only involve a
vielbein and a triplet of integrable complex structures. The resulting
geometry is that of a \emph{hypercomplex} manifold, which is a weakened
version of hyperk{\"a}hler geometry where the Ricci tensor is antisymmetric
and not necessarily zero.

Since the appearance of hypercomplex geometry is somehow new in the
physics literature, we have discussed their properties in
appendix~\ref{hyperappendix}. Group manifolds, e.g.\ $\SU(3)$, provide
examples of hypercomplex geometries that are not hyperk{\"a}hler, and we have
computed the non-vanishing components of the Ricci tensor for
hypercomplex group manifolds in appendix~\ref{app:exHypercplxGr}. The
main condition for a hypermultiplet action to exist, is the presence of a
target space metric. In that case, the target space becomes hyperk{\"a}hler.
Our results then coincide with the literature.

The results of our analysis, both with and without actions, are summarized
in table~\ref{tbl:action}, where we indicate the various geometrical
tensors and the restrictions they are subject to. The resulting scalar
potential is displayed in section~\ref{ss:potential}. After the analysis
for rigid conformal supersymmetry, we have extended our results to local
conformal supersymmetry. However, it turns out that the extra constraints
that are necessary for allowing rigid conformal symmetry are also
sufficient for the extension to local conformal supersymmetry. For this
formulation, we have used the previously obtained results on the Weyl
multiplet in five dimensions~\cite{Bergshoeff:2001hc,Fujita:2001kv}.

Note that in constructing these superconformal theories, we have allowed
kinetic terms for the scalars with arbitrary signature. This will be
important for the conformal gauge-fixing programme, where the
compensating scalars should have negative kinetic terms in order that the
full theory has positive kinetic energy. The couplings of superconformal
matter to the Weyl multiplet are gauge equivalent to matter-coupled
Poincar{\'e} supergravities. This involves a partial gauge fixing, which we
will investigate in a forthcoming paper, and which has been considered
for some cases in~\cite{Fujita:2001kv,Fujita:2001bd}. This should lead to
actions that can be compared with those
in~\cite{Gunaydin:1999zx,Ceresole:2000jd}.

However, not all our results can fall in the theories of the present
literature. We mentioned already above the extension to off-diagonal
vector-tensor couplings. The other extension is due to not requiring the
existence of an action.

{}From a string theory viewpoint, this is quite a natural thing to do. In
fact, string theory does not lead to an action, but it leads to field
equations, which in most cases can be integrated to an action. We should
point out that there are also other techniques for constructing matter
couplings that do not lead to an action. In many cases, the presence of
self-dual antisymmetric tensor fields makes the construction of actions
non-trivial. The gaugings of $N=8$ supergravity in 5 dimensions require in
some cases an odd number of antisymmetric tensors, which prohibits the
construction of an action~\cite{Andrianopoli:2000fi}. Its reduction to
$N=2$ theories should be in the class of the theories of this paper that
are not based on an action.

This interesting aspect of our paper is not confined to five spacetime
dimensions. A similar analysis can be done in other dimensions as well.
The results were obtained by emphasizing the distinction between
requirements from the algebra and requirements from action invariance,
which is especially interesting for multiplets with an `open' algebra,
where equations of motions are generated from the anticommutator of two
supersymmetries. E.g.\ the hypercomplex manifolds can be obtained in the
same way for $D=4$ and $D=6$ theories with 8 supersymmetries.

We conclude by remarking that it is likely that our newly found
matter-couplings will survive after gauge-fixing the local superconformal
symmetry to $N=2$ Poincar{\'e} supergravity. It will be of interest to see
the consequences of our results for studying domain walls,
renormalization group flows in the context of the AdS/CFT correspondence,
and for finding a supersymmetric Randall-Sundrum scenario.
\medskip
\section*{Acknowledgments}

\noindent We are grateful to Gary Gibbons, Dominic Joyce, Stefano
Marchiafava, George Papadopoulos and Walter Troost for interesting and
useful discussions. Special thanks go to the mathematicians Dmitri
Alekseevsky, Vicente Cort{\'e}s and Chand Devchand, who helped us in the
preparation of appendix~\ref{hyperappendix}. Part of the work was
performed while E.B., S.V. and A.V.P. were at the Isaac Newton institute
for Mathematical Sciences, whose hospitality we gratefully appreciated.
Work supported in part by the European Community's Human Potential
Programme under contract HPRN-CT-2000-00131 Quantum Spacetime, in which
E.B., T.d.W. and R.H. are associated with University Utrecht. J.G. is
Aspirant-FWO. The work of T.d.W. and R.H. is part of the research program
of the Stichting voor Fundamenteel Onderzoek der materie (FOM).

\newpage
\appendix
\section{The linear multiplet}

The significance of the linear multiplet appears when we introduce an
action for the vector multiplet, see~(\ref{eq:Vectoraction}) in
section~\ref{ss:actions}. This action contains a constant totally
symmetric tensor $C_{IJK}$. In section~\ref{ss:actions} we saw that this
tensor characterizes a special geometry. The linear multiplet is related
to this vector multiplet action in the sense that the equations of
motion~(\ref{eq:vectoreom}) that follow from the
action~(\ref{eq:Vectoraction}) transform precisely as a linear multiplet
in the adjoint representation.

The degrees of freedom of the linear multiplet are given in
table~\ref{tbl:multiplets}. We will consider a linear multiplet in the
background of an off-shell (non-abelian) vector multiplet. We take the
fields of the linear multiplet in an arbitrary representation of
dimension $m$. The rigid conformal supersymmetry transformation rules for
a linear multiplet in the background of a Yang-Mills multiplet are given
by
\begin{eqnarray}
\d L^{ij M}
&=& \rmi \bar{\e}^{(i} \vf^{j) M} \,, \nn \\
\d \vf^{i M} &=& -\frac 12 \rmi \slashed{\cal D} L^{ij M} \e_j - \frac 12
\rmi \g^a E_a^M \e^i + \frac 12 N^M \e^i + \frac 12 g \s^I  t_{IN}{}^M
L^{ij N} \e_j
+ 3 L^{ijM}\eta_j\,, \nn \\
\d E_a^M &=& -\frac 12 \rmi \bar{\e} \g_{ab} {\cal D}^b \vf^M -\frac 12 g
\bar{\e} \g_a t_{IN}{}^M \s^I  \vf^N  + \frac 12 g \bar{\e}^{(i}
t_{IN}{}^M \g_a \p^{j) I}  L_{ij}^N -2 \bar\eta\gamma_a
\varphi^M\,, \nn \\
\d N^M &=& \frac 12 \bar{\e} \slashed{\cal D} \vf^M  + \frac 12 \rmi g
\bar{\e}^{(i} t_{IN}{}^M \p^{j) I}  L_{ij}^N +\frac 32 \rmi \bar\eta
\varphi^M\,. \label{eq:linmultiplet}
\end{eqnarray}
The superconformal algebra closes provided the following constraint is
satisfied
\begin{equation}
{\cal D}_a E^{a M} + g t_{IN}{}^M \left(Y^{ij I} L_{ij}^N + \rmi \bar
\p^I \vf^N + \sigma^I N^N \right) = 0 \,. \label{eq:constrE}
\end{equation}

Note that the index $I$ refers to the adjoint representation of the
vector multiplet. To obtain the multiplet of equations of motion of the
vector multiplet one should also take for $M$ the adjoint representation
in which case all $t$ matrices become structure constants.

\section{Hypercomplex manifolds}\label{hyperappendix}
In this appendix we will present the essential properties of hypercomplex
manifolds, and show the relation with hyperk{\"a}hler and quaternionic
(K{\"a}hler) manifolds. We show how properties of the Nijenhuis tensor
determine whether suitable connections for these geometries can be
defined. We give the curvature relations, and finally the properties of
symmetry transformations of these manifolds.

Hypercomplex manifolds were introduced in~\cite{Salamon:1986}. A very
thorough paper on the subject is~\cite{AM1996}. Examples of homogeneous
hypercomplex manifolds that are not hyperk{\"a}hler, can be found
in~\cite{Spindel:1988sr,Joyce:1992}, and are further discussed in
section~\ref{app:exHypercplxGr}. Non-compact homogeneous manifolds are
dealt with in~\cite{Barberis:1996}. Various aspects have been treated in
two workshops with mathematicians and
physicists~\cite{QuatWorksh1,QuatWorksh2}. To prepare this appendix, we
used extensively~\cite{AM1996}, and some parts of this presentation use
original methods.

\subsection{The family of quaternionic-like manifolds}
Let $V$ be a real vector space of dimension $4r$, whose coordinates we
indicate as $q^X$ (with $X=1,\ldots ,4r$). We define a \emph{hypercomplex
structure} $H$ on $V$ to be a triple of complex structures $J^\alpha$,
(with $\alpha =1,2,3$) which realize the algebra of quaternions,
\begin{equation}
J^\alpha J^\beta =-\delta^{\alpha\beta}\unity
_{4r}+\varepsilon^{\alpha\beta\gamma} J^\gamma\,. \label{JJ}
\end{equation}
A \emph{quaternionic structure} is the space of linear combinations
$a_\alpha J^\alpha$ with $a_\alpha $ real numbers. In this case the
3-dimensional space of complex structures is globally defined, but the
individual complex structures do not have to be globally defined.

Let $\mathcal{M}$ be a $4r$ dimensional manifold. An \emph{almost
hypercomplex manifold} or \emph{almost quaternionic manifold} is defined
as a manifold $\mathcal{M}$ with a field of hypercomplex or quaternionic
structures.

The `almost' disappears under one extra condition. Different
terminologies are used to express this condition. Sometimes it is said
that the structure should be 1-integrable. The same condition is also
expressed as the statement that the structure should be covariantly
constant using some connections, and it is also sometimes expressed as
the `preservation of the structure' using that connection. The
connection\footnote{The word `connection' is by mathematicians mostly
used as the derivative including the `connection coefficients'. We use
here `connection' as a word denoting these coefficients, i.e.\ gauge
fields.} here should be a symmetric (i.e.\ `torsionless') connection
$\Gamma _{(XY)}{}^Z$ and possibly an $\SU(2)$ connection $\omega
_X{}^\alpha $. The condition is
\begin{equation}
  0=\covder_X\J\alpha YZ\equiv \partial_X\J\alpha YZ-\Gamma _{XY}{}^W\J\alpha WZ
  +\Gamma _{XW}{}^Z\J\alpha YW+2\varepsilon ^{\alpha \beta \gamma }\omega
  _X{}^\beta \J\gamma YZ\,.
 \label{covconstJ}
\end{equation}
If the $\SU(2)$ connection has non-vanishing curvature, the manifold is
called \emph{quaternionic}.\footnote{For $r=1$ there are subtleties in
the definition, to which we will return below.} If the
condition~(\ref{covconstJ}) holds with vanishing $\SU(2)$ connection,
i.e.\
\begin{equation}
  0=\covder_X\J\alpha YZ\equiv \partial_X\J\alpha YZ-\Gamma _{XY}{}^W\J\alpha WZ
  +\Gamma _{XW}{}^Z\J\alpha YW\,,
 \label{covconstJH}
\end{equation}
then the manifold is \emph{hypercomplex}. If there is a hermitian metric,
i.e.\ a metric such that
\begin{equation}
  J^\alpha{} _X{}^Zg_{ZY}= - J^\alpha{} _Y{}^Zg_{ZX}\,,
 \label{hermMetric}
\end{equation}
and if this metric is preserved using the connection $\Gamma $ (i.e.\ if
$\Gamma $ is the Levi-Civita connection of this metric) then the
hypercomplex and quaternionic manifolds are respectively promoted to
hyperk{\"a}hler and quaternionic-K{\"a}hler manifolds. Hence this gives rise to
the scheme\footnote{The table is essentially taken over
from~\cite{AM1996}, where there is also the terminology unimodular
hypercomplex or unimodular quaternionic if the $\Gl(r)$ is reduced to
$\Sl(r)$.} of table~\ref{tbl:quatlikeMan}.
\begin{table}[ht]
 \begin{center}
  \begin{tabular}{|c||c|c|}\hline
      & no preserved  metric & with a preserved metric \\ \hline\hline
    no $\SU(2)$ & hypercomplex & hyperk{\"a}hler \\
    curvature & $\Gl(r,\mathbb{H})$ & $\USp(2r)$ \\ \hline
    non-zero $\SU(2)$ & quaternionic & quaternionic-K{\"a}hler\\
    curvature & $\SU(2)\cdot \Gl(r,\mathbb{H})$  & $\SU(2)\cdot\USp(2r)$  \\ \hline
  \end{tabular}
 \caption{\it Quaternionic-like manifolds. These are the manifolds that have a quaternionic structure
  satisfying~(\ref{JJ}) and~(\ref{covconstJ}). The holonomy group is indicated. For the right column
  the metric may give another real form as e.g.\ $\USp(2,2(r-1))$.}\label{tbl:quatlikeMan}
\end{center}
\end{table}

We will show in section~\ref{ss:curvRel} that the spaces in the upper row
have a Ricci tensor that is antisymmetric, and those in the right column
have a Ricci tensor that is symmetric (and Einstein). It follows then
that the hyperk{\"a}hler manifolds are Ricci-flat. The restriction of
holonomy group when one goes to the right column, just follows from the
fact that the presence of a metric restricts the holonomy group further
to a subgroup of O$(4r)$.\footnote{The dot notation means that it is the
product up to a common factor in both groups that does not contribute. In
fact, one considers e.g.\ $\SU(2)$ and $\USp(2r)$ on coset elements as
working one from the left, and the other from the right. Then if both are
$-1$, they do not contribute. Thus: $\SU(2)\cdot
\USp(2r)=\frac{\SU(2)\times \USp(2r)}{\mathbb{Z}_2}$.}

A theorem of Swann~\cite{Swann} shows that all quaternionic-K{\"a}hler
manifolds have a corresponding hyperk{\"a}hler manifold which admit a
quaternionically extended homothety [a homothety extended to an $\SU(2)$
vector as in~(\ref{dil-su2})] and which has three complex structures that
rotate under an isometric $\SU(2)$ action. It has been shown
in~\cite{deWit:1999fp} that this can be implemented in superconformal
tensor calculus to construct the actions of hypermultiplets in any
quaternionic-K{\"a}hler manifold from a hyperk{\"a}hler cone. Similarly, it has
been proven in~\cite{PedersenPS1998,Pedersen2001Roma} that any
quaternionic manifold is related to a hypercomplex manifold.

Locally there is a vielbein $f_X^{iA}$ (with $i=1,2$ and $A=1,\ldots ,r$)
with reality conditions as in~(\ref{realfunctions}). In supersymmetry
(and thus in this paper), we always start from these vielbeins and the
integrability condition can be expressed as
\begin{equation}
  \partial _X f_Y^{iA}-\Gamma _{XY}{}^Zf_Z^{iA}+f_Y^{jA}\omega _{Xj}{}^i +f_Y^{iB}\omega
_{XB}{}^A=0\,.
 \label{covconstf}
\end{equation}

\subsection{Conventions for curvatures and lemmas} \label{app:convCurv}
We start with the notations for curvatures. The main conventions for
target space curvature, fermion reparametrization curvature and $\SU(2)$
curvature are
\begin{eqnarray}
 R_{XYZ}{}^W &\equiv & 2
\partial_{[X}\Gamma_{Y]Z}{}^W + 2 \Gamma_{V[X}{}^W \Gamma_{Y]Z}{}^V \,,\nonumber\\
{\cal R}_{XYB}{}^A &\equiv & 2\partial_{[X} \omega_{Y]B}{}^A
+ 2\omega_{[X|C|}{}^A \omega_{Y]B}{}^C\,, \nonumber\\
 {\cal R}_{XY i}{}^j  &\equiv & 2\partial_{[X}\omega _{Y]i}{}^j
+2\omega _{[X|k|}{}^j\omega _{Y] i}{}^k\,.
 \label{defCurv}
\end{eqnarray}
The $\SU(2)$ curvature and connection $\omega _{Xi}{}^j$ are hermitian
traceless,\footnote{This means symmetric if the indices are put at equal
height using the raising or lowering tensor $\varepsilon _{ij}$ (NW--SE
convention).} and one can make the transition to triplet indices $\alpha
=1,2,3$ by using the sigma matrices
\begin{eqnarray}
 {\cal R}_{XY i}{}^j     & = & \rmi (\sigma ^\alpha )_i{}^j {\cal R}_{XY}{}^\alpha\,, \nonumber\\
 {\cal R} _{XY}{}^\alpha & = &-\frac 12\rmi (\sigma ^\alpha )_i{}^j {\cal R}_{XY
 j}{}^i
 =2\partial _{[X}\omega _{Y]}{}^\alpha +2\varepsilon ^{\alpha \beta \gamma } \omega
 _X{}^\beta\omega _Y{}^\gamma \,.
 \label{ijtoalpha}
\end{eqnarray}
This transition between doublet and triplet notation is valid for any
triplet object as e.g.\ the complex structures. It is useful to know the
translation of the inner product: ${\cal R}_i{}^j{\cal R}_j{}^i=-2{\cal
R}^\alpha {\cal R}^\alpha$.

The curvatures by definition all satisfy the Bianchi identities that say
that they are closed 2-forms, e.g.\
\begin{equation}
   \covder_{[X} R_{YZ]V}{}^W=0\,.
 \label{BianchiR}
\end{equation}
Furthermore, due to the torsionless (symmetric) connection, also the
cyclicity property holds.
\begin{equation}
  R_{XYZ}{}^W+R_{ZXY}{}^W+R_{YZX}{}^W=0\,.
 \label{cyclicity}
\end{equation}
The Ricci tensor is defined as
\begin{equation}
  R_{XY}=R_{ZXY}{}^Z\,.
 \label{defRicci}
\end{equation}
This is not necessarily symmetric. When $\Gamma $ is the Levi-Civita
connection of a metric, then one can raise and lower indices,
$R_{WZXY}=R_{XYWZ}$ and the Ricci tensor is symmetric. Then one defines
the scalar curvature as $R=g^{XY}R_{XY}$.\bigskip

We now present three lemmas that are useful in connecting scalar manifold
indices with $\Gl(r,\mathbb{H})$ indices. These lemmas are used in
section~\ref{ss:hypermultiplet} and will simplify further derivations in
this appendix.

\begin{lemma}\label{lem:MnormJ}
If a matrix $M_X{}^Y$ satisfies
\begin{equation}
  [J^\alpha ,M]=2\varepsilon ^{\alpha \beta \gamma }J^\beta m^\gamma \,,
 \label{McommJ}
\end{equation}
for some numbers $m^\gamma $, then the latter are given by
\begin{equation}
  4\,r\,m^\alpha =\trace\left( J^\alpha M\right) ,
 \label{valuemalpha}
\end{equation}
and the matrix can be written as
\begin{equation}
  M= -m^\alpha J^\alpha + N\,, \qquad [N,J^\alpha ]=0\,.
 \label{Mdecomposed}
\end{equation}
A matrix $M$ of this type is said to `normalize the hypercomplex
structure'.
\end{lemma}
\noindent \textbf{Proof.} The first statement is proven by taking the
trace of~(\ref{McommJ}) with $J^\delta $. Inserting this value of
$m^\alpha $ in~(\ref{Mdecomposed}), it is obvious that the remainder $N$
commutes with the complex structures. \QED

\begin{lemma} \label{lem:McommJ}
If a matrix $M_X{}^Y$ commutes with the complex structures, then it can
be written~as
\begin{equation}
  M_X{}^Y=M_A{}^B f_X^{iA}f^Y_{iB}\,.
 \label{MtoAB}
\end{equation}
and vice-versa, any $M_A{}^B$ matrix can be transformed
with~(\ref{MtoAB}) to a matrix commuting with the complex structures.
\end{lemma}
\noindent \textbf{Proof.} The vice-versa statement is easy. For the other
direction, one replaces $J^\alpha $ with $J_i{}^j$ as
in~(\ref{doublet-J}). Then multiply this equation with $f_{jA}^Xf_Z^{kB}$
and consider the traceless part in~$AB$. \QED

\begin{lemma}\label{lem:curvABCD}
If a tensor $R_{[XY]Z}{}^W$ satisfies the cyclicity
condition~(\ref{cyclicity}) and commutes with the complex structures,
\begin{equation}
  R_{XYZ}{}^V \J\alpha VW- \J\alpha ZV  R_{XYV}{}^W=0\,,
 \label{commuteRJ}
\end{equation}
it can be written in terms of a tensor $W_{ABC}{}^D$ that is symmetric in
its lower indices. If $R_{XYZ}{}^Z=0$, then also $W$ is traceless.
\end{lemma}
\noindent \textbf{Proof.} By the previous theorem, we can write
\begin{equation}
 R_{XYW}{}^Z = f_W^{iA}f^Z_{iB}{\cal R}_{XYA} {}^B\,,\qquad
 {\cal R}_{XYA} {}^B = \frac 12 f^W_{iA}f_Z^{iB} R_{XYW}{}^Z\,.
 \label{RqinR}
\end{equation}
We can change all indices to tangent indices, defining
\begin{equation}
  R_{ij,CDB}{}^A \equiv f_{Ci}^X f_{jD}^Y{\cal R}_{XYB} {}^A = - R_{ji,DCB}{}^A\,.
 \label{defRijCDBA}
\end{equation}
The cyclicity property of $R$ can be used to obtain
\begin{equation}
0 = f_Z^{iA} R_{[WXY]}{}^Z = f^{iB}_{[Y} {\cal R}_{WX]B}{}^A\,.
 \label{cyclicityoncalR}
\end{equation}
We multiply this with $f^X_{iC}f^Y_{Dj}f^W_{kE}$, leading to
\begin{equation}
  R_{kj,ECD}{}^A + R_{kj,CDE}{}^A + 2 R_{jk,DEC}{}^A = 0 \,.
 \label{implcycl}
\end{equation}
The symmetric part in $(jk)$ of this equation implies that
$R_{(jk),ABC}{}^D=0$ [multiply the equation by 3, and subtract both
cyclicity rotated terms in $(CDE)$]. Thus we find
\begin{equation}
  R_{ij,CDB}{}^A = - \frac 12 \varepsilon_{ij} W_{CDB}{}^A\,,
 \label{RijinW}
\end{equation}
with
\begin{equation}
W_{CDB}{}^A \equiv \varepsilon ^{ij} f^X_{jC} f^Y_{iD} {\cal R}_{XYB}
{}^A = \frac 12 \varepsilon ^{ij} f^X_{jC} f^Y_{iD} f_{kB}^Z f_W^{Ak}
R_{XYZ}{}^W\,.\label{WfromRW}
\end{equation}
Now we prove that $W$ is completely symmetric in the lower indices. The
definition immediately implies symmetry in the first two. The $[jk]$
antisymmetric part of~(\ref{implcycl}) gives
\begin{equation}
  W_{ECD}{}^A + W_{DCE}{}^A - 2 W_{EDC}{}^A = 0\,.
 \label{Wsymm}
\end{equation}
Antisymmetrizing this in two of the indices gives the desired result.

Finally, it is obvious from~(\ref{WfromRW}) that the tracelessness of $R$
and $W$ are equivalent. \QED

The full result for such a curvature tensor is thus
\begin{equation}
 R_{XYW}{}^Z = - \frac 12 f^{Ai}_X \varepsilon_{ij}f^{jB}_Yf_W^{kC}f^Z_{kD} W_{ABC}{}^D
 \,.
 \label{RinW}
\end{equation}

\subsection{The connections}

In the definition of hypercomplex and quaternionic manifolds, appear the
affine connection $\Gamma _{XY}{}^Z$ and an $\SU(2)$ connection $\omega
_X{}^\alpha $. In this subsection we will show how they can be obtained.
The crucial ingredient is the Nijenhuis tensor.

\paragraph{Nijenhuis tensor.}
A Nijenhuis tensor $N_{XY}^{\alpha \beta \,Z}$ is defined for any
combination of two complex structures, but we will use only the
`diagonal' Nijenhuis tensor (normalization for later convenience)
\begin{eqnarray}
N_{XY}{}^Z&\equiv & \frac16\J{\alpha}XW \partial_{[W} \J{\alpha}{Y]}Z -
(X\leftrightarrow Y)=-N_{YX}{}^Z\,.
 \label{Nijenhuisdiag}
\end{eqnarray}
It satisfies a useful relation
\begin{equation}
  N_{XY}{}^Z=\J\alpha
  X{X'}N_{X'Y}{}^{Z'}\J\alpha{Z'}Z\,,
 \label{relNijenh}
\end{equation}
from which one can deduce that it is traceless.

\paragraph{Obata connection and hypercomplex manifolds.}
The torsionless \emph{Obata connection} \cite{Obata} is defined as
\begin{equation}
\Gamma^{\rm Ob}{}_{XY}{}^Z=-\frac16 \left( 2 \partial_{(X}\J\alpha
{Y)}W+\varepsilon ^{\alpha \beta \gamma }\J{\beta  }{(X}U
\partial_{|U|} \J{\gamma }{Y)}W\right) \J\alpha WZ\,. \label{Obataapp}
\end{equation}
First, note that if a manifold is hypercomplex, i.e.\
if~(\ref{covconstJH}) is satisfied, then by inserting the expression for
$\partial J$ from that equation in the right hand side
of~(\ref{Obataapp}), one finds that the affine connection of the
hypercomplex manifold should be the Obata connection, $\Gamma
=\Gamma^{\rm Ob}$. One may thus answer the question whether an almost
hypercomplex manifold [i.e.\ with three matrices satisfying~(\ref{JJ})],
defines a hypercomplex manifold [i.e.\ satisfies~(\ref{covconstJH})]. As
we now know that the affine connection in~(\ref{covconstJH}) should
be~(\ref{Obataapp}), this can just be checked. For that purpose, the
following equation is useful:
\begin{equation}
  \partial_X\J\alpha YZ-\left( \Gamma^{\rm Ob}{} _{XY}{}^W+N_{XY}{}^W\right) \J\alpha WZ
  +\left( \Gamma^{\rm Ob}{} _{XW}{}^Z+N _{XW}{}^Z\right) \J\alpha YW=0\,.
 \label{covconstJTorsion}
\end{equation}
It shows that any hypercomplex structure can be given a torsionful
connection such that the complex structures are covariantly constant. The
condition for a hypercomplex manifold is thus that this connection is
torsionless, i.e.\ that the Nijenhuis tensor vanishes. In conclusion,
\emph{a hypercomplex manifold consists of the following data: a manifold
$\mathcal{M}$, with a hypercomplex structure with vanishing Nijenhuis
tensor}. In the main text, we only use the Obata connection, and we thus
have $\Gamma =\Gamma ^{\rm Ob}$.

\paragraph{Oproiu connection and quaternionic manifolds.}
For the quaternionic manifolds, the affine connection and $\SU(2)$
connection can not be uniquely defined. Indeed, one can easily check
that~(\ref{covconstJ}) is left invariant when we change these two
connections simultaneously using an arbitrary vector $\xi _W$ as
\begin{equation}
  \Gamma _{XY}{}^Z\rightarrow \Gamma _{XY}{}^Z+S_{XY}^{WZ}\xi _W\,,\qquad
\omega_X {}^\alpha\rightarrow \omega_X {}^\alpha +\J\alpha XW\xi _W\,,
 \label{changeGammaomega}
\end{equation}
where $S$ is the tensor
\begin{equation}
    S^{XY}_{ZW}\equiv 2\delta ^X_{(Z}\delta ^Y_{W)}-2 \J\alpha Z{(X}\J\alpha
  W{Y)}\,,
 \label{defS}
\end{equation}
which satisfies the relation
\begin{equation}
  S^{XV}_{ZW}\J\alpha VY-\J\alpha WV S^{XY}_{ZV}=
2\varepsilon ^{\alpha \beta \gamma }\J\beta ZX\J\gamma WY\,.
 \label{Sprolong}
\end{equation}
An invariant $\SU(2)$ connection is
\begin{equation}
  \tilde \omega _X{}^\alpha =\omega _X{}^\alpha
  +\frac13\J\alpha XY\J\beta YZ\omega _Z{}^\beta =\frac23\omega
  _X{}^\alpha
  -\frac13\varepsilon ^{\alpha \beta \gamma }\J\beta XY\omega_Y{}^\gamma\,.
 \label{tilomega}
\end{equation}

If we use~(\ref{covconstJ}) in the expression for the Nijenhuis
tensor,~(\ref{Nijenhuisdiag}), we find that quaternionic manifolds do not
have a vanishing Nijenhuis tensor, but the latter should satisfy
\begin{equation}
  N_{XY}{}^Z=-\J\alpha{[X}Z\tilde \omega _{Y]}{}^\alpha \,.
 \label{NijenhQuat}
\end{equation}
This condition can be solved for $\tilde \omega$. We find
\begin{equation}
  (1-2\,r)\,\tilde \omega _X{}^\alpha =N_{XY}{}^Z\J\alpha ZY\,.
 \label{valueTilOmQ}
\end{equation}
Thus the condition for an almost quaternionic manifold to be quaternionic
is that the Nijenhuis tensor satisfies
\begin{equation}
  (1-2\,r)\,N_{XY}{}^Z=-\J\alpha{[X}ZN_{Y]V}{}^W\J\alpha WV\,.
 \label{conditQuat}
\end{equation}
On the other hand, one may also use~(\ref{covconstJ}) in the expression
for the Obata connection~(\ref{Obataapp}). Then we find that the affine
connection for the quaternionic manifolds is given by
\begin{equation}
  \Gamma _{XY}{}^Z=\Gamma^{\rm Ob}{}_{XY}{}^Z-\J\alpha{(X}Z
  \tilde \omega _{Y)}{}^\alpha -\frac13 S_{XY}^{ZU}\J\alpha UV\omega _V{}^\alpha
  \,,
 \label{GammaQuat}
\end{equation}
which exhibits the transformation~(\ref{changeGammaomega}).

One can take a gauge choice for the invariance. A convenient choice is to
impose
\begin{equation}
  \J\alpha YZ\omega _Z{}^\alpha=0\,.
 \label{gaugechoiceOproiu}
\end{equation}
With this choice $\tilde\omega_X{}^\alpha  =\omega_X{}^\alpha$. The
affine connection in~(\ref{GammaQuat}) simplifies, and this expression is
called the Oproiu connection~\cite{Oproiu1977}
\begin{eqnarray}
  \Gamma ^{\rm Op}{}_{XY}{}^Z&\equiv &
\Gamma^{\rm Ob}{}_{XY}{}^Z-\J\alpha {(X}Z\omega _{Y)}{}^\alpha \nonumber\\
&=&\Gamma^{\rm Ob}{}_{XY}{}^Z+N_{XY}{}^Z-
   \J\alpha YZ\omega _X{}^\alpha\,.
 \label{Oproiu}
\end{eqnarray}
The last expression shows that the Oproiu connection, which up to here
was only proven to be necessary for solving~(\ref{covconstJ}), gives
indeed rise to covariantly constant complex structures under the
condition~(\ref{NijenhQuat}). Indeed, the first two terms give already a
(torsionful) connection that gives rise to a covariantly constant
hypercomplex structure, see~(\ref{covconstJTorsion}), and the last term
cancels the $\SU(2)$ connection. The condition~(\ref{NijenhQuat}) is now
the condition that the connection $\Gamma ^{\rm Op}$ is torsionless.

In conclusion, \emph{a quaternionic manifold consists of the following
data: a manifold $\mathcal{M}$, with a hypercomplex structure with
Nijenhuis tensor satisfying~(\ref{conditQuat})}.

\paragraph{Levi-Civita connection and hyperk{\"a}hler or quaternionic-K{\"a}hler
manifolds.} For hyperk{\"a}hler manifolds, the Obata connection should
coincide with the Levi-Civita connection of a metric. For
quaternionic-K{\"a}hler manifolds, the connection that preserves the metric
can be one of the equivalence class defined from the Oproiu connection by
a transformation~(\ref{changeGammaomega}).

\paragraph{Final note on connections.}
Note that for a given $\mathcal{M}$ and $H$, it is possible to find
different connections which are all compatible with the hypercomplex
structures. The resulting curvatures are then also different, which
implies different (restricted) holonomy groups. An example on group
manifolds, where we use a torsionful and a torsionless connection, will
follow in section~\ref{app:exHypercplxGr}. Other examples can be found
in~\cite{Coles:1990hr,Gibbons:1997iy,Papadopoulos:2002gy}, which discuss
`HKT' manifolds, hypercomplex manifolds with torsion.

\subsection{Curvature relations} \label{ss:curvRel}
\paragraph{Splitting according to holonomy.}
There are two interesting possibilities of splitting the curvature on
quaternionic-like manifolds. First of all, the integrability condition
of~(\ref{covconstf}) yields that the total curvature on the manifold is
the sum of the $\SU(2)$ curvature and the $\Gl (r,\mathbb{H})$ curvature
which shows that the (restricted) holonomy splits in these two factors:
\begin{eqnarray}
  R_{XYW}{}^Z&=R^{\SU(2)}{}_{XYW}{}^Z&+\,R^{\Gl(r,\mathbb{H})}{}_{XYW}{}^Z  \label{RdecompJ}\\
&=-\J\alpha WZ\,{\cal R} _{XY}{}^\alpha & +\, L_W{}^Z{}_A{}^B\,{\cal
R}_{XYB}{}^A \,, \qquad \mbox{with}\qquad L_W{}^Z{}_A{}^B\equiv
f^Z_{iA}f_W^{iB} \,.\nonumber
\end{eqnarray}
The matrices $L_A{}^B$ and $J^\alpha $ commute and their mutual trace
vanishes
\begin{equation}
\J\alpha XY L_Y{}^Z{}_A{}^B=
  L_X{}^Y{}_A{}^B\J\alpha YZ \,,\qquad \J\alpha ZY L_Y{}^Z{}_A{}^B=0\,.
 \label{JLcommute}
\end{equation}
For hypercomplex (or hyperk{\"a}hler) manifolds, the $\SU(2)$ curvature
vanishes. Then the Riemann tensor commutes with the complex structures
and using the cyclicity, one may use lemmas~\ref{lem:McommJ}
and~\ref{lem:curvABCD} to write
\begin{equation}
  R_{XYW}{}^Z = - \frac 12 f^{Ai}_X \varepsilon_{ij}f^{jB}_Yf_W^{kC}f^Z_{kD}
  W_{ABC}{}^D\,.
 \label{RinWHC}
\end{equation}
This $W$ is symmetric in its lower indices. The Ricci tensor is then
\begin{equation}
  \Ric_{XY}= \frac 12  \varepsilon_{ij}f^{iB}_Xf_Y^{jC}
  W_{ABC}{}^A= -\Ric_{YX}\,.
 \label{Ricisas}
\end{equation}
Thus the Ricci tensor for hypercomplex manifolds is antisymmetric. In
general, the antisymmetric part can be traced back to the curvature of
the $\U(1)$ part in $\Gl(r,\mathbb{H})=\Sl(r,\mathbb{H})\times \U(1)$.
Indeed,  using the cyclicity condition:
\begin{equation}
  \Ric_{[XY]}=R_{Z[XY]}{}^Z=-\frac 12R_{XYZ}{}^Z=-{\cal R}_{XY}^{\U(1)}\,,
  \qquad {\cal R}_{XY}^{\U(1)}\equiv {\cal R}_{XYA}{}^A\,.
 \label{Ricas}
\end{equation}
\paragraph{Splitting in Ricci and Weyl curvature.}
The separate terms in~(\ref{RdecompJ}) for quaternionic manifolds do not
satisfy the cyclicity condition, and thus are not bona-fide curvatures.
We will now discuss another splitting
\begin{equation}
  R_{XYW}{}^Z=R^{\rm Ric}{}_{XYW}{}^Z+R^{(\rm W)}{}_{XYW}{}^Z\,.
 \label{RRBRW}
\end{equation}
Both terms will separately satisfy the cyclicity condition. The first
part only depends on the Ricci tensor of the full curvature, and is
called the `\emph{Ricci part}'. The Ricci tensor of the second part will
be zero, and this part will be called the `\emph{Weyl
part}'~\cite{AM1996}. We will prove that the second part commutes with
the complex structures. The lemmas of section~\ref{app:convCurv} then
imply that the second part can be written in terms of a tensor
$\mathcal{W}_{ABC}{}^D$, symmetric in the lower indices and traceless.
This tensor appears in supersymmetric theories, which is another reason
for considering this construction. The case $r=1$ needs a separate
treatment which will be discussed afterwards.

To define the splitting~(\ref{RRBRW}), we define the first term as a
function of the Ricci tensor, and $R^{(\rm W)}$ is just defined as the
remainder. The definition of $R^{\rm Ric}$ makes again use of the tensor
$S$ in~(\ref{defS}):
\begin{eqnarray}
  R^{\rm Ric}{}_{XYZ}{}^W&\equiv &2 S^{WV}_{Z[X}B_{Y]V}\,,\nonumber\\
  B_{XY}&\equiv &\frac{1}{4r}R_{(XY)}-\frac{1}{2r(r+2)}\Pi
  _{(XY)}{}^{ZW}R_{ZW}+\frac{1}{4(r+1)}R_{[XY]}\,.
 \label{BinRicB}
\end{eqnarray}
Here, $\Pi$ projects bilinear forms onto hermitian ones, i.e.\
\begin{equation}
  \Pi _{XY}{}^{ZW}\equiv \frac14\left( \delta _X{}^Z\delta _Y{}^W+\J\alpha XZ\J\alpha YW
  \right) .
 \label{ProjBilHerm}
\end{equation}
The Ricci part satisfies several properties that can be checked by a
straightforward calculation:
\begin{enumerate}
  \item The Ricci tensor of $ R^{\rm Ric}$ is just $R_{XY}$.
  \item The cyclicity property~(\ref{cyclicity}).
  \item Considered as a matrix in its last two indices, this matrix
normalizes the hypercomplex structure (see lemma~\ref{lem:MnormJ}).
\end{enumerate}
Especially to prove the last one, the property~(\ref{Sprolong}) can be
used (multiplying it with $B_{UX}$ and antisymmetrizing in $[ZU]$). The
relation is explicitly
\begin{eqnarray}
&&  \J\alpha ZW\,R^{\rm Ric}{}_{XYW}{}^V-R^{\rm
Ric}{}_{XYZ}{}^W\,\J\alpha WV=
  2\varepsilon^{\alpha \beta \gamma }\J\beta ZV\,{\cal R}^{\rm Ric}{}
  _{XY}{}^\gamma\,,\nonumber\\
  &&\mbox{with}\qquad {\cal R}^{\rm Ric}{}_{XY}{}^\alpha =\frac{1}{4r}\J\alpha WZ
  R^{\rm Ric}{}_{XYZ}{}^W=2\J\alpha {[X}ZB_{Y]Z}\,.
 \label{RRicnormJ}
\end{eqnarray}
The important information is now that the full curvature also satisfies
these 3 properties. The latter one is the integrability property
of~(\ref{covconstJ}):
\begin{equation}
  0=2\covder_{[X}\covder_{Y]}\J\alpha ZV=R_{XYW}{}^V\J\alpha ZW-R_{XYZ}{}^W\J\alpha WV
  -2\varepsilon ^{\alpha \beta \gamma }
  {\cal R}_{XY}{}^\gamma \J\beta ZV\,.
 \label{DDJ0}
\end{equation}
As in general for matrices normalizing the complex structure, we can also
express ${\cal R} _{XY}{}^\alpha$~as
\begin{equation}
  R_{XYZ}{}^W\J\alpha WZ=4r\,{\cal R}_{XY}{}^\alpha \,.
 \label{nearpropSU2J}
\end{equation}

This leads to properties of the Weyl part of the curvature. First of all,
it implies that this part is Ricci-flat. Secondly it also satisfies the
cyclicity property. Third, it also normalizes the hypercomplex structure,
defining some ${\cal R}^{({\rm W})\alpha }_{XY}$. We will now prove that
the latter is zero for $r>1$.

The expression for this tensor satisfies a property that can be derived,
starting from its definition, by first using the cyclicity of $R^{(\rm
W)}$, then the equation saying that it normalizes the hypercomplex
structure, and finally that it is Ricci-flat
\begin{eqnarray}\label{rRW}
 r\mathcal{R}^{({\rm W})\alpha}_{XY}&=&\frac 14 \J{\alpha}UV R^{(\rm W)}{}_{XYV}{}^U=-\frac 12\J{\alpha}UV
R^{(\rm W)}{}_{V[XY]}{}^U \nonumber\\&=&-\epsilon^{\alpha \beta \gamma}
\mathcal{R}^{({\rm W}) \beta}_{V[X} \J{\gamma}{Y]}V\,.
\end{eqnarray}
Multiplying with $\J\alpha VY$ and antisymmetrizing leads to
\begin{equation}\label{symJRW}
\J{\alpha}{[V}Y \mathcal{R}^{({\rm W})\alpha}_{X]Y}=0\,.
\end{equation}
Secondly,  multiplying~(\ref{rRW}) with $\J\delta ZX\J\delta WY$, and
using~(\ref{rRW}) again for multiplying the complex structures at the
right-hand side, leads to
\begin{equation}
\J{\beta}XZ \J{\beta}YV \mathcal{R}_{ZV}^{({\rm W})\alpha}=
-\mathcal{R}^{({\rm W}) \alpha}_{XY}\qquad \mbox{or}\qquad
\Pi_{XY}{}^{ZV}\mathcal{R}_{ZV}^{({\rm W})\alpha}=0 \,.
\label{secondconseqRW}
\end{equation}
Finally, multiplying~(\ref{rRW})  with $\varepsilon ^{\alpha \delta
\epsilon}\J\delta ZY$ leads to
\begin{equation}
{\cal R}^{({\rm W})\alpha} _{XY}=0\,,\qquad \mbox{if }\quad r>1\,.
\end{equation}
Therefore $R^{({\rm W})}{}_{XYZ}{}^V$ is a tensor that satisfies all
conditions of lemma~\ref{lem:curvABCD}, and we can thus write
\begin{equation}
  R_{XYZ}{}^W=R^{\rm Ric}{}_{XYZ}{}^W - \frac 12 f^{Ai}_X \varepsilon_{ij}f^{jB}_Yf_Z^{kC}f^W_{kD}
  \mathcal{W}_{ABC}{}^D  \,.
 \label{RdecompRBWA}
\end{equation}
For hypercomplex manifolds, we found that the full curvature can be
written in terms of a tensor $W_{ABC}{}^D$, see~(\ref{RinWHC}), which is
symmetric in the lower indices, but not necessarily traceless. One can
straightforwardly compute the corresponding $\mathcal{W}$, and find that
this is its traceless part, the trace determining the Ricci tensor:
\begin{equation}
  \mathcal{W}_{ABC}{}^D=W_{ABC}{}^D-\frac{3}{2(r+1)}\delta
  ^D_{(A}W_{BC)E}{}^E\,,\qquad R_{XY}=-{\cal R}_{XYA}{}^A=\frac 12\varepsilon
  _{ij}f^{iA}_Xf^{jB}_YW_{ABC}{}^C\,.
 \label{calWW}
\end{equation}

\paragraph{The 1-dimensional case.}
As
\begin{equation}
  \Gl(1,\mathbb{H})=\Sl(1,\mathbb{H})\times\U(1)=\SU(2)\times\U(1)\,,
 \label{decompr1}
\end{equation}
we have now two $\SU(2)$ factors in the full holonomy group. This can be
written explicitly by splitting $L$ in~(\ref{RdecompJ}) in a traceless
and trace part:
\begin{equation}
  L_X{}^Y{}_A{}^B   =  \frac 12\rmi(\sigma ^\alpha )_A{}^B\J{-\alpha }XY
 +\frac 12\delta _X^Y\delta _A^B\,.
 \label{J-}
\end{equation}
This leads to the $r=1$ form of~(\ref{RdecompJ}):
\begin{equation}
   R_{XYW}{}^Z=
-\J{+\alpha} WZ{\cal R} _{XY}^{+\alpha}-\J{-\alpha} WZ{\cal R}
_{XY}^{-\alpha} +\delta _W^Z{\cal R}_{XY}^{\U(1)}\,,
 \label{split+-U1}
\end{equation}
where for emphasizing the symmetry, we indicate the original complex
structures as $\J{+\alpha} XY$.\bigskip

We saw that for $r=1$ we could not perform all steps to get to the
decomposition~(\ref{RdecompRBWA}). However, some authors define
quaternionic and quaternionic-K{\"a}hler for $r=1$ as a more restricted class
of manifolds such that this decomposition is still
valid~\cite{Alekseevsky1975}. For quaternionic-K{\"a}hler manifolds, the
definition that is taken in general leads for $r=1$ to the manifolds with
holonomy $\SU(2)\times \USp(2)$, which is just $\SO(4)$. Thus with this
definition all 4-dimensional riemannian manifolds would be
quaternionic-K{\"a}hler. Therefore a further restriction is imposed. This
further restriction is also natural in supergravity, as it is equivalent
to a constraint that follows from requiring invariance of the
supergravity action.

In general, as ${\cal R}^{({\rm W})}$ normalizes the hypercomplex
structure, we can by lemma~\ref{lem:MnormJ} and lemma~\ref{lem:McommJ}
write
\begin{equation}
  R^{(\rm W)}{}_{XYZ}{}^W=-{\cal R}_{XY}^{({\rm W})\,\alpha }\J\alpha ZW+{\cal
  R}^{({\rm W})}_{XYA}{}^B\,L_Z{}^W{}_A{}^B=R^{(\rm W)+}{}_{XYZ}{}^W+R^{(\rm W)-}{}_{XYZ}{}^W\,.
 \label{Wsplit}
\end{equation}
\emph{We impose}
\begin{equation}
  {\cal R}_{XY}^{({\rm W})\,\alpha }=0\,,
 \label{RW+0}
\end{equation}
\emph{as part of the definition of quaternionic manifolds with }$r=1$.
This is thus the equation that is automatically valid for $r>1$. Using
lemma~\ref{lem:curvABCD}, this implies that~(\ref{RdecompRBWA}) is valid
for all quaternionic manifolds.\bigskip

In the 1-dimensional case, we can see that a possible metric is already
fixed up to a multiplicative function. Indeed, the $C_{AB}$ that is used
in~(\ref{defg}) can only be proportional to $\varepsilon _{AB}$.
Therefore, it is said that there is a \emph{conformal metric}, i.e.\ a
metric determined up to a (local) scale function $\lambda (q)$:
\begin{equation}
  g_{XY}\equiv \lambda(q)f_X^{iA}f_{Y}^{jB}\varepsilon _{ij}\varepsilon
  _{AB}\,.
 \label{Jindicesdown}
\end{equation}
One can check that this metric is hermitian for any $\lambda (q)$, i.e.\
$J^\alpha {}_{XY}=\J\alpha XZ g_{ZY}$ is antisymmetric. The remaining
question is whether this metric is covariantly constant, which boils down
to the covariant constancy of $C_{AB}$. This condition can be simplified
using the Schouten identity:
\begin{equation}
  \covder_XC_{AB}=\partial _XC_{AB}+2\omega _{X[A}{}^CC_{|C|B]}=
\partial _XC_{AB}+\omega _{XC}{}^CC_{AB}=\varepsilon _{AB}\left( \partial
_X\lambda(q)+\omega _{XC}{}^C\lambda(q)\right) .
 \label{DCr1}
\end{equation}
We can choose a function $\lambda (q)$ such that $C$ is covariantly
constant iff $\omega _{XC}{}^C$ is a total derivative, i.e.\ if the
$\U(1)$ curvature vanishes. Thus in the 1-dimensional case hypercomplex
manifolds become hyperk{\"a}hler, and quaternionic manifolds become
quaternionic-K{\"a}hler if and only if the $\U(1)$ factor in the curvature
part $G\ell (1,\mathbb{H})$ vanishes.

\paragraph{The curvature of Quaternionic-K{\"a}hler manifolds.}

In quaternionic-K{\"a}hler manifolds, the affine connection is the
Levi-Civita connection of a metric. Therefore, the Ricci tensor is
symmetric. As we have already proven that in the hypercomplex case the
symmetric part vanishes, hyperk{\"a}hler manifolds have vanishing Ricci
tensor. Now we will prove that the quaternionic-K{\"a}hler spaces are
Einstein, and that moreover the $\SU(2)$ curvatures are proportional to
the complex structures with a proportionality factor that is dependent on
the scalar curvature.

We start again from the integrability property~(\ref{DDJ0}). Multiplying
with $\J\delta VX$ gives
\begin{eqnarray}
  \Ric_{YZ}\delta ^{\alpha \delta }-\varepsilon ^{\alpha \delta \beta
  }R_{XYZ}{}^W\J\beta WX+\J\alpha ZWR_{XYW}{}^V\J\delta VX-\qquad\qquad
  \nonumber\\
\qquad\qquad-2\varepsilon ^{\alpha \beta \delta }{\cal R} _{ZY}{}^\beta
  +2\delta ^{\alpha \delta }{\cal R}
  _{XY}{}^\beta\J\beta ZX-2{\cal R} _{XY}{}^\delta\J\alpha ZX=0\,.
 \label{conseq}
\end{eqnarray}
The second and third term can be rewritten
\begin{eqnarray}
 R_{XYW}{}^V\J\delta VX  & = & -  R_{YWX}{}^V\J\delta VX
 -  R_{WXY}{}^V\J\delta VX \nonumber\\
 & = &  -  R_{YWX}{}^V\J\delta VX +  R_{YXW}{}^V\J\delta
 VX\,,\nonumber\\
 2 R_{XYW}{}^V\J\delta VX  & = &-   4r{\cal R}_{YW}{}^\delta\,.
 \label{contrRJ}
\end{eqnarray}
In the first line, the cyclicity property of the Riemann tensor is used.
Then, the symmetry in interchanging the first two and last two indices
(here we use that the curvature originates from a Levi-Civita connection)
and finally interchanging the indices on the last complex structure, using
its antisymmetry (Hermiticity of the metric). This leads to
\begin{equation}
  \Ric_{YZ}\delta ^{\alpha \delta }+\varepsilon ^{\alpha \delta \beta
  }2(r-1){\cal R}_{YZ}{}^\beta -2(r-1){\cal R}_{YX}{}^\delta \J\alpha ZX+2\delta ^{\alpha \delta }
  {\cal R}^\beta _{XY}\J\beta ZX=0\,.
 \label{cons3}
\end{equation}
Multiplying with $\delta ^{\alpha \delta }$ gives
\begin{equation}
  \Ric_{YZ}=-\frac23(r+2)\J\beta ZX {\cal R}_{XY}{}^{\beta}\,.
 \label{RicinRJ}
\end{equation}
On the other hand, multiplying~(\ref{cons3}) with $\varepsilon ^{\alpha
\delta \gamma}$ gives only a non-trivial result for $r\neq 1$, in which
case we find
\begin{equation}
\mbox{for}\quad r>1\ :\qquad   2{\cal R}_{YZ}{}^{\alpha }=\varepsilon
^{\alpha \beta \gamma }\J\beta YX{\cal R} _{XZ}{}^\gamma\,.
 \label{auxeq}
\end{equation}
\emph{We impose the same equation for $r=1$. We will connect this
equation to another requirement below.}

By replacing $\varepsilon ^{\alpha \beta \gamma }\J\beta YX$ by
$-(J^\alpha J^\gamma)_Y{}^X-\delta _Y^X\delta ^{\alpha \gamma }$ we get
\begin{equation}
  {\cal R} _{XY}{}^\alpha=-\frac13 \J\alpha XZ\J\beta ZV{\cal R} _{VY}{}^\beta=\frac{1}{2(r+2)}
\J\alpha XZ \Ric_{ZY}\,.
 \label{cRisJRic}
\end{equation}
We also have
\begin{equation}
\J\alpha XZ {\cal R}_{ZY}{}^\beta =\varepsilon ^{\alpha \beta \gamma
}{\cal R}_{XY}{}^\gamma -\frac{1}{2(r+2)}\delta ^{\alpha \beta }\Ric_{XY}
\,.
 \label{conclcJcR}
\end{equation}

The final step is obtained by using~(\ref{DDJ0}) once more. Now multiply
this equation with $\varepsilon ^{\alpha \beta \gamma }J^{\beta
YX}\J\gamma VU$, and use for the contraction of the Riemann curvature
tensor with $J^{\beta YX}$ that we may interchange pairs of indices such
that~(\ref{nearpropSU2J}) can be used. Then everywhere appears $J^\alpha
{\cal R}^\beta $, for which we can use~(\ref{conclcJcR}). This leads to
the equation expressing that the manifold is Einstein:
\begin{equation}
  \Ric_{XY}=\frac{1}{4r} g_{XY} R\,.
 \label{Einstein}
\end{equation}
With~(\ref{cRisJRic}), the $\SU(2)$ curvature is proportional to the
complex structure:
\begin{equation}
   {\cal R}_{XY}{}^\alpha =\frac 12\nu
J^\alpha{}_{XY}\,,\qquad \nu
  \equiv \frac{1}{4r(r+2)}R\,.
 \label{RlambdaJ}
\end{equation}
\bigskip

The Einstein property drastically simplifies the expression for $B$
in~(\ref{BinRicB}) to
\begin{equation}
  B_{XY}=\frac14\nu g_{XY}\,.
   \label{BinqK}
\end{equation}
The Ricci part of the curvature then becomes proportional to the
curvature of a quaternionic projective space of the same dimension:
\begin{equation}
\left(R^{{\mathbb H}P^r}\right)_{XYWZ} =\frac 12 g_{Z[X}g_{Y]W}+\frac 12
J^\alpha _{XY}J^\alpha _{ZW}-\frac 12 J^\alpha _{Z[X}J^\alpha _{Y]W}
=\frac 12 J^\alpha _{XY}J^\alpha _{ZW}+L_{[ZW]}{}^{AB}L_{[XY]AB}\,.
 \label{RHPn}
\end{equation}
The full curvature decomposition is then
\begin{equation}
 R_{XYWZ}  = \nu (R^{{\mathbb H}P^r})_{XYWZ}+ \frac 12 L_{ZW}{}^{AB}\mathcal{W}
 _{ABCD}L_{XY}{}^{CD}\,,
 \label{RdecompHPnW}
\end{equation}
with $\mathcal{W}_{ABCD}$ completely symmetric. The constraint appearing
in supergravity fixes the value of $\nu $ to $-1$. The
quaternionic-K{\"a}hler manifolds appearing in supergravity thus have
negative scalar curvature, and this implies that all such manifolds that
have at least one isometry are non-compact.
\bigskip

Finally, we should still comment on the extra constraint~(\ref{auxeq})
for $r=1$. In the mathematics literature~\cite{Alekseevsky1975} the extra
constraint is that the quaternionic structure annihilates the curvature
tensor, which is the vanishing of
\begin{eqnarray}
(J^\alpha \cdot R)_{XYWZ}  &\equiv&
\J\alpha XV R_{VYWZ}+\J\alpha YV R_{XVWZ}+\J\alpha ZV R_{XYWV}+\J\alpha WV R_{XYVZ}\nonumber\\
&=& \varepsilon ^{\alpha \beta \gamma }\left( {\cal R} _{XY}{}^\beta
J^\gamma _{ZW}+{\cal R}_{ZW}{}^\beta J^\gamma_{XY}\right) ,
 \label{JR}
\end{eqnarray}
where the second expression is obtained using once more\,(\ref{DDJ0}). We
have proven that\,(\ref{auxeq}) was sufficient extra input to have ${\cal
R}^\alpha _{XY}$ proportional to $J^\alpha _{XY}$ implying $J^\alpha\cdot
R=0$. Vice versa: multiplying~(\ref{JR}) with $\varepsilon ^{\alpha
\delta \epsilon }J^\epsilon{}_{YZ}$ leads to~(\ref{auxeq}) if
$J^\alpha\cdot R=0$. Thus indeed the vanishing of~(\ref{JR}) is an
equivalent condition that can be imposed for $r=1$ and that is
automatically satisfied for $r>1$.

\subsection{Symmetries}\label{app:symmetries}
Symmetries of manifolds are most known as isometries for riemannian
manifolds (i.e.\ when there is a metric). They are transformations
$\delta q^X=k^X_I(q)\Lambda ^I$, where $\Lambda ^I$ are infinitesimal
parameters. They are determined by the Killing equation\footnote{See also
`conformal Killing vectors' in section~\ref{ss:defnRigidConf}.}
\begin{equation}
  \covder_{(X} k_{Y)I}=0\,, \qquad  k_{XI}\equiv g_{XY}k^Y{}_I\,.
 \label{killing}
\end{equation}
This definition can only be used when there is a metric. However, there
is a weaker equation that can be used for defining symmetries also in the
absence of a metric, but when parallel transport is defined. Indeed, the
Killing equation implies that
\begin{equation}
 -R_{YZX}{}^Wk_{WI}=\covder_Y\covder_Z k_{XI} -\covder_Z \covder_Yk_{XI}=\covder_Y\covder_Z k_{XI} +\covder_Z \covder_Xk_{YI}\,.
 \label{RK}
\end{equation}
Using the cyclicity condition on the left hand side to write
\begin{equation}
  R_{YZX}{}^W =\frac 12\left(   R_{YZX}{}^W
  -  R_{ZXY}{}^W-  R_{XYZ}{}^W\right)\,,
 \label{RKisDDKprep}
\end{equation}
we obtain
\begin{equation}
  \covder_X\covder_Y k^Z_I=R_{XWY}{}^Zk_I^W\,.
 \label{RKisDDK}
\end{equation}
This equation does not need a metric any more. We will use it as
definition of symmetries when there is no metric available. We will see
that it leads to the group structure that is known from the riemannian
case.

Of course, we will require also that the symmetries respect the
quaternionic structure. This is the statement that the vector $k_I^X$
normalizes the quaternionic structure:
\begin{equation}
{\cal L}_{k_I} \J\alpha XY\equiv   k_I^Z\partial _Z \J\alpha XY +
\left(\partial_X k_I^Z\right)\J\alpha ZY -\J\alpha XZ\left(\partial _Z
k^Y_I\right)= b_I^{\alpha\beta}\J\beta XY\,,
 \label{KnormQ}
\end{equation}
for some functions $b_I^{\alpha\beta}(q)$. This $b_I$ is antisymmetric,
as can be seen by multiplying the equation with $\J\gamma YX$.

\emph{Thus we define symmetries in quaternionic-like manifolds as those
$\delta q^X=k^X_I(q)\Lambda ^I$, such that the vectors $k^X_I$
satisfy~(\ref{RKisDDK}) and~(\ref{KnormQ}).}

We first consider~(\ref{KnormQ}). One can add an affine torsionless
connection to the derivatives, because they cancel. As a total covariant
derivative on $J$ vanishes, we add in case of quaternionic manifolds the
$\SU(2)$ connection to the first derivative. This addition is of the form
of the right-hand side. Thus defining $P^\gamma _I$ by $b_I^{\alpha \beta
}-2\varepsilon ^{\alpha \beta \gamma }\omega _X{}^\gamma k_I^X
=2\varepsilon ^{\alpha \beta \gamma }\nu P^\gamma _I$, the remaining
statement is that there is a $P_I^\alpha (q)$ (possibly zero) such
that\footnote{Here we introduce in fact $\nu P$. The factor $\nu $ is
included for agreement with other papers and allows a smooth limit $\nu
=0$ to the hypercomplex or hyperk{\"a}hler case. In fact, we have seen
in~(\ref{intJ}) that supersymmetry in the setting of hypercomplex
manifolds demands that the right-hand side of~(\ref{KnormQ}) is zero. We
will see below that this is unavoidable for hypercomplex manifolds even
outside the context of supersymmetry.}
\begin{equation}
\J\alpha XZ\left(\covder_Z k^Y_I\right)-\left(\covder_X
k_I^Z\right)\J\alpha ZY = -2\varepsilon ^{\alpha\beta\gamma }\J\beta XY
\nu P^\gamma _I\,.
 \label{KnormQ2}
\end{equation}
The equation now takes on the form of~(\ref{McommJ}) in
lemma~\ref{lem:MnormJ}. Thus, using this lemma, as well as
lemma~\ref{lem:McommJ}, we have
\begin{equation}
  \covder_Xk^Y_I=\nu \J\alpha XY\,P^\alpha _I + L_X{}^Y{}_A{}^B t_{IB}{}^A\,.
 \label{decompDK}
\end{equation}
$t_{IB}{}^A$ is the matrix that we saw in the fermion gauge
transformation law~(\ref{gauge-tr}). The rule~(\ref{valuemalpha}) gives
an expression for $P^\alpha _I$, which is called the \emph{moment map}:
\begin{equation}
4r\,\nu \,  P^\alpha _I= -\J\alpha XY \left(\covder_Yk^X_I\right).
 \label{valuePDk}
\end{equation}

Using the second equation,~(\ref{RKisDDK}) we now find
\begin{equation}
  R_{ZWX}{}^Yk^W_I=\covder_Z\covder_Xk^Y_I=\nu \J\alpha XY\,(\covder_ZP^\alpha _I) + L_X{}^Y{}_A{}^B \left(\covder_Zt_{IB}{}^A\right).
 \label{Rkdecomp}
\end{equation}
Using the curvature decomposition~(\ref{RdecompJ}) and projecting onto the
complex structures and $L$, we find two equations
\begin{equation}
  {\cal R}_{ZW}{}^\alpha k^W_I= -\nu \covder_Z P^\alpha _I\,, \qquad
  {\cal R}_{ZWB}{}^Ak^W_I= \covder_Z t_{IB}{}^A\,.
 \label{Rkin2}
\end{equation}
\bigskip

The algebra that the vectors $k_I^X$ define is
\begin{equation}
  2k_{[I}^Y\covder_Y k_{J]}^X + f_{IJ}{}^Kk_K^X=0\,,
 \label{commrelf}
\end{equation}
where $f_{IJ}{}^K$ are structure constants. Multiplying this relation
with $\J\alpha XZ \covder_Z$, and using (\ref{RKisDDK}),
and~(\ref{valuePDk}) gives
\begin{equation}
  2\J\alpha XZ (\covder_Zk_{[I}^Y)(\covder_Y k_{J]}^X) + 2\J\alpha
  XZ\,R_{ZWY}{}^Xk_{[I}^Yk_{J]}^W-4r\nu f_{IJ}{}^K P_K^\alpha =0\,.
 \label{step1equiv}
\end{equation}
The trace that appears in the first term can be evaluated by
using~(\ref{KnormQ2}) and once more (\ref{valuePDk}), while in the second
term we can use the cyclicity condition of the curvature
and~(\ref{nearpropSU2J}) to obtain
\begin{equation}
  -2\nu ^2 \varepsilon ^{\alpha \beta \gamma }P_I^\beta P_J^\gamma + {\cal
  R} _{YW}{}^\alpha k_I^Y k_J^W-\nu f_{IJ}{}^K P_K^\alpha=0\,.
 \label{equivariance}
\end{equation}

We thus found that the moment maps, defined in~(\ref{valuePDk})
satisfy~(\ref{Rkin2}) and~(\ref{equivariance}). The first of these shows
that we can take $\nu =0$ for the hypercomplex or hyperk{\"a}hler manifolds.
Both these two relations vanish identically in this case. However, for
quaternionic-K{\"a}hler and hyperk{\"a}hler manifolds, we can
use~(\ref{RlambdaJ}), and dividing by $\nu $ leads to
\begin{eqnarray}
 J^\alpha{}_{ZW} k^W_I&=& -2 \covder_Z P^\alpha _I\,,
 \label{JkisDP} \\
-2\nu  \varepsilon ^{\alpha \beta \gamma }P_I^\beta P_J^\gamma +
  \frac 12J^\alpha{} _{YW} k_I^Y k_J^W-f_{IJ}{}^K P_K^\alpha&=&0\,.
 \label{equivariance2}
\end{eqnarray}
These equations are thus equivalent to the previous ones for $\nu \neq 0$
if there is a metric. This is thus the quaternionic-K{\"a}hler case, for
which these relations appear already in~\cite{Galicki:1987ja}. But we did
not \emph{derive} these equations for the $\nu =0$ (hyperk{\"a}hler) case.
Rather, the first one is taken as the definition of $P$ for this case.
This equation also follows from supersymmetry requirements, where the
moment map $P^\alpha _I$ is an object that is needed to define the action,
see~(\ref{momentmap}). The moment map is then determined up to constants.
As we saw in section~\ref{isoaction}, the constants are fixed when
conformal symmetry is imposed. Similarly, the second equation appears in
supersymmetry as a requirement, see~(\ref{constraintfP}). For a conformal
invariant theory, the constants in $P_I^\alpha$ are determined and the
moment map again satisfies~(\ref{equivariance2}) automatically due to a
similar calculation as the one that we did above for $\nu \neq 0$. Note,
however, that for the quaternionic manifolds that are not
quaternionic-K{\"a}hler, we can only use~(\ref{Rkin2})
and~(\ref{equivariance}), as~(\ref{JkisDP}) and~(\ref{equivariance2})
need a metric. For hypercomplex manifolds, on the other hand, the moment
maps are not defined.

\section{Examples: hypercomplex group manifolds} \label{app:exHypercplxGr}
In this appendix we illustrate explicit examples of hypercomplex
manifolds. Specifically, we demonstrate the non-vanishing of the
antisymmetric Ricci tensor for some of these manifolds. The examples that
we have in mind are group manifolds, or cosets thereof. These have two
connections preserving the complex structures, one with and one without
torsion. The torsionful connection preserves a metric, which is on the
group manifolds the Cartan-Killing metric. First we consider the generic
setup which has such two connections.

\subsection{Hypercomplex manifolds with metric and torsionful connection}
\label{ss:hctorsion}

We consider a space with a metric $g_{XY}$ and torsionful connection
coefficients
\begin{equation}
\Gamma_{{\pm}YZ}{}^X=\gamma _{YZ} {}^X \pm  T_{YZ}{}^X\,, \label{Gamma+-}
\end{equation}
where $\gamma _{YZ}{}^X$ are the Levi-Civita connection coefficients with
respect to this metric, and where $T_{YZ}{}^X=-T_{ZY}{}^X$ is the torsion.

We assume that there are hypercomplex structures that are covariantly
constant with respect to the connection~(\ref{Gamma+-}). We also assume
the Nijenhuis condition and therefore have an Obata connection $\Gamma
_{XY}{}^Z$. Taking the plus sign in~(\ref{Gamma+-}) we have
\begin{eqnarray}
 0=\covder_X\J\alpha YZ & = & \partial _X\J\alpha YZ-(\gamma +T)_{XY}{}^W
 \J\alpha WZ+
 (\gamma +T)_{XW}{}^Z\J\alpha YW\,, \nonumber\\
   & = & \partial _X\J\alpha YZ-\Gamma _{XY}{}^W\J\alpha WZ+
 \Gamma _{XW}{}^Z\J\alpha YW\,.
\end{eqnarray}
Then the Obata connection can be related to the Levi-Civita connection
and torsion by
\begin{equation}
 \Gamma _{XY}{}^Z=\gamma _{XY}{}^Z+
 \frac16\varepsilon ^{\alpha\beta\gamma}\J\alpha XU\J\beta YV T_{UV}{}^W\J\gamma WZ
 +\frac23\J\alpha {(X}V T_{Y)V}{}^W\J\alpha WZ  \,.
\end{equation}

The antisymmetric part of the Ricci tensor of the Obata connection is
\begin{eqnarray}
R_{[XY]}&=&\partial_{[Y} \Gamma_{X]Z}{}^Z \nonumber\\
&=&\frac23 \J{\alpha}WZ \J{\alpha}{[X}V
\covder_{Y]}T_{ZV}{}^W+\frac23T_{YX}{}^U
\J{\alpha}WZ \J{\alpha}UV T_{ZV}{}^W + \nonumber\\
&& +\covder_{[X}T_{Y]Z}{}^Z+T_{YX}{}^UT_{WU}{}^W\,,
\end{eqnarray}
where $\covder_X$ is the torsionful connection. If the torsion is
covariant constant and traceless, as it is in group manifolds, then
\begin{equation}
  R_{[XY]}=\frac 23 T_{XY}{}^ZV_Z\qquad \mbox{with}\qquad V_Z= \J\alpha ZW T_{WV}{}^U\J\alpha UV\,.
 \label{RiGr}
\end{equation}
This is the only surviving part of the Ricci tensor in hypercomplex
manifolds, and will be used below.

The Nijenhuis condition can be written as a condition on the torsion
(using the metric to lower indices) as~\cite{Spindel:1988sr}
\begin{equation}
  3\J{(\alpha}{[X}U\J{\beta )}YV T_{Z]UV}=\delta ^{\alpha \beta }T_{XYZ}\,.
 \label{Nijenhuisgroup}
\end{equation}
Using the quaternionic algebra $J^1J^2=J^3=-J^2J^1$ and the Nijenhuis
condition for one of the complex structures, one can show that the
contributions from $\alpha =1,2$ and $3$ in~(\ref{RiGr}) are all equal.

\subsection{Group manifolds}

In~\cite{Spindel:1988sr}, 2-dimensional sigma models with extended
supersymmetry on group manifolds were studied. In the case of $N=4$, it
was shown to be possible to construct three globally defined, covariantly
constant complex structures, on certain groups. Using cohomology, one
argues\footnote{We thank George Papadopoulos for pointing this out to
us.} that these manifolds are in fact hypercomplex. For these arguments
one makes use the fact that the second de Rham cohomology vanishes for
all simple groups, whereas K{\"a}hler manifolds have a non-trivial K{\"a}hler
2-form.

%

We will explicitly construct the Ricci tensor on the group manifolds
considered in~\cite{Spindel:1988sr}, and show that there are cases with
non-vanishing Ricci tensor. As this is an antisymmetric tensor, there is
no invariant metric for the Obata connection.

In~\cite{Spindel:1988sr}, the complex structures were first constructed
in one fibre, and then used to form a field of complex structures with
the help of the left- or right-invariant vector fields, giving rise to
($J^{\alpha}_{\pm}$). As the sigma models included an antisymmetric tensor
field in their action, the connection used in the equations of motion had
torsion, which could be written in terms of the structure constants of
the groups.  The connections $\Gamma_\pm$ corresponding to $J^\alpha_{\pm}$
differed in a sign, in the sense of~(\ref{Gamma+-}).  The torsion
$T_{XYZ}$ is completely anti-symmetric, and defined as (denoting the flat
indices on the group manifold with $\hat{\Lambda} ,\hat{\Sigma} , \ldots
$)
\begin{equation}
g^{ZV} T_{VXY}\equiv T_{XY}{}^Z=\frac 12 e_X^{\hat{\Lambda}}
e_Y^{\hat{\Sigma}} f_{\hat{\Lambda}\hat{\Sigma}}^{\hat{\Gamma}}
e_{\hat{\Gamma}}^Z
\end{equation}
where the $e_X^{\hat{\Lambda}}$ are vielbeins, and dual to the left- or
right invariant vector fields. The vielbeins (and the torsion) are
covariantly constant with respect to the connection $\Gamma_{\pm}
{}_{YZ}{}^X$.

We will now construct the vector $V$ of~(\ref{RiGr}) explicitly, using
the connection $\Gamma_+$. This means that the complex structures,
defined in one fibre, define a field of complex structures using the
\emph{left}-invariant vector fields.

A key concept in the construction of hypercomplex group manifolds, are
the so-called stages. This is because the 3 complex structures in fact
act within any such `stage'. A stage consists of a subset of the
generators of a group on which a hypercomplex structure is defined. One
can start from any simple group $G$ to define a stage. One starts by
picking out a highest root $\theta $. One adds $-\theta $, all the roots
that are \emph{not} orthogonal to $\theta $ and two more generators. One
of these is the generator in the Cartan subalgebra (CSA) in the direction
of $\theta $ and $-\theta $. If the subspace of roots orthogonal to
$\theta $ form a root space of dimension $(\rank G-2)$, then the second
one is the other element in the CSA that does not belong to the simple
group defined with the roots orthogonal to $\theta $. This happens only
for $G=\SU(n)$ with $n\geq 3$. In all other cases one has to consider
$G\times \U(1)$ in order to be able to define a hypercomplex structure.
The roots $\theta $ and $-\theta $ and the two generators of the CSA
define an algebra $\SU(2)\oplus\U(1)$. The stage can thus be written as
\begin{equation}
 \SU(2)\oplus \U(1)\oplus  W\,,
 \label{stage}
\end{equation}
where $W$ are all the roots not orthogonal to $\theta $. These form a
`Wolf space'. The Wolf spaces
\begin{equation}
 \begin{array}{ll}
   W=\frac{G}{H\times \SU(2)}\,, & \quad G\neq \SU(n)\,, \\[4pt]
   W=\frac{\SU(n)}{\SU(n-2)\times \SU(2)\times \U(1)}\,, &\quad n\geq 3\,,
 \end{array} \qquad  \dim W=4(\tilde h_g-2)\,,
 \label{Wolf}
\end{equation}
where $\tilde h_g$ is the dual Coxeter number\footnote{Tables are given
in~\cite{Spindel:1988sr}, e.g.\ $\tilde h_g=n$ for $\SU(n)$.} of the
group $G$, are the quaternionic symmetric spaces. So far, we considered
compact groups. The only non-compact groups that are allowed are those
real forms where just the generators in $W$ are non-compact, and all the
others are compact. Hereafter, the group generated by the roots
orthogonal to $\theta$, together with the remaining elements in the
Cartan subalgebra [being $H$ or $\SU(n-2)$ in~(\ref{Wolf})], is used to
construct a new stage in the same way. By this procedure, one constructs
the complex structures in one fibre of the group. For more details we
refer to~\cite{Spindel:1988sr} or~\cite{Joyce:1992}.

We will now give explicitly the hypercomplex structures (in one stage) in
a language adapted to this paper. As we use flat space indices on a Lie
group, these take values in the Lie algebra. The base for our Lie algebra
is taken to be Cartan-Weyl. We will use hatted Greek capitals to denote
all Lie algebra elements. $\theta $ and $-\theta $ are the chosen highest
root and its negative. Greek capital letters denote the positive
generators in $W$. The full set in $W$ consists thus of those indicated by
$\Delta$ and those by $-\Delta $. Small Roman letters $k,\ell $ indicate
elements of the Cartan subalgebra. The full set of generators is thus
\begin{equation}
  \hat{\Delta }=\left\{-\theta ,-\Delta ,k,\Delta ,\theta  \right\}\oplus \mbox{other
  stages}\,,
 \label{indicesDelta}
\end{equation}
where $\Delta $ runs over $2(\tilde h_g-2)$ values and $k=1,2$.

First, it is useful to give some more information about the structure of
the algebra in a stage.  The root vectors are indicated as $\vec \theta $
or $\vec \Delta$ and particular components as $\theta _k$ or $\Delta _k$.
The following properties of structure constants, Cartan-Killing metric and
root vectors are useful:
\begin{eqnarray}
  f_{k,\pm \Delta }^{\pm \Delta}  & = & \pm \Delta _k \,,\quad \qquad f_{k,\pm\theta }^{\pm \theta }=\pm\theta _k\,,\quad
f_{\Delta ,-\Delta }^k= \Delta _k\,,\quad  f_{\theta ,-\theta }^k  =
\theta _k\,, \quad
 f_{\pm\Delta  ,\mp \theta }^{\pm\Delta  \mp \theta}=
\pm  \frac{1}{2x}\alpha _\Delta\,,\nonumber\\
2x^2&\equiv &\tilde h_g=\frac{1}{\vec \theta ^2}\,,\quad \ \ \vec \Delta
\cdot \vec \theta = \frac 12\vec \theta ^2\,,\quad \quad \ \alpha _\Delta
=-\alpha _{\theta -\Delta }=\pm 1\,,\nonumber\\  g_{k,\ell }&=&-\delta
_{k\ell }\vec \theta ^2\,, \qquad
  \ g_{\theta ,-\theta }= g_{\Delta ,-\Delta }= -\vec \theta ^2\,.
 \label{feqns}
\end{eqnarray}
These relations fix a normalization for the generators.

We can now write the non-zero elements of the complex structures as
\begin{equation}
 \begin{array}{lll}
   \J1k\ell   = \varepsilon _{k\ell }\,,\qquad & \J1{\pm\theta  }{\pm\theta  }=\pm
   \rmi\,,\qquad &
\J1{\pm\Delta  }{\pm \Delta}  = \pm \rmi   \\[2mm]
\J2{\pm \theta\mp\Delta   }{\mp \Delta  }  =  \mp \rmi\alpha
_\Delta\,,\qquad &\J2k{\pm \theta }=x\left(\pm \rmi\theta _k-\varepsilon
_{k\ell }\theta_\ell \right)\,, \qquad &\J2{\pm \theta }k=x\left(\pm \rmi
\theta_ k +\varepsilon _{k\ell }\theta
_\ell \right),\\[2mm]
\J3{ \pm \theta\mp\Delta }{\mp\Delta } =   \alpha _\Delta\,,\qquad
&\J3k{\pm \theta }=x \left(\theta _k\pm \rmi\varepsilon _{k\ell
}\theta_\ell\right)\,, \qquad &\J3{\pm \theta }k=x\left(-\theta _k\pm
\rmi \varepsilon _{k\ell }\theta _\ell \right).
  \end{array}
 \label{JonStage}
\end{equation}
These satisfy the Nijenhuis conditions~(\ref{Nijenhuisgroup}).

As written at the end of section~\ref{ss:hctorsion}, we can limit the
calculation of $V$ to the contribution of one of the complex structures.
The torsion is proportional to the structure constants, and as $J^1$ is
diagonal in the roots, the vector $V_{\hat{\Sigma}}$ has only non-zero
components along the Cartan subalgebra:
\begin{equation}
V_{k}=\frac32\J1{k}{\ell } f_{\ell,  \hat{\Delta}}{}^{\hat{\Gamma }}
\J1{\hat{\Gamma }}{\hat{\Delta}}=  3 \varepsilon_{k\ell } \rmi \left(
\theta _\ell +\sum_{\Delta } \Delta _\ell \right) =3 \rmi
\varepsilon_{k\ell }\theta _\ell (\tilde h_g-1) \,. \label{VkGroup}
\end{equation}
Though this is non-zero for all the groups under consideration, the Ricci
tensor is only non-vanishing for $G=\SU(n)$ with $n\geq 3$. Indeed, in
all other cases, the generator corresponding to the index $k$
in~(\ref{VkGroup}) corresponds to the extra $\U(1)$ factor that was added
to $G$, and there are thus no non-vanishing
$R_{XY}=\frac23T_{XY}{}^kV_k$.

The only case in which we find a non-vanishing Ricci tensor, is when the
Wolf space is
\begin{equation}
  W=\frac{\SU(n)}{\SU(n-2)\times \SU(2)\times \U(1)}\,, \qquad  n\geq 3\,,
 \label{WSUn}
\end{equation}
Then the non-vanishing components of the Ricci tensor are of the form
\begin{equation}
  R_{\Delta ,-\Delta }=-R_{-\Delta ,\Delta }=
  \rmi \Delta _k\varepsilon _{k\ell } \theta _\ell (\tilde h_g-1)\,,
 \label{Ricciresult}
\end{equation}
and one can see again that $\Delta _k\varepsilon _{k\ell }\theta _\ell $
vanishes for all other cases than $G=\SU(n)$. In this case, it is simply
a function of $n$.

The group manifolds that have a non-zero Ricci tensor are those that have
a stage with the Wolf spaces~(\ref{WSUn}). Checking the list
in~\cite{Spindel:1988sr}, these are $\SU(2n-1)$, $\SU(2n)\times \U(1)$
(both for $n\geq 2$) and $E_6\times \U(1)^2$. The other cases are Ricci
flat, and one may wonder whether there is a metric whose Levi-Civita
tensor is the Obata connection. This can not be the Cartan-Killing metric
as its Levi-Civita tensor has a non-vanishing Ricci tensor and we just
proved that the Obata connection has vanishing Ricci tensor. One may try
to use cohomological arguments to exclude also any other metric.

After obtaining this result, we can understand it from the geometrical
structure of the stages. We see that the origin of a non-zero Ricci
tensor sits in the fact that there are non-zero roots in the direction of
the $\U(1)$ factor in the decomposition~(\ref{stage}). Thus, we see that
we obtain a non-zero Ricci tensor if this $\U(1)$ is already present in
the structure of the Wolf space, i.e.\ the origin sits in the $\U(1)$
factor in the structure of the coset~(\ref{WSUn}).

\providecommand{\href}[2]{#2}\begingroup\raggedright\endgroup

\end{document}